\documentclass[aps, prd, amsmath, amssymb, amsfonts, floats, %
floatfix, superscriptaddress, nofootinbib, twocolumn, showpacs]%
{revtex4}

\allowdisplaybreaks[2]

\usepackage{graphicx}
\usepackage[usenames, dvipsnames]{color}
\usepackage[colorlinks, pdfborder={0 0 0}, plainpages=false]{hyperref}
\usepackage{breakurl}
\usepackage{amsmath, amssymb, amsfonts}
\usepackage{xspace} 
\usepackage{dcolumn}
\usepackage{bm}
\usepackage{multirow} 

\definecolor{CiteColor}{rgb}{0, 0.5, 0}
\hypersetup{citecolor=CiteColor} %
\definecolor{RefColor}{rgb}{0.55, 0, 0}
\hypersetup{linkcolor=RefColor} %

\usepackage{ulem}
\definecolor {darkgreen}{rgb}{0.2, 0.7, 0.2}

\newcommand{\Caltech}{\affiliation{Theoretical Astrophysics 130-33,
    California Institute of Technology, Pasadena, CA 91125, USA}}
\newcommand{\CITA}{\affiliation{Canadian Institute for Theoretical
    Astrophysics, 60 St. George Street,\\ University of Toronto,
    Toronto, ON M5S 3H8, Canada}}
\newcommand{\Cornell}{\affiliation{Center for Radiophysics and Space
    Research, Cornell University, Ithaca, New York, 14853, USA}}
\newcommand{\Maryland}{\affiliation{Maryland Center for Fundamental
    Physics \& Joint Space-Science Institute,\\ Department of Physics,
    University of Maryland, College Park, MD 20742, USA}}

\newcommand{\OrbitalPhase}{\Phi}

\newcommand{\OrbitalFreq}{\Omega}

\newcommand{\abs}[1]{\left\lvert #1 \right\rvert}
\newcommand{\define}{\equiv}

\newcommand{\MM}{\mathcal{M}}

\newcommand{\Pade}{Pad\'{e}\xspace}

\newcommand{\etal}{et al.}

\renewcommand{\v}{{\bar{v}}}

\begin{document}

\title{Inspiral-merger-ringdown multipolar waveforms of nonspinning
  black-hole binaries using the effective-one-body formalism}

\author{Yi Pan} \Maryland %
\author{Alessandra Buonanno} \Maryland %
\author{Michael Boyle} \Cornell %
\author{Luisa T. Buchman} \Caltech %
\author{Lawrence E. Kidder} \Cornell %
\author{Harald P. Pfeiffer} \CITA %
\author{Mark A. Scheel} \Caltech %

\begin{abstract} 
  We calibrate an effective-one-body (EOB) model to
  numerical-relativity simulations of mass ratios $1, 2, 3, 4$, and
  $6$, by maximizing phase and amplitude agreement of the leading
  $(2,2)$ mode and of the subleading modes $(2,1)$, $(3,3)$, $(4,4)$
  and $(5,5)$.  Aligning the calibrated EOB waveforms and the
  numerical waveforms at low frequency, the phase difference of the
  $(2,2)$ mode between model and numerical simulation remains below 
  $\sim 0.1$ rad throughout the evolution for all mass ratios
  considered. The fractional amplitude difference at peak amplitude of
  the $(2,2)$ mode is $2\%$ and grows to $12\%$ during the
  ringdown. Using the Advanced LIGO noise curve we study the
  effectualness and measurement accuracy of the EOB model, and stress
  the relevance of modeling the higher-order modes for parameter
  estimation.  We find that the effectualness, measured by the
  mismatch, between the EOB and numerical-relativity polarizations
  which include only the $(2,2)$ mode is smaller than $0.2\%$ for
  binaries with total mass $20\mbox{--}200 M_\odot$ and mass ratios $
  1, 2, 3, 4$, and $6$.  When numerical-relativity polarizations
  contain the strongest seven modes, and stellar-mass black holes 
with masses less than $50M_\odot$ are
  considered, the mismatch for mass ratio $6$ ($1$) can be as high as
  $7\%$ ($0.2\%$) when only the EOB $(2,2)$ mode is included, and an
  upper bound of the mismatch is $0.5\%$ ($0.07\%$) when all the four
  subleading EOB modes calibrated in this paper are taken into
  account. For binaries with intermediate-mass black holes with masses
  greater than $50M_\odot$ the
  mismatches are larger. We also determine for which signal-to-noise
  ratios the EOB model developed here can be used to measure binary
  parameters with systematic biases smaller than statistical errors
  due to detector noise.
\end{abstract}

\date{\today}


\maketitle

\section{Introduction}
\label{sec:introduction}

Binary systems composed of black holes and/or neutron stars, spiraling in
toward each other and losing energy through the emission of
gravitational waves, are among the most promising detectable sources
of gravitational waves with the Laser Interferometer
Gravitational-wave Observatory (LIGO)~\cite{Abbott:2007}, Virgo~\cite{Acernese:2008}, 
GEO~\cite{Grote:2008zz}, the Large Cryogenic Gravitational Telescope (LCGT)~\cite{Kuroda:2010}, 
and future space-based detectors.  The
detectors' noise level and the weakness of the waves prevent observing
the waveforms directly. For this reason the search for gravitational
waves from binary systems and the extraction of parameters, such as
the masses and spins, are based on the matched-filtering technique,
which requires accurate knowledge of the waveform of the incoming
signal.

The post-Newtonian (PN) expansion is the most powerful approximation
scheme in analytical relativity capable of describing the two-body
dynamics and gravitational-wave emission of inspiraling compact binary
systems~\cite{Blanchet2006, Sasaki:2003xr, Futamase:2007zz,
  Goldberger:2004jt}.
The PN approach expands the Einstein equations in the ratio of the
characteristic velocity of the binary $v$ to the speed of light or the
characteristic size of the compact body to the relative distance
between the two bodies. However, as the bodies 
approach each other towards merger, we expect the PN expansion to lose accuracy
because the velocity of the bodies approaches the speed of light, and
the relative distance becomes comparable to the size of the compact
body. The difficulty in analytically solving the Einstein equations in
the merger regime lies mainly in its nonlinear structure. Solving the
Einstein equations numerically overcomes this problem.

Prior to the numerical-relativity
breakthroughs~\cite{Pretorius2005a, Baker2006a, Campanelli2006a}, a new
and unique method was proposed in analytical relativity to describe the dynamics and gravitational-wave emission of binary black holes during inspiral,
merger and ringdown: the effective-one-body (EOB)
approach~\cite{Buonanno00, Buonanno99, DJS00, Damour01c, Buonanno06}. This
approach uses the very accurate results of PN theory. However, it does
not use those results in their original Taylor-expanded form (i.e., as
polynomials in $v/c$), but instead in some appropriate resummed form.
In particular, the effective-one-body
approach~\cite{Buonanno00, Damour01c, DJS00, Damour:024009, Barausse:2009xi}
maps the dynamics of two compact objects of masses $m_1$ and $m_2$,
and spins $\mathbf{S}_1$ and $\mathbf{S}_2$, into the dynamics of one
test particle of mass $\mu = m_1\,m_2/(m_1+m_2)$ and spin
$\mathbf{S}_*$ moving in a deformed Kerr metric with mass $M =
m_1+m_2$ and spin $\mathbf{S}_\mathrm{Kerr}$. The deformation parameter
is the symmetric mass ratio $m_1\,m_2/(m_1+m_2)^2$ which ranges
between $0$ (test particle limit) and $1/4$ (equal-mass limit). The
other crucial aspect of the EOB approach is the way it builds the full
waveform, including merger and ringdown. The EOB approach assumes that
the merger is very short in time, although broad in frequency, and
builds the merger-ringdown signal by attaching to the plunge signal a
superposition of quasinormal modes. This match happens at the EOB
light ring (or photon orbit) where the peak of the potential barrier
around the merged black hole sits.

The analyses and theoretical progress made in
Refs.~\cite{Buonanno-Cook-Pretorius:2007,
  Buonanno2007, Pan2007, Boyle2008a, Buonanno:2009qa, Racine2008, Barausse:2009xi, Pan:2009wj, Pan2010hz,
  Damour2007a, DN2007b, DN2008, DIN, Damour2009a, Bernuzzi:2010xj} have
demonstrated that it is possible to devise and calibrate analytical
EOB waveforms for use in detection searches. This is crucial, since
thousands of waveform templates need to be computed to extract the
signal from the noise, an impossible demand for numerical relativity
alone. For example the EOB waveforms calibrated to
numerical-relativity waveforms in Ref.~\cite{Buonanno2007} have been
used in LIGO and Virgo to search for the first time for high-mass
merging black holes~\cite{Abadie:2011kd}.

This paper is a step forward in building more faithful EOB waveforms
to be used for detection and parameter estimation.  We calibrate the
EOB model to accurate numerical-relativity simulations of mass ratios
$1, 2, 3, 4$, and $6$, so that the phase and amplitude agreement of
the leading $(2,2)$ mode and also the subleading modes $(2,1)$,
$(3,3)$, $(4,4)$ and $(5,5)$ are minimized throughout inspiral, merger
and ringdown. The numerical simulations are produced by the
pseudospectral code SpEC of the Caltech-Cornell-CITA collaboration~\cite{Scheel2009, 
Szilagyi:2009qz, Boyle2007, Pfeiffer-Brown-etal:2007, Lindblom2006, Scheel2006, SpECwebsite}
(see particularly Ref.~\cite{Buchman-etal-in-prep} for details).  The waveforms are
extracted as Regge-Wheeler-Zerilli data~\cite{Rinne2008b}, and extrapolated to infinite
extraction radius~\cite{Boyle-Mroue:2008}.  Since the
numerical-relativity modes satisfy the relation $h_{\ell
  m}=(-1)^{\ell}\,h_{\ell\,-m}^*$ with high accuracy, where $*$
denotes complex conjugate, we assume its validity also for the
analytical modes.  As a consequence, any statement in the paper
concerning an $(\ell,m)$ mode automatically holds for its complex
conjugate $(\ell,-m)$ mode. We find that the $(3,2)$ mode has a
distinct feature that currently cannot be accounted for with the EOB
model used in this paper. Moreover, we find that the $(6,6)$ mode (and
very likely other modes with $m=\ell$) can be calibrated in the same
way as the $(4,4)$ and $(5,5)$ modes.  However, we do not consider the
$(6,6)$ mode since, for the range of mass ratios considered here, its
amplitude is much lower than the other subleading-mode amplitude, and
therefore contributes little to the full polarization waveforms.

The paper is organized as follows. In Sec.~\ref{sec:EOB}, we describe
the EOB dynamics, the waveforms and its adjustable parameters. In
Sec.~\ref{sec:calibrations} we discuss the numerical-relativity
simulations produced by the pseudospectral code
SpEC~\cite{Buchman-etal-in-prep} and estimate the phase and amplitude
errors. Then, we calibrate EOB to numerical-relativity modes and
discuss its effectualness and measurement accuracy when searching for
gravitational waves with Advanced LIGO detectors.  In
Sec.~\ref{sec:prevwork} we compare our EOB model and its performance
in matching numerical-relativity results to previous work. We
summarize our main conclusions in Sec.~\ref{sec:conclusions}. In
Appendix~\ref{app:h32} we discuss some interesting features of the
$(3,2)$ mode. In Appendix ~\ref{app:eob} we list several quantities
which enter the EOB waveforms and energy flux.

\section{Effective-one-body model}
\label{sec:EOB}

In Secs.~\ref{sec:EOBdynamics}--\ref{sec:EOBmergerRDwaveform} (see
also Appendix~\ref{app:eob}) we shall discuss in detail all the
building blocks of the EOB dynamics and waveforms, and its adjustable
parameters.  The EOB model used in this paper is presented in a
self-contained way to allow readers to reproduce it if desired.
We note that many important features of the EOB model have
been developed in several
papers~\cite{Buonanno99, DJS00, Buonanno00, Damour03,
  Buonanno-Cook-Pretorius:2007, Buonanno2007, Damour2007, Damour2007a, DN2007b,
  DN2008, Boyle2008a, DIN, Damour2009a, Buonanno:2009qa, Bernuzzi:2010xj}.

If the EOB model were compared to the numerical-relativity simulations
used in this paper \emph{without any calibration}, i.e., at the PN
order currently known, 3PN in the conservative dynamics and 3.5PN in
the radiation-reaction sector, we would find at merger a phase
difference for the $(2,2)$ mode of up to $3.6\,\text{rad}$ over the
mass-ratio range $q = 1, 2, 3, 4, 6$.  Moreover, the EOB amplitude
would peak around $30M$ before the numerical-relativity peak, with a
fractional amplitude difference at the peak of $\sim 8\%$. A
straightforward way of reducing the differences is to insert in the
dynamics and radiation-reaction force higher-order (pseudo) PN terms
(or EOB adjustable parameters) and \textit{calibrate} them to the
numerical results.  The advantage of the EOB approach is that the
dynamics and radiation-reaction force (and modes) are written in a way
which isolates the crucial functions that determine the evolution. As
we shall see below, these functions are the EOB radial potential
$A(r)$, or time-time component of the EOB metric, and some phase and
amplitude functions appearing in the EOB (factorized) gravitational modes.

\subsection{Effective-one-body dynamics}
\label{sec:EOBdynamics}

We set $M=m_1+m_2$, $\mu = m_1\,m_2/M = \nu \,M$, $q = m_1/m_2$, and
use natural units $G=c=1$. The EOB effective metric
reads~\cite{Buonanno99}
\begin{equation}
  ds_\mathrm{eff}^2 = -A(r)\,dt^2 + \frac{D(r)}{A(r)}\,dr^2 +
  r^2\,\Big(d\Theta^2+\sin^2\Theta\,d\OrbitalPhase^2\Big) \,,
  \label{eq:EOBmetric}
\end{equation}
where we use dimensionless polar coordinates $(r,\OrbitalPhase)$ and
their conjugate momenta $(p_r,p_\OrbitalPhase)$. Replacing the radial
momentum $p_r$ with $p_{r_*}$ which is the conjugate momentum to the
EOB \textit{tortoise} radial coordinate $r_*$,
\begin{equation}
  \frac{dr_*}{dr}=\frac{\sqrt{D(r)}}{A(r)}\,,
\end{equation}
we obtain the EOB effective
Hamiltonian~\cite{Buonanno99, DJS00, Damour:2007cb}
\begin{multline}
  \label{eq:genexp}
  H^\mathrm{eff}(r,p_{r_*},p_\OrbitalPhase) \define \mu\,\widehat{H}^\mathrm{eff}(r,p_{r_*},p_\Phi)  \\
  = \mu\,\sqrt{p^2_{r_*}+A (r) \left[ 1 +
      \frac{p_\OrbitalPhase^2}{r^2} +
      2(4-3\nu)\,\nu\,\frac{p_{r_*}^4}{r^2} \right]} \,,
\end{multline}
where we have neglected the factor $D(r)^2/A(r)^4$ in front of the
term $p_{r_*}^4$ which would introduce PN terms higher than 3PN order,
but more importantly would cause the EOB gravitational frequency to
grow too quickly near merger.

The real EOB Hamiltonian reads~\cite{Buonanno99}
\begin{multline}
  \label{himpr}
  H^\mathrm{real}(r,p_{r_*},p_\OrbitalPhase) \define \mu\hat{H}^\mathrm{real}(r,p_{r_*},p_\Phi)  \\
  = M\,\sqrt{1 + 2\nu\,\left ( \frac{H^\mathrm{eff} - \mu}{\mu}\right )}
  -M\,.
\end{multline}
The Taylor approximants to the coefficients $A(r)$ and $D(r)$ can be
written as~\cite{Buonanno99, DJS00}
\begin{equation}
  \label{coeffA}
  A_{k}(r) = \sum_{i=0}^{k+1} \frac{a_i(\nu)}{r^i}\,, \quad D_{k}(r) =
  \sum_{i=0}^k \frac{d_i(\nu)}{r^i}\,.
\end{equation}
The functions $A(r)$, $D(r)$, $A_k(r)$ and $D_k(r)$ all depend on the
symmetric mass ratio $\nu$ through the $\nu$-dependent coefficients
$a_i(\nu)$ and $d_i(\nu)$ [see Eqs.~(47) and (48) in
Ref.~\cite{Boyle2008a}]. The functions $A_k(r)$ and $D_k(r)$ are
currently known through 3PN order, i.e., $k=3$.  During the last
stages of inspiral and plunge, the EOB dynamics can be adjusted closer
to the numerical simulations by including in the radial potential
$A(r)$ a few adjustable parameters of the EOB dynamics. Notably, the 4PN
coefficient
$a_5(\nu)$~\cite{Damour03, Buonanno2007, Boyle2008a, Buonanno:2009qa, Damour2007a, DN2007b, DN2008}
and even the 5PN coefficient
$a_6(\nu)$~\cite{Damour2009a}.\footnote{The radial potential $A(r)$
  may contain logarithmic terms at 4PN and 5PN
  orders~\cite{Damour:2009sm, Blanchet:2010zd} which we do not try to
  model here.}

To enforce the presence of the EOB innermost stable circular orbit
(ISCO), Ref.~\cite{DJS00} suggested using the \Pade expansion of the
function $A(r)$. For $A(r)$ we employ the \Pade expression $A_5^1(r)$
at 5PN order, while for $D(r)$ we use the \Pade expression $D_3^0(r)$
at 3PN order. We could also introduce EOB adjustable parameters at 4PN
and 5PN order in $D(r)$, say $d_4(\nu)$ and $d_5(\nu)$. However, this
modification would affect mainly the radial motion [see
Eq.~\eqref{eq:eobhamone} below] which becomes important only at the
very end of the evolution.  For the EOB model developed in this paper
we find that these other adjustable parameters are not needed. The
quantity $D_3^0(r)$ reads
\begin{equation}
  D_3^0(r)=\frac{r^{3}} {(52\,\nu - 6\,\nu^{2}) + 6\, \nu\, r +
    r^{3}}\,,
\end{equation}
while $A_5^1(r)$ reads
\begin{equation}
  A_5^1(r) = \frac{\mathrm{Num}(A_5^1)}{\mathrm{Den}(A_5^1)}\,,
\end{equation}
with
\begin{eqnarray} \mathrm{Num}(A_5^1) &=& r^4\,\left[-64 +
    12\,a_4+4\,a_5+a_6+64 \nu-4 \nu ^2 \right]
  \nonumber\\
  &+& r^5\,\left[32-4\,a_4-a_5-24 \nu \right]\,,
\end{eqnarray}
and
\begin{eqnarray}
  && \mathrm{Den}(A_5^1) = 4\,a_4^2+4\,a_4\,a_5+a_5^2-a_4\,a_6+16\,a_6+ (32\,a_4 \nonumber \\
  &&\qquad + 16\,a_5-8\,a_6)\,\nu + 4\,a_4\,\nu^2+32\,\nu^3 + r\,\left[4\,a_4^2+a_4\,a_5 \right. 
  \nonumber\\
  &&\qquad \left. +16\,a_5+8\,a_6+(32\,a_4 -2\,a_6)\,\nu + 32\,\nu^2+8\,\nu^3\right] 
  \nonumber\\
  &&\qquad + r^2\,\left[16\,a_4+8\,a_5+4\,a_6+(8\,a_4+2\,a_5)\,\nu +32\,\nu^2\right] \nonumber \\
  &&\qquad + r^3\,\left[8\,a_4+4\,a_5+2\,a_6+32\,\nu-8\,\nu^2\right] 
  \nonumber\\
  &&\qquad + r^4\,\left[4\,a_4+2\,a_5+a_6+16\,\nu-4\,\nu^2\right] \nonumber \\
  &&\qquad + r^5\,\left[32-4\,a_4-a_5-24\,\nu\right]\,,
\end{eqnarray}
where $a_4=[94/3-(41/32)\,\pi^2]\,\nu$ and to ease the notation we
have omitted the $\nu$ dependence of $a_5$ and $a_6$ in the
expressions above. The quantities $a_5$ and $a_6$ are the
adjustable parameters of the EOB dynamics~\cite{Buonanno:2009qa} (see
Table~\ref{tab:adjparams}). They will be determined below when
calibrating the EOB to numerical-relativity waveforms. Their explicit
expressions are given in Eq.~\eqref{a5a6cal}.

\begin{table}
  \begin{tabular}{c c}
    \hline\hline EOB dynamics & EOB waveform\\
    adjustable parameters \qquad & \qquad adjustable parameters\\\hline
    $a_5, a_6$ & $\Delta t_\mathrm{match}^{\ell m}$ \\
    & $\rho_{\ell m}^{(p)}$\\
    & $\delta_{\ell m}^{(q)}$\\
    & $\omega^\mathrm{pQNM}_{\ell m}$\\\hline\hline
  \end{tabular}
  \caption{\label{tab:adjparams} Summary of adjustable parameters of the EOB model 
    considered in this paper. We notice that to calibrate the EOB $(2,2)$ mode, we \emph{only} need 
    $a_5, a_6$ and $\Delta t_\mathrm{match}^{22}$. To calibrate each subleading mode $(2,1)$, $(3,3)$, 
    $(4,4)$, and $(5,5)$, we need four adjustable parameters.
    The values of the adjustable parameters used 
    in this paper are given 
    in Eqs.~\eqref{a5a6cal} to~\eqref{deltacal} and~\eqref{modecal}.}
\end{table}

The EOB Hamilton equations are written in terms of the reduced, i.e.,
dimensionless quantities $\widehat{H}^\mathrm{real}$ [defined in
Eq.~\eqref{himpr}]~\cite{Buonanno00}.  They read\footnote{We notice
  that the second term on the right-hand side of
  Eq.~\eqref{eq:eobhamthree} is generated when taking the nonspinning
  limit of the spinning EOB model of Ref.~\cite{Buonanno06}}
\begin{subequations} \label{eq-eob}
  \begin{align}
    \frac{dr}{d \widehat{t}} &=
    \frac{A_5^1(r)}{\sqrt{D_3^0(r)}}\frac{\partial
      \widehat{H}^\mathrm{real}} {\partial
      p_{r_*}}(r,p_{r_*},p_\OrbitalPhase)\,,
    \label{eq:eobhamone} \\
    \frac{d \OrbitalPhase}{d \widehat{t}} &= \frac{\partial
      \widehat{H}^\mathrm{real}} {\partial
      p_\OrbitalPhase}(r,p_{r_*},p_\OrbitalPhase)\,,
    \label{eq:eobhamtwo}\\
    \frac{d p_{r_*}}{d \widehat{t}} &=
    -\frac{A_5^1(r)}{\sqrt{D_3^0(r)}}\,\frac{\partial \widehat{H}^\mathrm{
        real}} {\partial r}(r,p_{r_*},p_\OrbitalPhase) +{}^\mathrm{
      nK}\widehat{\cal F}_\OrbitalPhase \,
    \frac{p_{r_*}}{p_\OrbitalPhase}\,, \label{eq:eobhamthree}\\
    \frac{d p_\OrbitalPhase}{d \widehat{t}} &= {}^\mathrm{
      nK}\widehat{\cal F}_\OrbitalPhase\,,
    \label{eq:eobhamfour}
  \end{align}
\end{subequations}
with the definition $\widehat{\OrbitalFreq}\define d \OrbitalPhase/d
\widehat{t} \define M\Omega$.  The initial conditions for the EOB
Hamilton equations will be discussed in Sec.~\ref{sec:eobinit}.
Furthermore, for the $\OrbitalPhase$ component of the
radiation-reaction force we use a non-Keplerian (nK)
radiation-reaction force:\footnote{Note that Eq.~\eqref{RadReacForce}
  is only \emph{implicitly} non-Keplerian, in contrast to similar
  expressions in other papers which explicitly introduce non-Keplerian
  terms.  In this case, the non-Keplerian behavior is hidden in the
  wave amplitudes $h_{\ell m}$, as described in the following
  section.}
\begin{equation}\label{RadReacForce} {}^\mathrm{nK}\widehat{\cal
    F}_\OrbitalPhase = -\frac{1}{\nu\,v_\Omega^3}\,\frac{dE}{dt}\,,
\end{equation}
where $v_\Omega \define \widehat{\OrbitalFreq}^{1/3}$, and $dE/dt$ is
the gravitational-wave energy flux for quasi-circular orbits obtained
by summing over the gravitational-wave modes $(l,m)$.  We use
\begin{equation}\label{resflux}
  \frac{dE}{dt}=\frac{v_\Omega^6}{8\pi}\,
  \sum_{\ell=2}^7\,\sum_{m=\ell-2}^{\ell}\, m^2\, \abs{
    \frac{\mathcal{R}}{M}\, h_{\ell m}}^2~.
\end{equation}
Note that because $\abs{h_{\ell, m}}=\abs{h_{\ell,-m}}$, we extend the
sum over positive $m$ modes only.  Moreover, for the cases studied in
this paper, including more modes in the summation has a negligible
effect on the energy flux. Specifically, if we sum $\ell$ through
$\ell=8$ and sum $m = 0, \dots, \ell$, the gravitational-wave phase of
the EOB $(2,2)$ mode changes by less than $0.01$ rad at merger, which
is negligible compared to the phase error of numerical-relativity
modes.  We find that the dominant computational cost in generating the
EOB waveforms is the calculation of the energy flux. The choice of
modes $(\ell,m)$ in Eq.~\eqref{resflux} saves us about a third of the
computational time when compared to the case where the sum extends up
to $\ell =8$, and runs over $m = 0, \dots, \ell$.  The explicit
expression of the modes $h_{\ell m}$ is given below, in
Secs.~\ref{sec:EOBinspiralwaveforms} and
~\ref{sec:EOBmergerRDwaveform}.

\subsection{EOB waveform: Inspiral \& Plunge}
\label{sec:EOBinspiralwaveforms}

Having described the inspiral and plunge dynamics, we now turn to the
gravitational-wave modes $h_{\ell m}$. The latter can be employed to
compute consistently the inspiral dynamics through the
radiation-reaction force~\cite{Damour2009a} [see Eq.~\eqref{resflux}].
The inspiral and plunge EOB modes are given by
\begin{equation}\label{hip}
  h^\mathrm{insp-plunge}_{\ell m} = h^\mathrm{F}_{\ell m}\,N_{\ell m}\,,
\end{equation}
where the $N_{\ell m}$ describe effects that go beyond the
quasi-circular assumption and will be defined below [see
Eq.~\eqref{Nlm}], and the $h^\mathrm{F}_{\ell m}$ are the factorized
resummed modes. In the nonspinning case,
Refs.~\cite{Damour2009a, Buonanno:2009qa} have shown that the resummed,
factorized modes proposed in Ref.~\cite{DIN} are in excellent
agreement with the numerical waveforms~\cite{DN2008, Scheel2009}. We have~\cite{DIN},
\begin{equation}\label{hlm}
  h^\mathrm{F}_{\ell m}=h_{\ell m}^{(N,\epsilon)}\,\hat{S}_\mathrm{
    eff}^{(\epsilon)}\, T_{\ell m}\, e^{i\delta_{\ell
      m}}\,\left(\rho_{\ell m}\right)^\ell\,,
\end{equation}
where $\epsilon$ denotes the parity of the multipolar waveform. In the
circular-orbit case, $\epsilon$ is the parity of $\ell+m$:
\begin{equation}
\epsilon = \begin{cases} 0\,, & \ell+m\mbox{ is even} \\
 1\,, & \ell+m\mbox{ is odd} \end{cases}\,.
\end{equation}
The leading term in Eq.~\eqref{hlm}, $h_{\ell
  m}^{(N,\epsilon)}$, is the Newtonian contribution
\begin{equation}\label{hlmNewt}
  h_{\ell m}^{(N,\epsilon)}=\frac{M\nu}{\cal{R}}\, n_{\ell
    m}^{(\epsilon)}\, c_{\ell+\epsilon}(\nu)\, V^{\ell}_\Phi\,
  Y^{\ell-\epsilon,-m}\,\left(\frac{\pi}{2},\Phi\right)\,,
\end{equation}
where $\cal{R}$ is the distance from the source; the $Y^{\ell
  m}(\Theta,\Phi)$ are the scalar spherical harmonics; the functions
$n_{\ell m}^{(\epsilon)}$ and $c_{\ell+\epsilon}(\nu)$ are explicitly
given in Appendix~\ref{app:eob} [see Eqs.~\eqref{n0n1} and~\eqref{cl}].
Moreover, for reasons that will be explained in
Sec.~\ref{sec:eobneq22}, we choose
\begin{subequations}
  \label{Veq}
  \begin{eqnarray}
    V^{\ell}_\Phi &=& v_\Phi^{(\ell+\epsilon)} \qquad \qquad \; (\ell, m) \neq (2, 1)\,, (4, 4)\,,\\
    V^{\ell}_\Phi &=& \frac{1}{r_\Omega}\, v_\Phi^{(\ell+\epsilon-2)} \qquad (\ell, m) = (2, 1)\,, (4, 4)\,;
  \end{eqnarray}
\end{subequations}
\begin{figure}
  \includegraphics[scale=0.53]{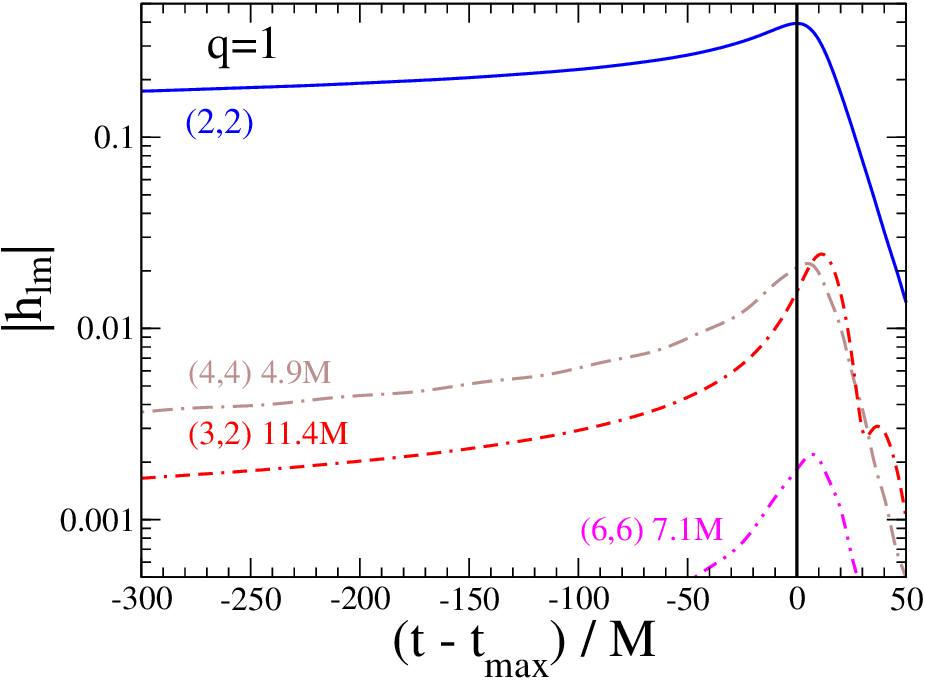} \\[1em]
  \includegraphics[scale=0.53]{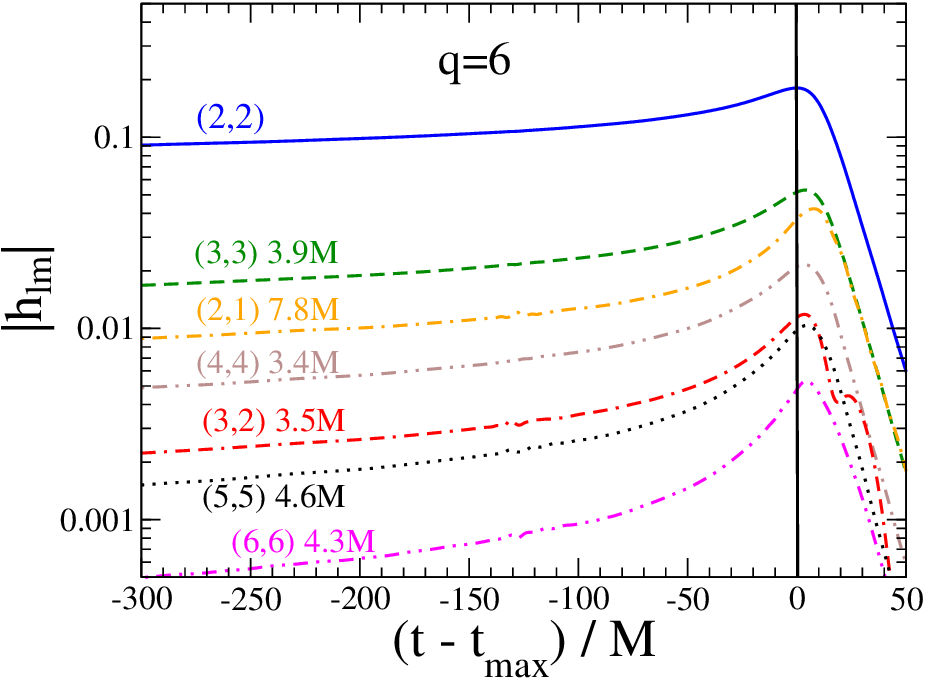}
  \caption{\label{fig:hierarchy} Amplitude of extrapolated
    numerical-relativity waveforms for the dominant modes. The
    two panels from top to bottom are for mass ratios $q=1$ and $q=6$,
    respectively. Each curve is labeled with its respective $(l,m)$ mode, and the time-delay $\Delta t_\mathrm{
      peak}^{\ell m}$ between the extrema of
    $|h_{22}|$ and $|h_{lm}|$. The horizontal axis measures the time-difference to the peak of $|h_{22}|$.}
\end{figure}
with $v_\Phi$ and $r_\Omega$ defined by~\cite{Boyle2008a}
\begin{equation}
  \label{vPhi}
  v_\OrbitalPhase \equiv \widehat{\OrbitalFreq}\, r_\Omega \equiv
  \widehat{\OrbitalFreq}\, r\,[\psi(r, p_\OrbitalPhase)]^{1/3}
\end{equation}
and
\begin{equation}
  \psi(r, p_\Phi)=\frac{2\left \{1+2\nu\,\left
        [\sqrt{A_5^1(r)\,(1+{p_\Phi^2}/{r^2})}-1\right ]\right
    \}}{r^2\, dA_5^1(r)/dr}\,.
\end{equation}
The quantity $p_\Phi$ in the above equation is the dynamical $p_\Phi$
being used in the evolution, contrary to the choice made in
Ref.~\cite{Boyle2008a} where $p_\Phi$ was chosen to satisfy the
circular-orbit condition.  The function $\hat{S}_\mathrm{
  eff}^{(\epsilon)}$ in Eq.~\eqref{hlm} is an effective source term
that in the circular-motion limit contains a pole at the EOB light
ring. It is given in terms of the EOB dynamics as
\begin{equation}
  \hat{S}_\mathrm{eff}^{(\epsilon)}(r, p_{r_*}, p_\Phi) =
  \begin{cases}
    \hat{H}^\mathrm{eff}(r, p_{r_*}, p_\Phi)\,, & \epsilon = 0\,, \\
    \hat{L}_\mathrm{eff}=p_\Phi\, v_\Omega\,, & \epsilon = 1\,,
  \end{cases}
\end{equation}
where $\hat{H}^\mathrm{eff}(r, p_{r_*}, p_\Phi)$ can be read from
Eq.~\eqref{eq:genexp}.  The factor $T_{\ell m}$ in Eq.~\eqref{hlm}
resums the leading order logarithms of tail effects, it reads
\begin{equation}\label{eq:tailterm}
  \begin{split}
    T_{\ell m} &=\frac{\Gamma(\ell+1-2i\, m\, H^\text{real}\, \Omega)}
    {\Gamma(\ell+1)}\, \exp \left[ \pi\, m\,\Omega\, H^\text{real}
    \right] \\ & \qquad \times \exp \left[
      2i\, m\,\Omega\, H^\text{real}\, \log(2\, m\,\Omega\, r_0) \right]~,
  \end{split}
\end{equation}
where $r_0=2M/\sqrt{e}$~\cite{Pan2010hz} and $H^\mathrm{real}$ can be
read from Eq.~\eqref{himpr}.

The factor $e^{i\delta_{\ell m}}$ in Eq.~\eqref{hlm} is a phase
correction due to subleading order logarithms, while the factor
$(\rho_{\ell m})^{\ell}$ in Eq.~\eqref{hlm} collects the remaining PN
terms. The full expressions for $\delta_{\ell m}$ and $\rho_{\ell m}$ are given in Appendix~\ref{app:eob}. To improve the
agreement with the numerical-relativity higher modes, we introduce and
calibrate in $\delta_{\ell m}$ and $\rho_{\ell m}$ a few higher-order,
yet unknown, PN terms (see also
Refs.~\cite{Yunes:2009ef, Yunes:2010zj}). Details of this are given
below [see Eqs.~\eqref{rhocal} and~\eqref{deltacal}], so here we merely remark that in the spirit of Ref.~\cite{Buonanno:2009qa}, those coefficients should be considered adjustable parameters of the EOB waveform (see
Table~\ref{tab:adjparams}). The introduction of these higher-order PN
terms in the modes $\neq (2, 2)$ is not surprising, because these modes
are known at lower PN order than the $(2, 2)$ mode. We note that 
the adjustable parameters in $\delta_{\ell m}$ and $\rho_{\ell m}$ will be used 
only to improve the agreement between the EOB and numerical-relativity modes. 
They will not be included in the energy flux entering the dynamics through Eq.~\eqref{resflux}.

Finally, the function $N_{\ell m}$ entering Eq.~\eqref{hip} reads
\begin{equation}
  \label{Nlm}
  \begin{split}
    N_{\ell m} &= \left [ 1 + \frac{p_{r^*}^{2}}{(r
        \,\hat{\Omega})^{2}} \, \left( a^{h_{\ell m}}_{1}\,
    + \frac{a^{h_{\ell m}}_{2}}{r} + \frac{a^{h_{\ell
          m}}_{3}}{r^{3/2}} \right) \right ] 
  \\ &\quad \times \exp\left[i\,\left(b^{h_{\ell m}}_{1}\,\frac{p_{r^*}}{r \,\hat{\Omega}}+b^{h_{\ell m}}_{2}\,\frac{p_{r^*}^3}{r \,\hat{\Omega}}\right)\right]\,,
  \end{split}
\end{equation}
where the quantities $a^{h_{\ell m}}_{i}$ and $b^{h_{\ell m}}_{i}$ are
the non-quasicircular (NQC) orbit
coefficients~\cite{Damour2007, Damour2007a, DN2007b, DN2008,
  Buonanno:2009qa, Pan:2009wj, Bernuzzi:2010xj}.

To better understand how we fix the parameters $a^{h_{\ell m}}_{i}$
and $b^{h_{\ell m}}_{i}$, we plot in Fig.~\ref{fig:hierarchy} the
amplitudes of the dominant numerical $h_{\ell m}$. As already
observed in Refs.~\cite{Buonanno2007, Schnittman2007, Baker2008a,
  Bernuzzi:2010ty, Bernuzzi:2010xj}, the peaks of the modes occur at
different times. In Fig.~\ref{fig:hierarchy} we have indicated these
times relative to the peak of $h_{22}$. Our goal is to model the EOB
modes, through the parameters $a^{h_{\ell m}}_{i}$ and $b^{h_{\ell
    m}}_{i}$, in such a way to (i) reproduce the \emph{shape} of the
numerical-relativity amplitudes close to the peak, and (ii) preserve
the time differences between the modes.

\begin{figure*}
  \includegraphics[scale=0.38]{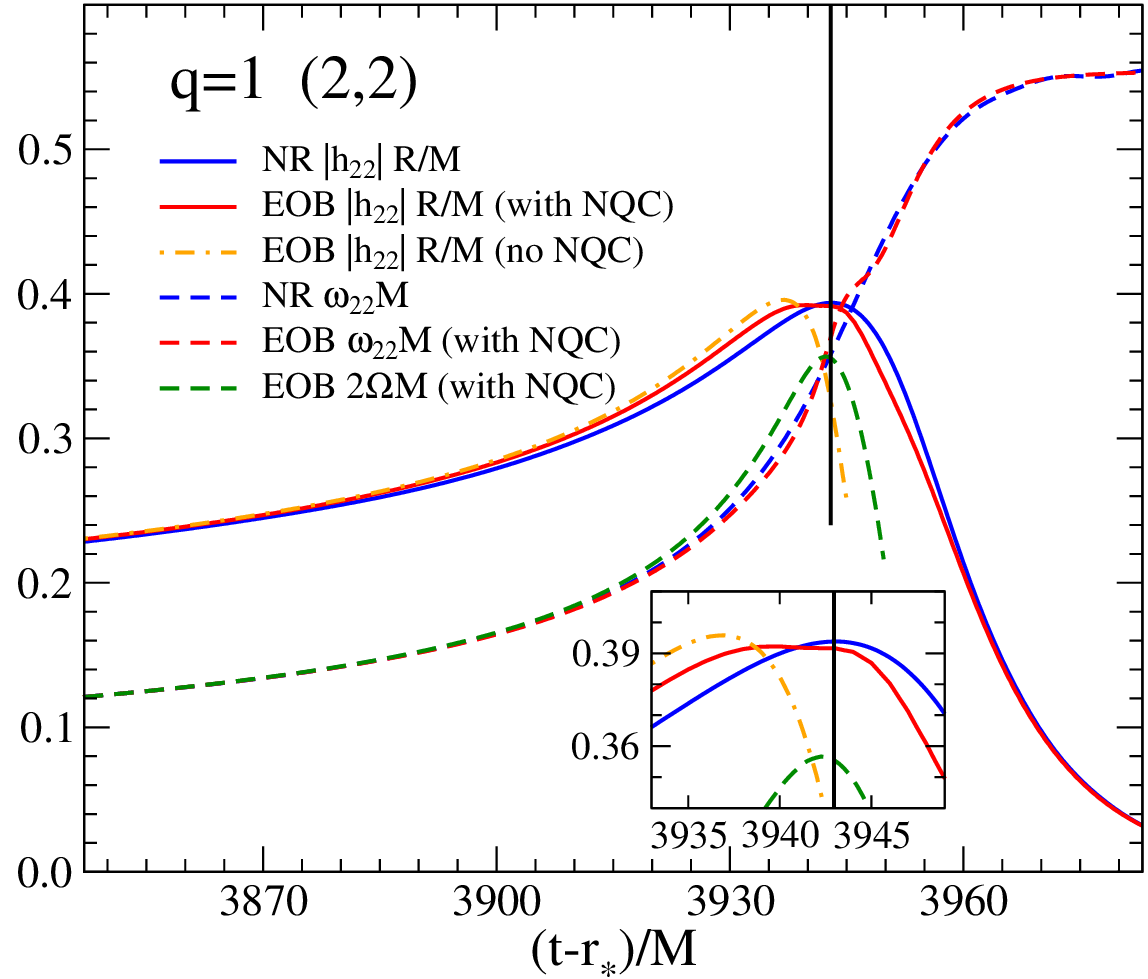} \qquad \qquad
  \includegraphics[scale=0.38]{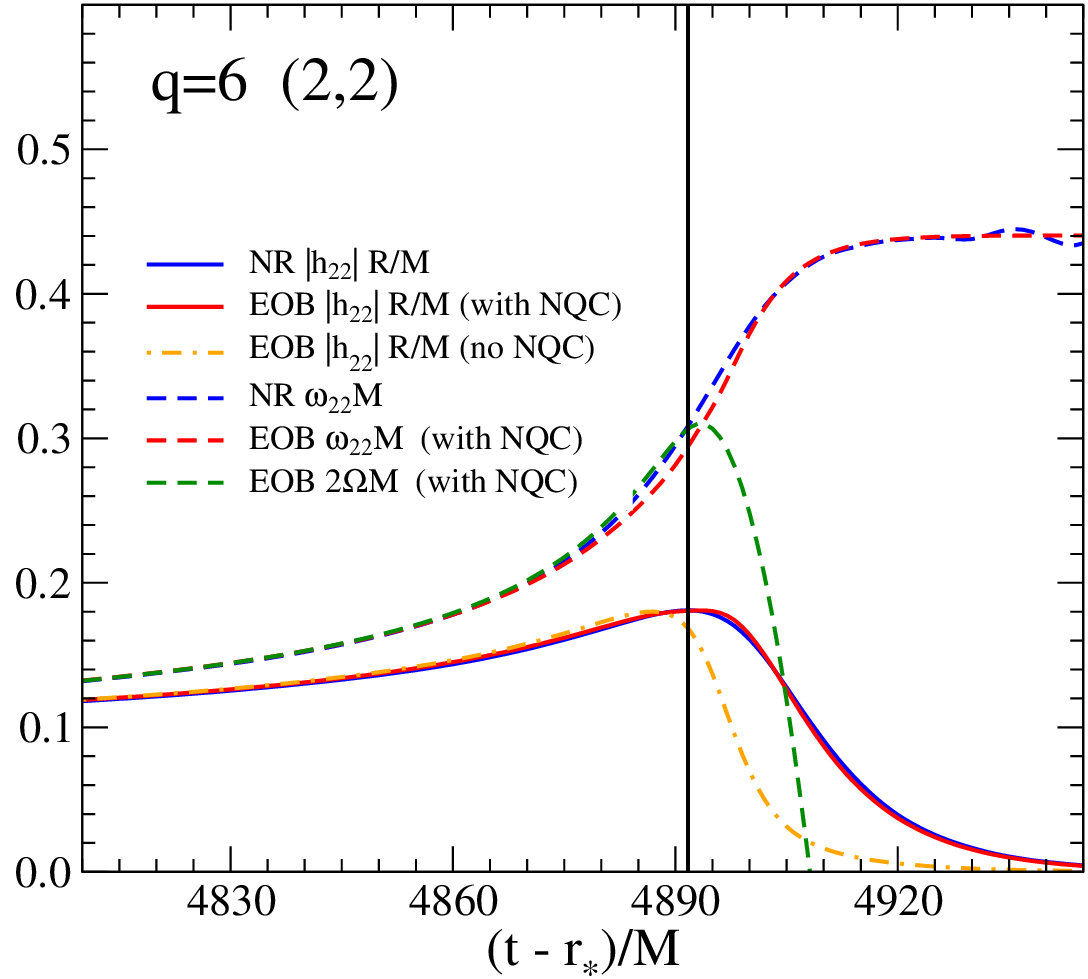}
  \caption{ We compare the numerical-relativity and EOB $h_{22}$
    amplitudes with and without the NQC corrections $N_{\ell m}$ given
    in Eq.~\eqref{Nlm}. We also plot the numerical and EOB
    gravitational frequency of the $(2, 2)$ mode and twice the EOB
    orbital frequency. The left panel refers to $q = 1$ and the right
    panel to $q =6$. The horizontal axis is the retarded time in the
    numerical-relativity simulation. The vertical lines mark the peaks
    of the numerical-relativity $h_{22}$
    amplitudes.\label{fig:h22AmpOmega}}
\end{figure*}

So, for each mode, we fix the three parameters $a_i^{h_{\ell m}}$ in
the amplitude and the two coefficients $b_i^{h_{\ell m}}$ in the phase
by requiring that the peaks of the numerical and EOB $h_{\ell m}$,
as well as their frequencies at the peaks, coincide.\footnote{During
  the course of this work we noticed that Ref.~\cite{Bernuzzi:2010xj}
  had independently developed the same procedure of us in fixing the
  NQC phase coefficients using $\omega^\mathrm{NR}_{\ell m}$ and
  $\dot{\omega}^\mathrm{NR}_{\ell m}$.  However,
  Ref.~\cite{Bernuzzi:2010xj} computes those numerical quantities at
  the light-ring position $3 M$ instead of the peak of the $(\ell, m)$
  modes as we do. Reference~\cite{Bernuzzi:2010xj} focused on the EOB
  modeling in the extreme mass-ratio limit.}  Specifically, we impose
five conditions listed below for each mode.
\begin{enumerate}
 \item The time at which the EOB $h_{22}$ reaches its peak should
  coincide with the time at which the EOB orbital frequency $\Omega$
  reaches its peak. We denote this time with $t_\mathrm{
    peak}^{\Omega}$. It was
  observed~\cite{Damour2009a, Buonanno:2009qa, Pan2010hz} that, once
  the EOB and numerical phases are aligned at low frequency and
  calibrated, the time at which the numerical $h_{22}$ reaches its
  peak coincides with the EOB light-ring time. Moreover, the latter
  occurs immediately ($<1M$ in time) after the peak of $\Omega$.  The
  peaks of higher-order numerical modes differ from the peak of the
  numerical $h_{22}$ mode by a few $M$ in time. We define this time
  difference as
  \begin{eqnarray}
    \Delta t_\mathrm{peak}^{\ell m} &=& t_\mathrm{peak}^{\ell m} - t_\mathrm{peak}^{22}
    \nonumber\\
    &=& t_{d|h^\mathrm{NR}_{\ell m}|/dt=0} - t_{d|h^\mathrm{NR}_{22}|/dt=0}\,,
  \end{eqnarray}
  and require that the peaks of the EOB $h_{\ell m}$ occur at the time
  $t_\mathrm{peak}^{\Omega}+\Delta t_\mathrm{peak}^{\ell m}$.
 \item The peak of the EOB $h_{\ell m}$ should have the same amplitude
  as the peak of the numerical $h_{\ell m}$, that is
  \begin{equation}
    \left|h_{\ell m}^\mathrm{EOB}(t_\mathrm{peak}^{\Omega}+\Delta t_\mathrm{
        peak}^{\ell m})\right|=\left|h_{\ell m}^\mathrm{NR}(t_\mathrm{
        peak}^{\ell m})\right|.
  \end{equation}
 \item The peak of the EOB $h_{\ell m}$ should have the same
  second-order time derivative as the peak of the numerical $h_{\ell
    m}$, that is
  \begin{equation}
    \left. \frac{d^2\left|h_{\ell m}^\mathrm{
            EOB}\right|}{dt^2}\right|_{t_\mathrm{peak}^{\Omega}+\Delta
      t_\mathrm{peak}^{\ell m}} = \left. \frac{d^2\left|h_{\ell m}^\mathrm{
            NR}\right|}{dt^2} \right|_{t_\mathrm{peak}^{\ell m}}.
  \end{equation}
  This condition guarantees that the local extremum of $|h_{\ell
    m}^\mathrm{EOB}|$ at $t=t_\mathrm{peak}^{\Omega}+\Delta t_\mathrm{
    peak}^{\ell m}$ is a local maximum.
 \item The frequency of the numerical and EOB $h_{\ell m}$ waveforms
  should coincide at their peaks, that is
  \begin{equation}
    \omega^\mathrm{EOB}_{\ell m}(t_\mathrm{peak}^{\Omega}+\Delta t_\mathrm{
      peak}^{\ell m})=\omega^\mathrm{NR}_{\ell m}(t_\mathrm{peak}^{\ell m})
  \end{equation}
 \item Time derivative of the frequency of the numerical and EOB
  $h_{\ell m}$ waveforms should coincide at their peaks, that is
  \begin{equation}
    \dot{\omega}^\mathrm{EOB}_{\ell m}(t_\mathrm{peak}^{\Omega}+\Delta
    t_\mathrm{peak}^{\ell m})=\dot{\omega}^\mathrm{NR}_{\ell m}(t_\mathrm{
      peak}^{\ell m})
  \end{equation}
\end{enumerate}
In Sec.~\ref{sec:NRwaveforms} we shall find that the functions $\Delta
t_\mathrm{peak}^{\ell m}$, $|h_{\ell m}^\mathrm{NR}(t_\mathrm{peak}^{\ell
  m})|$, $\left. d^2|h_{\ell m}^\mathrm{NR}|/dt^2 \right|_{t_\mathrm{
    peak}^{\ell m}}$, $\omega^\mathrm{NR}_{\ell m}(t_\mathrm{peak}^{\ell
  m})$, and $\dot{\omega}^\mathrm{NR}_{\ell m}(t_\mathrm{peak}^{\ell m})$
are reasonably approximated by smooth functions of $\nu$ (see
Table~\ref{tab:NRnufit}).

The NQC coefficients $a_i^{h_{\ell m}}$ are calculated within the EOB
model using the fitting formulas in Table~\ref{tab:NRnufit}.  The
calculation involves a computationally expensive iterative procedure.
Basically, in each round of the iteration, $a_i^{h_{\ell m}}$ and
$b_i^{h_{\ell m}}$ are calculated to satisfy the five conditions
listed above.  The amplitude
corrections $a_i^{h_{\ell m}}$ then enter the dynamics through the
energy flux given in Eq.~\eqref{resflux}.  The new EOB dynamics
generate new modes and, thus, new $a_i^{h_{\ell m}}$. We stop the
iteration when $a_i^{h_{\ell m}}$ converge on successive runs.  In
this paper, for technical convenience, we include \emph{only}
$a_i^{h_{22}}(\nu)$ in the energy flux \eqref{resflux} and ignore the
effect of higher-order-mode NQC corrections on the inspiral
dynamics. However, we do include the higher-order-mode NQC corrections
when building the EOB waveforms and compare them to the
numerical-relativity ones. Neglecting higher-order-mode NQC
corrections in the energy flux \eqref{resflux} is insignificant for
three reasons: The higher-order-mode contribution to the energy flux
is about an order of magnitude smaller than that of the dominant
$(2, 2)$ mode; the NQC corrections in the amplitude are a relatively
small correction, typically $\sim 10\%$ at merger; and the NQC
correction is most important close to merger where the radiation
reaction has little effect on the plunging dynamics.

The iterative procedure explained above usually takes 4 to 5
iterations to converge, bringing a factor of 4 to 5 to the
computational cost of generating an EOB waveform. Therefore, to reduce
computational cost, we give fitting formulas of $a_i^{h_{22}}$ in
Eq.~\eqref{acal}. We shall emphasize however that $a_i^{h_{22}}$ are
not adjustable parameters of the EOB model and are not required as
inputs to generate the $(2, 2)$ modes. 

Finally, to demonstrate the effect of the NQC corrections \eqref{Nlm},
we show in Fig.~\ref{fig:h22AmpOmega} the EOB amplitude for the $(2, 2)$
mode without and with NQC corrections, that is without and with the
factor $N_{\ell m}$ in Eq.~\eqref{hip}. We notice that for $q=1$ and
$q=6$, the amplitude without the NQC corrections differs from the
numerical one \emph{only} by $-0.23$\% and $-0.67$\%, at $t =6.2 M$ and
$t =5.7 M$ before the peak, respectively. However, even this small
difference needs to be removed to minimize the error when attaching
the merger-ringdown waveform. The latter will be discussed in the next
section.

\subsection{EOB waveform: Merger \& Ringdown}
\label{sec:EOBmergerRDwaveform}

The procedure of building the merger-ringdown waveform in the EOB
approach has improved over the
years~\cite{Buonanno00, Damour06, Buonanno-Cook-Pretorius:2007, Buonanno2007,
  DN2007b, DN2008, Buonanno:2009qa}. For each mode $(\ell, m)$ we have
\begin{equation}
  \label{RD}
  h_{\ell m}^\mathrm{merger-RD}(t) = \sum_{n=0}^{N-1} A_{\ell mn}\,
  e^{-i\sigma_{\ell mn} (t-t_\mathrm{match}^{\ell m})},
\end{equation}
where $n$ is the overtone number of the Kerr quasinormal mode (QNM),
$N$ is the number of overtones included in our model, and $A_{\ell
  mn}$ are complex amplitudes to be determined by a matching procedure
described below. The quantity $\sigma_{\ell mn} = \omega_{\ell m n} -
i/\tau_{\ell m n}$, where the oscillation frequencies $\omega_{\ell m
  n}>0$ and the decay times $\tau_{\ell m n}>0$, are numbers
associated with each QNM. The complex frequencies are known functions
of the final black-hole mass and spin and can be found in
Ref.~\cite{Berti2006a}.

\begin{figure*}
  \includegraphics[scale=0.39]{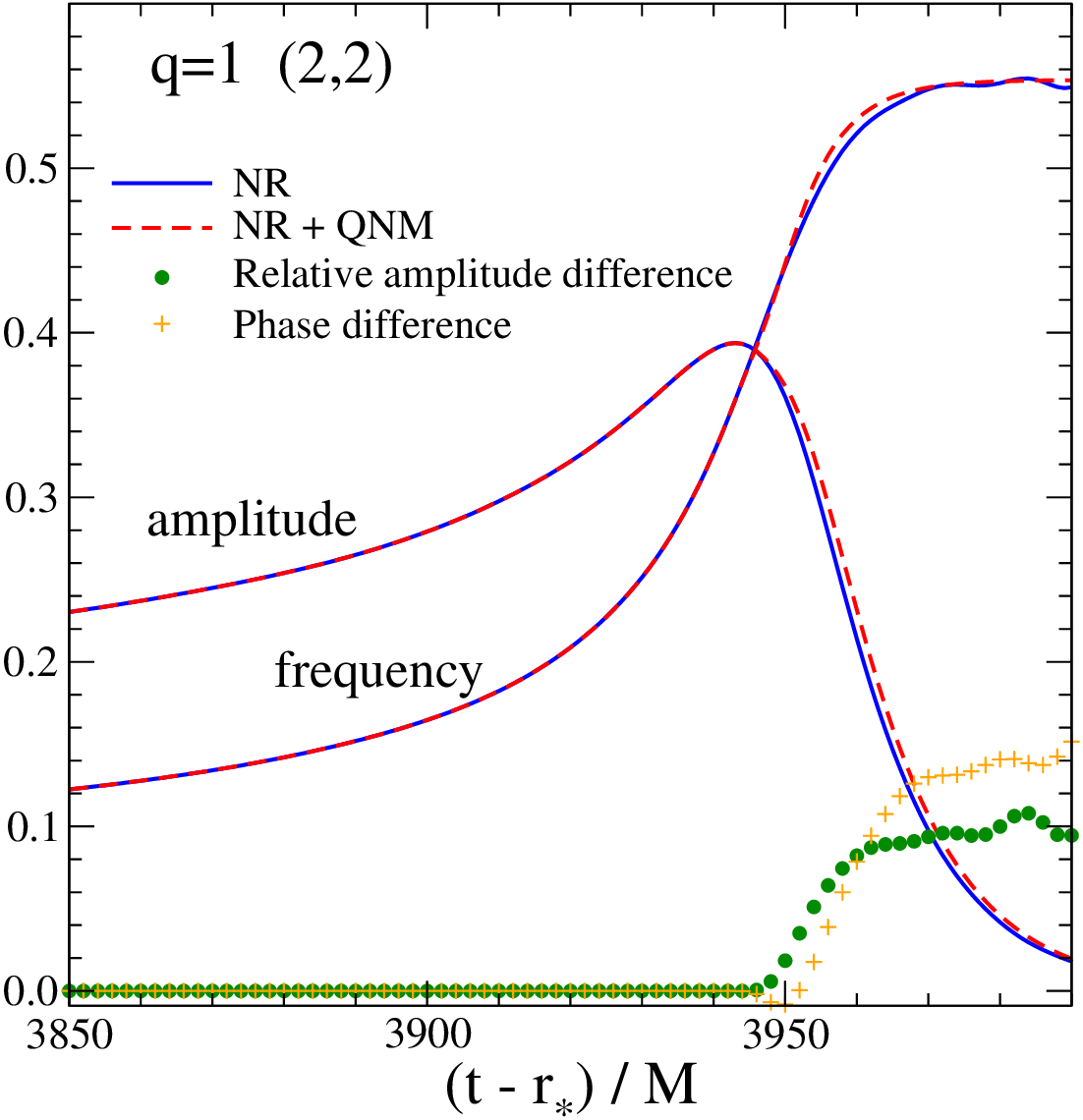} \qquad \qquad
  \includegraphics[scale=0.39]{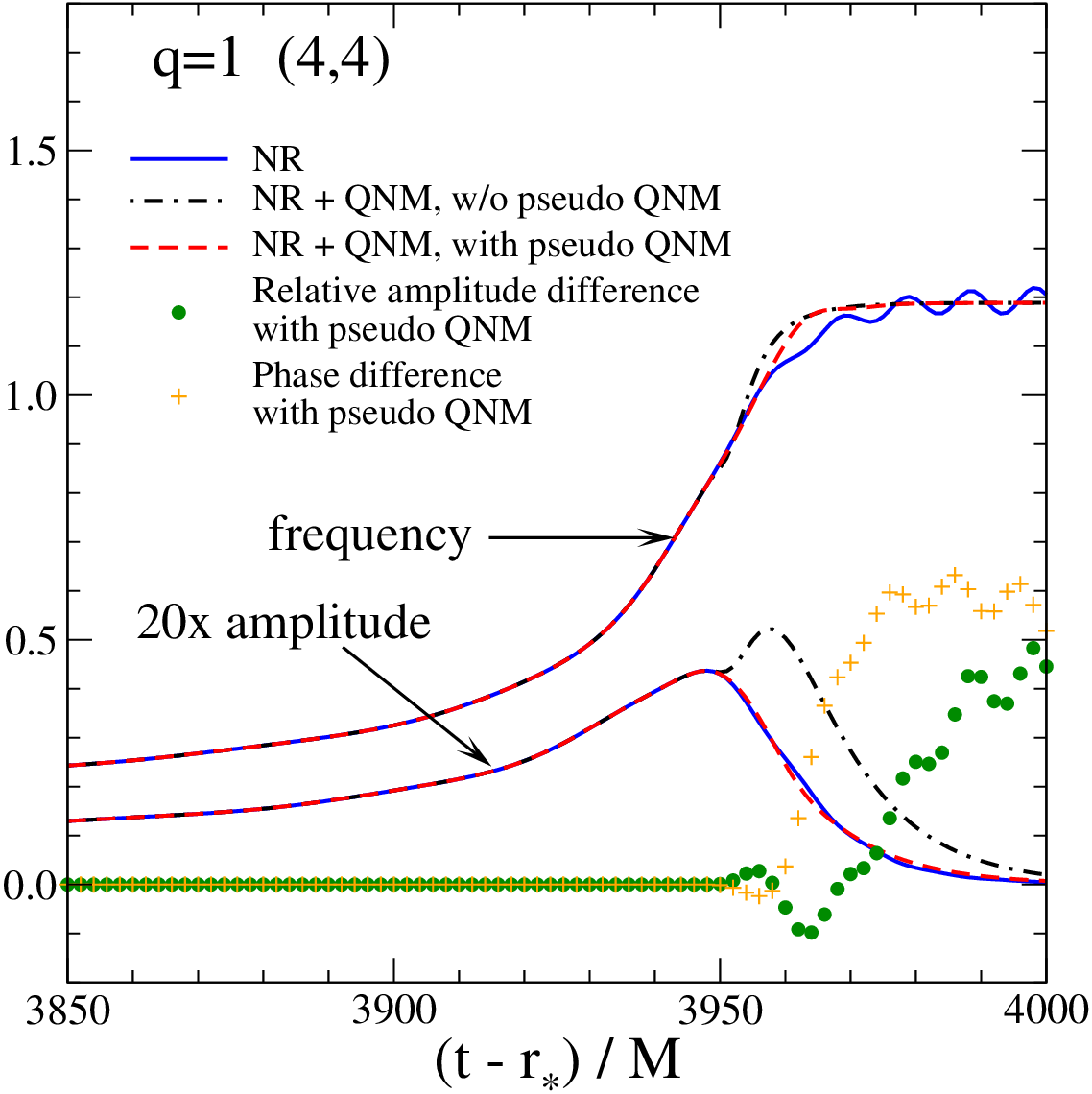}
  \caption{\label{fig:NRattQNM} Amplitude (in units of $M/{\cal R}$) 
    and frequency (in units of $1/M$) comparison
    between the full ``NR'' waveform and the ``NR$+$QNM'' waveform
    generated by attaching QNMs to the inspiral-plunge numerical
    waveform. We show also the relative amplitude and phase
    differences. In the left panel, we compare $h_{22}$.  
    In the right
    panel, we compare the numerical $h_{44}$ mode with two
    ``NR$+$QNM'' mode. One of them is generated by attaching the
    physical QNMs, the other is generated by attaching both the
    physical QNMs and the pseudo QNM.  The former is very different
    from the numerical-relativity mode and we do not show their
    amplitude and phase differences. All $h_{44}$ amplitudes have been
    multiplied by a factor of 20, so that they are more visible. The
    horizontal axis is the retarded time in the numerical-relativity
    simulation.}
\end{figure*}
In this paper we model the ringdown modes as a linear combination of
eight QNMs, i.e., $N =8$. Mass and spin of the final black hole $M_f$
and $a_f$ are computed from numerical data for mass ratios
$q=1, 2, 3, 4,$ and $6$. Notably, we employ the fitting formula obtained 
by fitting the numerical results of $M_f$ and $a_f$
\begin{subequations}
\label{finalMS}
  \begin{eqnarray}
    \!\!\!\!\!\!\frac{M_f}{M} &=& 1+\left (\sqrt{\frac{8}{9}}-1\right )\nu-0.4333 \nu^2-0.4392 \nu^3, \\
    \!\!\!\!\!\!\!\!\frac{a_f}{M_f} &=& \sqrt{12}\nu-3.871 \nu^2+4.028 \nu^3.
  \end{eqnarray}
\end{subequations}
The above formula differs from the analogous fitting formula given in
Ref.~\cite{Buonanno2007} by $<0.3\%$ in $M_f$ and $<2\%$ in $a_f$,
because of the more accurate numerical data used in this paper.

The complex amplitudes $A_{\ell mn}$ in Eq.~\eqref{RD} are determined
by matching the EOB merger-ringdown waveform \eqref{RD} with the EOB
inspiral-plunge waveform \eqref{hip}. In order to do this, $N$
independent complex equations are needed. In
Ref.~\cite{Buonanno:2009qa} we introduced the hybrid-comb matching in
which $N$ equations are obtained at $N-4$ points evenly sampled in a
small time interval $\Delta t_\mathrm{match}^{\ell m}$ ended at $t_\mathrm{
  match}^{\ell m}$, and we imposed the condition that the inspiral-plunge and
merger-ringdown waveforms coincide at the $N-4$ points and their first
and second order time derivatives coincide at the first and the last
points. Unlike in Ref.~\cite{Buonanno:2009qa}, we now no longer require
second order time derivatives of the waveforms to coincide anywhere in
order to improve the numerical stability of the matching procedure. 
Instead, we impose the continuity of the waveform at $N-2$ points evenly sampled
from $t_\mathrm{match}^{\ell m}-\Delta t_\mathrm{match}^{\ell m}$ to $t_\mathrm{
  match}^{\ell m}$, and we require continuity of the first time
derivative of the waveforms at $t_\mathrm{match}^{\ell m}-\Delta t_\mathrm{
  match}^{\ell m}$ and $t_\mathrm{match}^{\ell m}$, i.e.,
\begin{multline}
  h_{\ell m}^\mathrm{insp-plunge}(t_\mathrm{match}^{\ell m} -
  \frac{k}{N-3}\Delta t_\mathrm{match}^{\ell m})\\= h_{\ell m}^\mathrm{
    merger-RD}(t_\mathrm{match}^{\ell m}
  - \frac{k}{N-3}\Delta t_\mathrm{match}^{\ell m} )\,,\\
  (k=0, 1, 2,\cdots, N-3)\,,
\end{multline}
and
\begin{multline}
  \dot{h}_{\ell m}^\mathrm{insp-plunge}(t_\mathrm{match}^{\ell m} -
  \frac{k}{N-3}\Delta t_\mathrm{match}^{\ell m})\\= \dot{h}_{\ell m}^\mathrm{
    merger-RD}(t_\mathrm{match}^{\ell m}
  - \frac{k}{N-3}\Delta t_\mathrm{match}^{\ell m} )\,,\\
  (k=0 \mbox{ and } N-3)\,.
\end{multline}
The matching time $t_\mathrm{match}^{\ell m}$ is fixed to be the peak of
the EOB $h_{\ell m}$ mode, i.e., $t_\mathrm{match}^{\ell m}=t_\mathrm{
  peak}^{\Omega}+\Delta t_\mathrm{peak}^{\ell m}$.  The matching interval
$\Delta t_\mathrm{match}^{\ell m}$ is an adjustable parameter that we fix
by reducing the difference against numerical merger-ringdown modes
(see Table~\ref{tab:adjparams}).

Finally, the full (inspiral-plunge-merger-ringdown) EOB waveform reads
\begin{equation}
  \label{eobfullwave}
  h_{\ell m} = h_{\ell m}^\mathrm{insp-plunge}\, \theta(t_\mathrm{
    match}^{\ell m} - t) + h_{\ell m}^\mathrm{merger-RD}\,\theta(t-t_\mathrm{
    match}^{\ell m})\,.
\end{equation}
It was noticed in Ref.~\cite{Buonanno:2009qa} that when the lowest QNM
frequency is substantially larger than the EOB mode frequency at
$t_\mathrm{match}^{\ell m}$, the EOB mode frequency will generally grow
very rapidly to the QNM frequency immediately after $t_\mathrm{
  match}^{\ell m}$. Such growth in the EOB frequency is much more
rapid than what is seen in numerical-relativity frequencies. Moreover,
when this happens, the EOB amplitude shows an unphysical ``second
peak'' shape where the ringdown amplitude grows for a while before
eventually decaying. The growth of the EOB frequency can be slowed
down by including a pseudo QNM~\cite{Buonanno:2009qa} that has a
frequency close to the EOB mode frequency at $t_\mathrm{match}^{\ell m}$
and a decay time comparable but smaller than the decay time of the
least damped $n=0$ QNM. As we shall discuss below, we find it
necessary to introduce a pseudo QNM in modeling the EOB $(4, 4)$ and
$(5, 5)$ modes.  The pseudo QNM should be counted as another
adjustable parameter of the EOB waveform (see Table~\ref{tab:adjparams}). The
frequency and decay time of this pseudo QNM mode are given in
Eq.~\eqref{modecal}.

We have outlined the procedure to match the inspiral waveform to the
merger-ringdown waveform. We would like now to understand what is the
\emph{intrinsic} error that this procedure introduces.  To answer this
question, we build an inspiral-merger-ringdown waveform where the
inspiral part coincides with the numerical-relativity waveform, and
the merger-ringdown part is built using the EOB procedure. We then
extract the intrinsic error by comparing it to the
numerical-relativity full waveform. The results for the $(2, 2)$ and
$(4, 4)$ modes are shown in Fig.~\ref{fig:NRattQNM}.  By construction,
the two waveforms agree exactly before the matching time $t_\mathrm{
  match}^{\ell m}$, i.e., the time of the peak amplitude. For
$h_{22}$, the relative amplitude difference and phase difference
during ringdown are about $10\%$ and $0.1$ rad. These are
reasonable intrinsic errors for the EOB model and are comparable to
systematic errors in the best existing analytical
models~\cite{Buonanno:2009qa, Damour2009a}.  For $h_{44}$, the pseudo
QNM reduces the amplitude and phase differences substantially to the
level of $50\%$ (when the ringdown amplitude is below 10\% of the peak
amplitude) and $0.6$ rad. Although the differences are not as
small as those of $h_{22}$, they are for now acceptable considering
the relatively small amplitude of $h_{44}$ compared to $h_{22}$, at
least for the mass ratios considered in this paper. So, in the
following we shall not attempt to over-calibrate the EOB $h_{44}$
model to obtain smaller differences against the numerical results.

The intrinsic error depends on the procedure to match the inspiral
waveform to the merger-ringdown waveform. In particular, it depends on
the choice of $t_\mathrm{match}^{\ell m}$ and $\Delta t_\mathrm{match}^{\ell
  m}$, as well as the continuity conditions we impose on the points
sampled from $t_\mathrm{match}^{\ell m}-\Delta t_\mathrm{match}^{\ell m}$ to
$t_\mathrm{match}^{\ell m}$. In Fig.~\ref{fig:NRattQNM}, the results are
optimized only over $\Delta t_\mathrm{match}^{\ell m}$. To find the best
matching procedure, in principle, we should optimize over both $t_\mathrm{
  match}^{\ell m}$ and $\Delta t_\mathrm{match}^{\ell m}$, and consider
options that sample points and impose continuity conditions in
different ways. Since the relation between the intrinsic error and all
these parameters is not straightforward, we decide to refrain from
fine-tuning the ringdown waveform by assigning different matching
procedures to different mass ratios or modes. We prefer to use a
single, simple prescription that works well for all mass ratios and
modes. Thus, we fix $t_\mathrm{match}^{\ell m}$ to be the peak of the EOB
$h_{\ell m}$ waveform and find that this prescription works fairly
well.

In summary, there is an intrinsic error introduced by the current
procedure to match inspiral to merger-ringdown EOB waveforms.  This
error cannot be improved by better calibrating the EOB
inspiral-plunge dynamics and waveforms. It can be overcome only by
improving and/or changing the matching procedure. We leave to the
future this important work.

\begin{table}
  \begin{tabular}{|c|c|c|c|c|c|}
    \hline
    & $\; q=1 \;$ & $\; q=2 \;$ & $\; q=3 \;$ & $\; q=4\;$ & $\; q=6 \;$ \\
    \hline
    $\; t_1/M \;$ & 820 & 770 & 570 & 670 & 870 \\
    \hline
    $\; t_2/M \;$ & 2250 & 2255 & 1985 & 1985 & 2310 \\
    \hline
  \end{tabular}
  \caption{\label{tab:alignwindow} The range of integration $(t_1, t_2)$ for waveform alignment and mass ratios $q=1, 2, 3, 4$ and $6$.}
\end{table}

\subsection{Initial conditions for the EOB dynamics}
\label{sec:eobinit}

Before completing this section, we briefly review the way in which
initial conditions of the EOB Hamilton equations \eqref{eq-eob} are
implemented.

In Ref.~\cite{Buonanno06}, quasi-spherical initial conditions are given
for generic precessing black hole binaries. We adopt the nonspinning
limit of those conditions
\begin{subequations}
  \begin{eqnarray}
    \frac{\partial\hat{H}^\mathrm{real}}{\partial r} &=& 0\,, \label{IC1}\\
    \frac{\partial\hat{H}^\mathrm{real}}{\partial p_{r_*}} &=& \frac{1}{\nu}\frac{dE}{dt}\frac{(\partial^2\hat{H}^\mathrm{real}/\partial r\partial p_\Phi)}{(\partial\hat{H}^\mathrm{real}/\partial p_\Phi) (\partial^2\hat{H}^\mathrm{real}/\partial r^2)}\,, \label{IC2}\\
    \frac{\partial\hat{H}^\mathrm{real}}{\partial p_\Phi} &=& \Omega_0 \,. \label{IC3}
  \end{eqnarray}
\end{subequations}
Given the initial orbital frequency $\Omega_0$, we solve the initial
$r$, $p_{r_*}$ and $p_\Phi$ from Eqs.~\eqref{IC1}-\eqref{IC3} and set
the initial orbital phase $\Phi=0$. We always start the orbital
evolution at initial orbital frequency $M \Omega_0\leq0.0025$,
corresponding to an initial radius $r\gtrsim 50M$, such that the
binary orbits are sufficiently circularized at the frequency where
numerical simulations start.\footnote{Note that Eq.~\eqref{IC2} is
  derived in Ref.~\cite{Buonanno06} for $p_r$. At large initial
  separations, the difference between $p_{r_*}$ and $p_r$ is
  negligible and we do not distinguish them when setting initial
  conditions.}

\section{Calibrations and comparisons}
\label{sec:calibrations}

In this section we calibrate the EOB model against the numerical
simulations, and compare numerical and EOB $(2, 2)$, $(2, 1)$, $(3, 3)$,
$(4, 4)$ and $(5, 5)$ modes.

\subsection{Numerical-relativity simulations of unequal-mass binary
  black holes}
\label{sec:NRwaveforms}

The numerical simulations themselves are described in a separate
paper~\cite{Buchman-etal-in-prep}. We extract both the Newman-Penrose
(NP) modes $\Psi_4^{\ell m}$ and the strain modes $h_{\ell m}$ from
the simulations. The strain modes are extracted with the
Regge-Wheeler-Zerilli (RWZ)
formalism~\cite{ReggeWheeler1957, Zerilli1970b, Sarbach2001, Rinne2008b}
(see Appendix A of Ref.~\cite{Buonanno:2009qa} for details of the
numerical implementation used to obtain $h_{\ell m}$).  The waveforms
are then extrapolated to infinite extraction radius
with order $N=5$ polynomials in the $q=1$ case,
and $N=3$ polynomials in other cases~\cite{Boyle-Mroue:2008}.
In this section, we use the RWZ $h_{\ell m}$ to calibrate the EOB
adjustable parameters and to determine the EOB NQC coefficients as
functions of the mass ratio. We use the NP $\Psi_4^{\ell m}$ only to
align numerical and/or EOB waveforms at low frequency where
numerical errors of the former are smaller than those of the RWZ $h_{\ell m}$.

\begin{table}
  \begin{tabular}{|c|c|c|c|c|c|}
    \hline
    & $\;(\ell, m)\;$ & Fit formula \\
    \hline
    \multirow{5}{*}{$\Delta t_\mathrm{peak}^{\ell m}$} 
    & $(2, 2)$ & 0 \\
    & $(3, 3)$ & $ 3.383+3.847\,\nu+8.979\,\nu^2 $ \\
    & $(2, 1)$ & $ 10.67-41.41\,\nu+76.1\,\nu^2 $ \\
    & $(4, 4)$ & $ 5.57-49.86\,\nu+154.3\,\nu^2 $ \\
    & $(5, 5)$ & $ 6.693-34.47\,\nu+102.7\,\nu^2 $ \\
    \hline
    \multirow{5}{*}{$|h_{\ell m}^\mathrm{NR}(t_\mathrm{peak}^{\ell m})|$} 
    & $(2, 2)$ & $ \nu\,(1.422+0.3013\,\nu+1.246\,\nu^2)$ \\
    & $(3, 3)$ & $ \nu\,\delta m\,(0.5761-0.09638\,\nu+2.715\,\nu^2) $ \\
    & $(2, 1)$ & $ \nu\,\delta m\,(0.4832-0.01032\,\nu) $ \\
    & $(4, 4)$ & $ \nu\,(0.354-1.779\,\nu+2.834\,\nu^2) $ \\
    & $(5, 5)$ & $ \nu\,\delta m\,(0.1353-0.1485\,\nu) $ \\
    \hline
    \multirow{5}{52pt}{$100 \times$ $\left. \frac{d^2 \abs{h_{\ell
              m}^\mathrm{NR}}} {dt^2} \right|_{t_\mathrm{peak}^{\ell m}}$}
    & $(2, 2)$ & $ -\nu\,(0.1679+1.44\,\nu-2.001\,\nu^2) $ \\
    & $(3, 3)$ & $ -\nu\,\delta m\,(0.2518-0.8145\,\nu+5.731\,\nu^2) $ \\
    & $(2, 1)$ & $ -\nu\,\delta m\,(0.1867+0.6094\,\nu) $ \\
    & $(4, 4)$ & $ -\nu\,(0.1813-0.9935\,\nu+1.858\,\nu^2) $ \\
    & $(5, 5)$ & $ -\nu\,\delta m\,(0.09051-0.1604\,\nu) $ \\
    \hline
    \multirow{5}{*}{$\omega^\mathrm{NR}_{\ell m}(t_\mathrm{peak}^{\ell m})$} 
    & $(2, 2)$ & $ 0.2733 + 0.2316\,\nu + 0.4463\,\nu^2 $ \\
    & $(3, 3)$ & $ 0.4539 + 0.5376\,\nu + 1.042\,\nu^2 $ \\
    & $(2, 1)$ & $ 0.2907 - 0.08338\,\nu + 0.587\,\nu^2 $ \\
    & $(4, 4)$ & $ 0.6435 - 0.05103\,\nu + 2.216\,\nu^2 $ \\
    & $(5, 5)$ & $ 0.8217 + 0.2346\,\nu + 2.599\,\nu^2 $ \\
    \hline
    \multirow{5}{*}{$\dot{\omega}^\mathrm{NR}_{\ell m}(t_\mathrm{peak}^{\ell m})$} 
    & $(2, 2)$ & $ 0.005862 + 0.01506\,\nu + 0.02625\,\nu^2 $ \\
    & $(3, 3)$ & $ 0.01074 + 0.0293\,\nu + 0.02066\,\nu^2 $ \\
    & $(2, 1)$ & $ 0.00149 + 0.09197\,\nu - 0.1909\,\nu^2 $ \\
    & $(4, 4)$ & $ 0.01486 + 0.08529\,\nu - 0.2174\,\nu^2 $ \\
    & $(5, 5)$ & $ 0.01775 + 0.09801\,\nu - 0.1686\,\nu^2 $ \\
    \hline
  \end{tabular}
  \caption{\label{tab:NRnufit} 
    We list in the third column $\nu$-fits of $\Delta t_\mathrm{peak}^{\ell m}$, 
    $\left|h_{\ell m}^\mathrm{NR}(t_\mathrm{peak}^{\ell m})\right|$, $\left.d^2\left|h_{\ell m}^\mathrm{NR}\right|/dt^2\right|_{t_\mathrm{peak}^{\ell m}}$, $\omega^\mathrm{NR}_{\ell m}(t_\mathrm{peak}^{\ell m})$, and $\dot{\omega}^\mathrm{NR}_{\ell m}(t_\mathrm{peak}^{\ell m})$ against 
    numerical data of mass ratios $q = 1, 2, 3, 4$ and $6$ for modes $(2, 2)$, $(3, 3)$, $(2, 1)$, $(4, 4)$ and $(5, 5)$. 
    In the fitting formulas, the relative mass difference is $\delta m\define(m_1-m_2)/(m_1+m_2)=\sqrt{1-4\,\nu}$.}
\end{table}

We adopt the same waveform alignment procedure used in
Refs.~\cite{Boyle2008a, Buonanno:2009qa, Pan2010hz}.  That is, we align
waveforms at low frequency by minimizing the quantity
\begin{equation}\label{waveshifts}
  \Xi(\Delta t,\Delta\phi)=\int_{t_1}^{t_2}\left[\phi_1(t)-
    \phi_2(t-\Delta t)-\Delta\phi\right]^2\,dt\,,
\end{equation}
over a time shift $\Delta t$ and a phase shift $\Delta\phi$, where
$\phi_1(t)$ and $\phi_2(t)$ are the phases of the two $\Psi_4^{\ell
  m}$ waveforms.  The range of integration $(t_1, t_2)$ is chosen to be
as early as possible to maximize the length of the waveform, but late
enough to avoid the contamination from junk radiation present in the
numerical initial data. The range of integration should also be large
enough to average over numerical noise. Moreover, this range should
extend from peak to peak or trough to trough of oscillations (if
visible) in the gravitational-wave frequency due to residual
eccentricity in the initial data.  For different mass ratios, the
lengths of numerical simulations and waveforms are different, thus we
must also choose different $t_1$ and $t_2$ in
Eq.~\eqref{waveshifts}. Ref.~\cite{MacDonald:2011ne}
suggested a minimal length of the integration interval $(t_1, t_2)$, 
which is satisfied by our choices as listed in Table~\ref{tab:alignwindow}.

The numerical uncertainties are estimated by combining convergence
errors with extrapolation errors. The convergence estimates also use
the matching procedure described above.  Specifically, the resolution
convergence uncertainty is obtained by matching data from the highest
and second-highest resolutions over the relevant region from
Table~\ref{tab:alignwindow}, then taking the relative amplitude and
phase difference between them.  The same process is repeated for the
extrapolation uncertainty, comparing waveforms extrapolated with two
orders---the order used in this paper, and the next higher order.
Specifically, those orders are $N=5$ and $N=6$ for the $q=1$ case, and
$N=3$ and $N=4$ for $q = 2, 3, 4, 6$.  The absolute values of those
uncertainties are then added to give the final uncertainty estimate.
These uncertainties are shown as dotted lines in every figure of this
paper where phase and fractional amplitude differences are shown.

\subsection{Extracting information from numerical-relativity
  waveforms}
\label{sec:NRinfo}

As discussed in Secs.~\ref{sec:EOBinspiralwaveforms},
\ref{sec:EOBmergerRDwaveform} we need to extract specific information
from the numerical data that we use to determine the NQC coefficients
$a_i^{h_{\ell m}}$ and $b_i^{h_{\ell m}}$ in Eq.~\eqref{Nlm}, and the
QNM coefficients $A_{\ell mn}$ in Eq.~\eqref{RD}.

\begin{figure}
  \includegraphics[width=7.5cm]{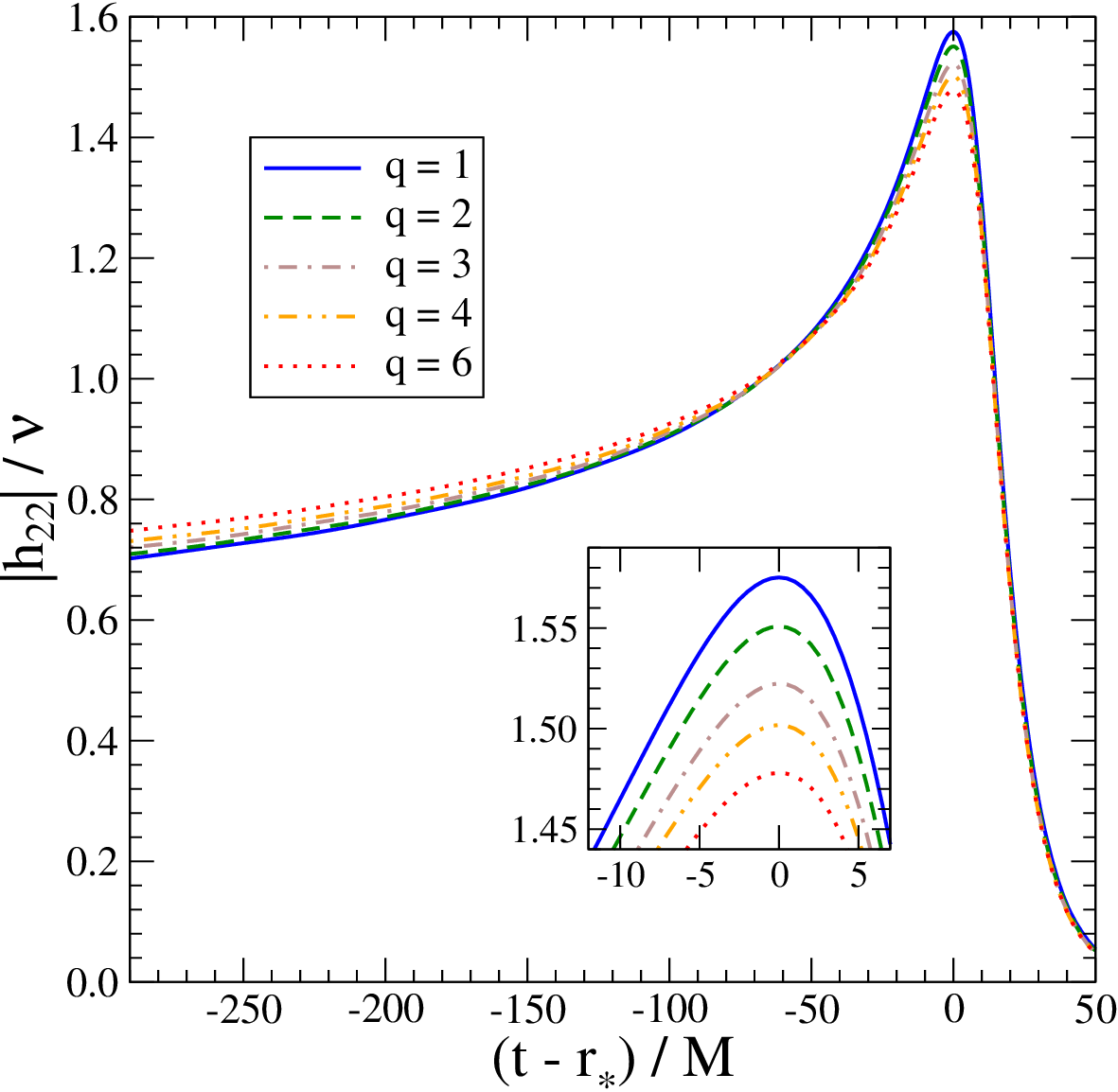}
  \caption{We show the amplitude of the numerical-relativity $h_{22}$
    for mass ratios $q = 1, 2, 3, 4, 6$. We have time shifted the modes so
    that their peaks are aligned. We have also rescaled them by $\nu$.
    The horizontal axis is the retarded time in the
    numerical-relativity simulation. The inset shows an
      enlargement of the merger
      region.\label{fig:h22amp_eta_comparison}}
\end{figure}

In Table~\ref{tab:NRnufit} we have listed the fitting formulas for the
relevant quantities $\Delta t_\mathrm{peak}^{\ell m}$, $|h_{\ell
  m}^\mathrm{ NR}(t_\mathrm{peak}^{\ell m})|$, $\left. d^2|h_{\ell
    m}^\mathrm{NR}|/dt^2 \right|_{t_\mathrm{peak}^{\ell m}}$,
$\omega^\mathrm{NR}_{\ell m}(t_\mathrm{ peak}^{\ell m})$, and
$\dot{\omega}^\mathrm{NR}_{\ell m}(t_\mathrm{ peak}^{\ell
  m})$. These formulas are least-square fits of
  numerical-relativity results at mass ratios $q=1,2,3,4$ and $6$ and
  numerical results in the test-particle limit $\nu=0.001$ generated
  by a time-domain Teukolsky code \cite{Barausse:2011a}. The
errors of the fitting formulas are worst for the $(5, 5)$ mode and for
the quantity $\left. d^2|h_{\ell m}^\mathrm{ NR}|/dt^2
\right|_{t_\mathrm{peak}^{\ell m}}$. The $(5, 5)$ mode generally has
the lowest amplitude among the modes being studied and is therefore
mostly contaminated by numerical artifacts. Specifically, we find
oscillations in the amplitude of numerical waveforms that are
amplified in the process of extrapolating the extraction radii to
infinity.  Such oscillations modify the position and shape of the peak
amplitude and they start to become significant for the $(5, 5)$ mode
and the other modes with lower amplitudes. The errors suggest that we
can barely model the merger amplitude of the $(5, 5)$ mode for generic
mass ratios. Numerical errors prevent us from modeling the merger and
ringdown of any other mode with smaller amplitude.  The relative error
on $\left. d^2|h_{\ell m}^\mathrm{NR}|/dt^2
\right|_{t_\mathrm{peak}^{\ell m}}$ may reach a few percent for the
$(2, 2)$, $(2, 1)$, $(3, 3)$ and $(4, 4)$ modes. Nevertheless, we find
the fitted values accurate enough for constraining the shape of the
amplitude peaks. In fact, the error in the EOB merger-ringdown
waveform of these modes are dominated by the intrinsic error of the
matching procedure discussed in Sec.~\ref{sec:EOBmergerRDwaveform}.
The error on $\Delta t_\mathrm{ peak}^{\ell m}$ is the error in
determining the peak of $|h_{\ell m}|$ in the EOB model. Since the
peak of the orbital frequency is used as the reference time of merger
in the EOB model, and since it coincides with the peak of the
numerical amplitude $h_{22}$ to within $1.8M$ (see
Sec.~\ref{sec:EOBinspiralwaveforms}), the $0.5M$ error in fitting
$\Delta t_\mathrm{peak}^{33}$ and $\Delta t_\mathrm{peak}^{21}$ is
sufficiently small.  The relative errors in fitting all other
quantities are within $1\%$ and we expect these fitting formulas to
work with such accuracy for any mass ratio $q \le 6$.

It is interesting to observe that the reason why the shape and
characteristics of the numerical amplitude peaks can be easily
reproduced by polynomials in $\nu$ (see Table~\ref{tab:NRnufit}) rests
on the fact that when the modes for different $\nu$ are aligned at the
peak and rescale by $\nu$, their peaks differ by less than $7\%$ and
the width at half peak by less than $18\%$.  This is illustrated in
Fig.~\ref{fig:h22amp_eta_comparison}, and was initially pointed out in
Ref.~\cite{Baker2008a}.

\subsection{Calibrating the EOB adjustable parameters}

We carry out the calibration of the adjustable parameters as
follows.  First, we fix the EOB-dynamics parameters by
minimizing the phase difference between the numerical and EOB $(2, 2)$
modes during the inspiral. Second, we shall evolve the EOB dynamics
with the calibrated parameters and fixed the EOB-waveform parameters
by minimizing the difference between the numerical and EOB full
waveforms of all relevant modes. Moreover, the adjustable parameters
might be functions of the mass ratio $\nu$. So, we first calibrate
them for individual mass ratios then fit the calibrated values with
quadratic functions of $\nu$. We find that many parameters depend
weakly on $\nu$ and can be set as constants.

\begin{figure}
  \includegraphics[width=8cm, clip=true]{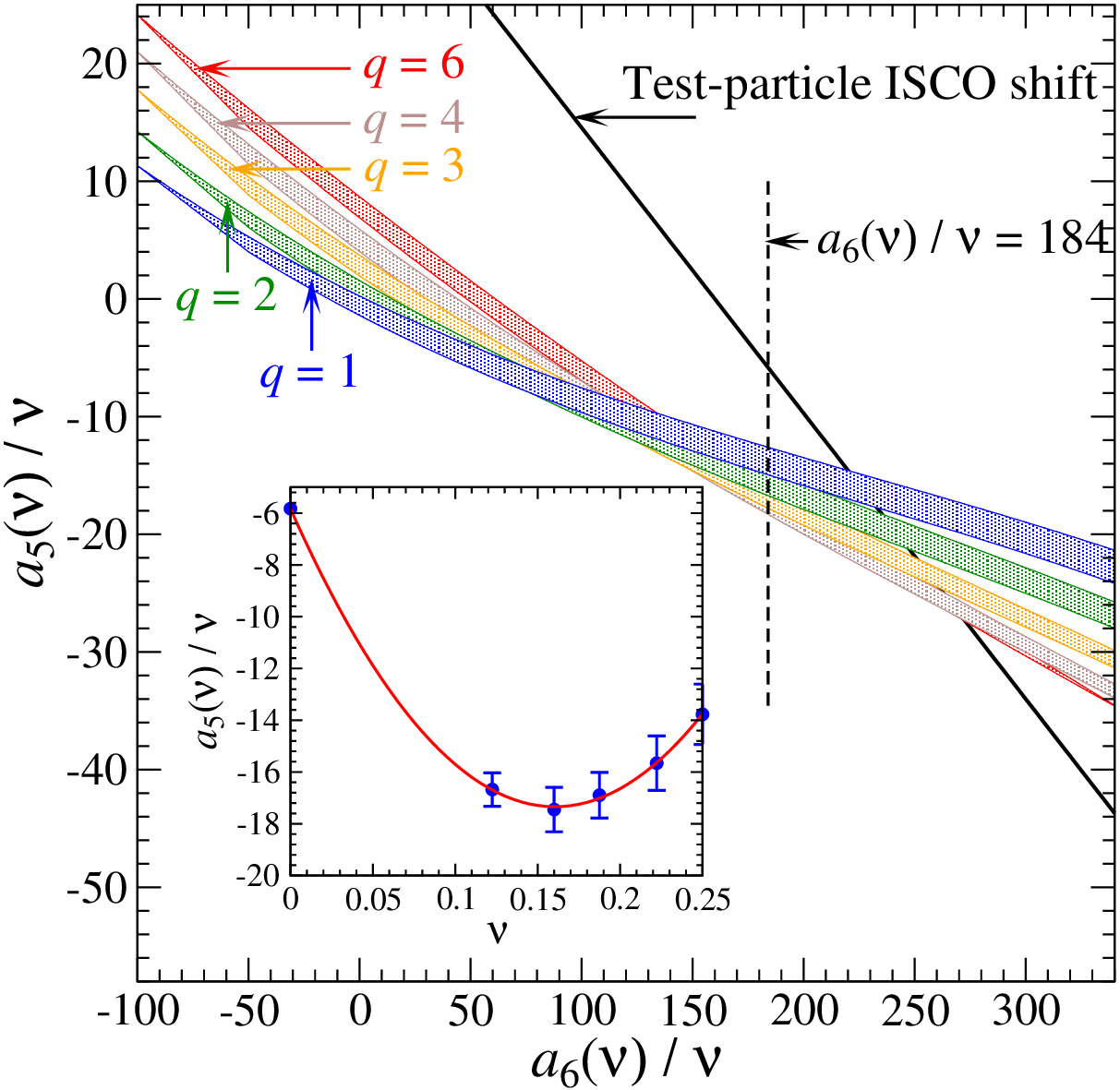}
  \caption{\label{fig:eob2contour} We calibrate
    adjustable parameters of the EOB dynamics. For mass ratios $q=1, 2, 3, 4$ and $6$, the
    shaded regions correspond to $(a_5, a_6)$ values for which the EOB
    and numerical-relativity $h_{22}$ agree within $0.2$ rad at
    merger, i.e., at the peak of the numerical $h_{22}$.  In the
    inserted subplot, for $a_6(\nu)/\nu = 184$, we show $a_5(\nu)/\nu$
    values constrained by the shaded regions and by the test-particle
    ISCO-shift result~\cite{BarackSago09}, and also the quadratic fit
    (red curve) given by Eq.~\eqref{a5cal}.}
\end{figure}

Figure~\ref{fig:eob2contour} summarizes the calibration result of the
inspiral dynamics and our choice of the $a_5(\nu)$ and $a_6(\nu)$
values.  We calibrate the EOB model to five numerical $h_{22}$
waveforms of mass ratios $q=1, 2, 3, 4$ and $6$. For each mass ratio, we
show a contour in the $a_5(\nu)/\nu$--$a_6(\nu)/\nu$ parameter space
in which the numerical and EOB $h_{22}$ waveforms agree in phase to
within $0.2$ rad at merger, that is at the peak of the (2, 2) mode.
As observed in Ref.~\cite{Damour2009a} (and also evident from Fig.~\ref{fig:eob2contour}), 
there is strong degeneracy between $a_5(\nu)$ and $a_6(\nu)$. Furthermore, 
there are many ways to choose $a_5(\nu)$ and $a_6(\nu)$ so that
the numerical and EOB $h_{22}$ waveforms agree for $q=1, 2, 3, 4$ and $6$.
For instance, we could choose a point near $a_5(\nu)/\nu=-10$ and $a_6(\nu)/\nu =126$ where the contours of
different mass ratios approximately cross. However, we find that those
values do not satisfy the constraint imposed by the self-force result
of the ISCO shift of Ref.~\cite{BarackSago09}. More importantly, even
if we were not keen at imposing the ISCO shift prediction, the EOB
model obtained by choosing $a_5(\nu)/\nu$ and $a_6(\nu)/\nu$ around
that crossing region will not be very satisfactory, because the point
will not lie in the middle of the contours for each mass ratio.

By contrast, we follow a more satisfactory route.  We decide to model
$a_5(\nu)/\nu$ and $a_6(\nu)/\nu$ as generic quadratic functions of
$\nu$, i.e., $a_5(\nu)/\nu=\sum_{i=0}^2a_5^{(i)}\,\nu^i$ and
$a_6(\nu)/\nu=\sum_{i=0}^2a_6^{(i)}\,\nu^i$.  In the test-particle
limit $\nu\rightarrow 0$, we impose the constraint on $a_5^{(0)}$ and
$a_6^{(0)}$ that the conservative EOB dynamics incorporates the exact
ISCO-frequency shift of the self-force calculation
\cite{BarackSago09}, that is~\cite{Damour:2009sm}
\begin{equation}
  M\,\Omega_\mathrm{
    ISCO}=6^{-3/2}\,\left[1+1.2513\,\nu+\mathcal{O}(\nu^2)\right]\,.
\end{equation}
The remaining five coefficients of $a_5^{(i)}$ and $a_6^{(i)}$ are
determined by minimizing the distances of the
$a_5(\nu)/\nu$--$a_6(\nu)\nu$ points to the center of the contours in
Fig.~\ref{fig:eob2contour} (i.e., the points at equal distance from
the top and bottom boundaries of the contours) for mass ratios
$q=1, 2, 3, 4$ and $6$. This procedure is not unique; 
we optimize $a_5^{(i)}$ and $a_6^{(i)}$ over a
relatively coarse grid and find reasonable results setting
$a_6(\nu)/\nu$ as a constant and
\begin{subequations} \label{a5a6cal}
  \begin{align}
    \label{a5cal}
    a_5(\nu)&= (-5.828-143.5\,\nu+447\,\nu^2)\,\nu\,,\\
    \label{a6cal}
    a_6(\nu)&= 184\,\nu\,.
  \end{align}
\end{subequations}
For this constant choice of $a_6(\nu)/\nu$, we show in a subplot of
Fig.~\ref{fig:eob2contour} the $a_5(\nu)/\nu$ values constrained by
the contours, as well as the quadratic fit of them.  The error bars
correspond to the width of the contours. More constraints imposed by
new numerical waveforms may further lift the degeneracy between
$a_5(\nu)$ and $a_6(\nu)$.  We shall show in the next section that
this model calibrated to the five numerical waveforms, even though not
carefully optimized with a fine global grid, is good enough for
detection with Advanced LIGO, and fair for parameter estimation
purposes.

Although $a_5(\nu)/\nu$ and $a_6(\nu)/\nu$ can not both be constants
in this calibrated model, we shall emphasize that it does not imply
that the physical 4PN and 5PN coefficients shall depend on $\nu$
beyond the linear order. The optimal choice of $a_5(\nu)$ and
$a_6(\nu)$ depends on other elements of the dynamics, for instance,
the \Pade expression of $A(r)$ or $D(r)$, or the way we factorize or
resum $h^\mathrm{F}_{\ell m}$ in the inspiral waveforms that enter the
energy flux, etc. Therefore, it is possible that when a different EOB
model is calibrated to the same set of numerical waveforms subject to
the constraint imposed by the self-force result, the optimal choice of
$a_5(\nu)/\nu$ and $a_6(\nu)/\nu$ are constants. It might be the case
that a minor change in the dynamics of the EOB model being calibrated
in this paper could bring all contours and the self-force constraint to
cross at exactly the same point.

Having calibrated the adjustable parameters of the EOB dynamics, we calibrate
the adjustable parameters of the EOB waveform listed in
Table~\ref{tab:adjparams}, notably the width of the comb $\Delta
t_\mathrm{match}^{\ell m}$, a few higher-order PN terms in $\rho_{\ell
  m}$ and $\delta_{\ell m}$ (see Appendix~\ref{app:eob}), and a pseudo
QNM. For technical reasons we do not include the adjustable parameters
$\rho_{21}$, $\rho_{33}$, $\rho_{44}$, and $\rho_{55}$ in the energy
flux \eqref{resflux}, but only in the EOB modes \eqref{hip}. The
energy flux being used in the dynamical evolution is therefore
slightly different from the energy flux defined in
Eq.~\eqref{resflux}. For the mass ratios and length of waveforms being
considered in this paper, the fractional difference in the energy flux
grows from $\sim 10^{-4}$ during low frequency to $\sim 10^{-2}$ at
merger. Such a difference in the dynamics generates a phase difference
at merger that increases from $0.02$ rad to $0.35$ rad with mass ratio
increasing from $q=1$ to $q=6$. In principle, to have an energy flux
in the dynamics that is exactly consistent with the definition in
Eq.~\eqref{resflux}, we need to calibrate those adjustable parameters
in $\rho_{\ell m}$ as EOB-dynamics parameters, together with $a_5$ and
$a_6$. Calibrating a large number of EOB-dynamics parameters has two
consequences: the calibration becomes computationally expensive and
all parameters become highly degenerate. Because of these technical
difficulties and the relatively small fractional difference in the
flux, we choose not to include the adjustable parameters in
$\rho_{\ell m}$ in the energy flux.

\begin{figure}
  \includegraphics[scale=0.39]{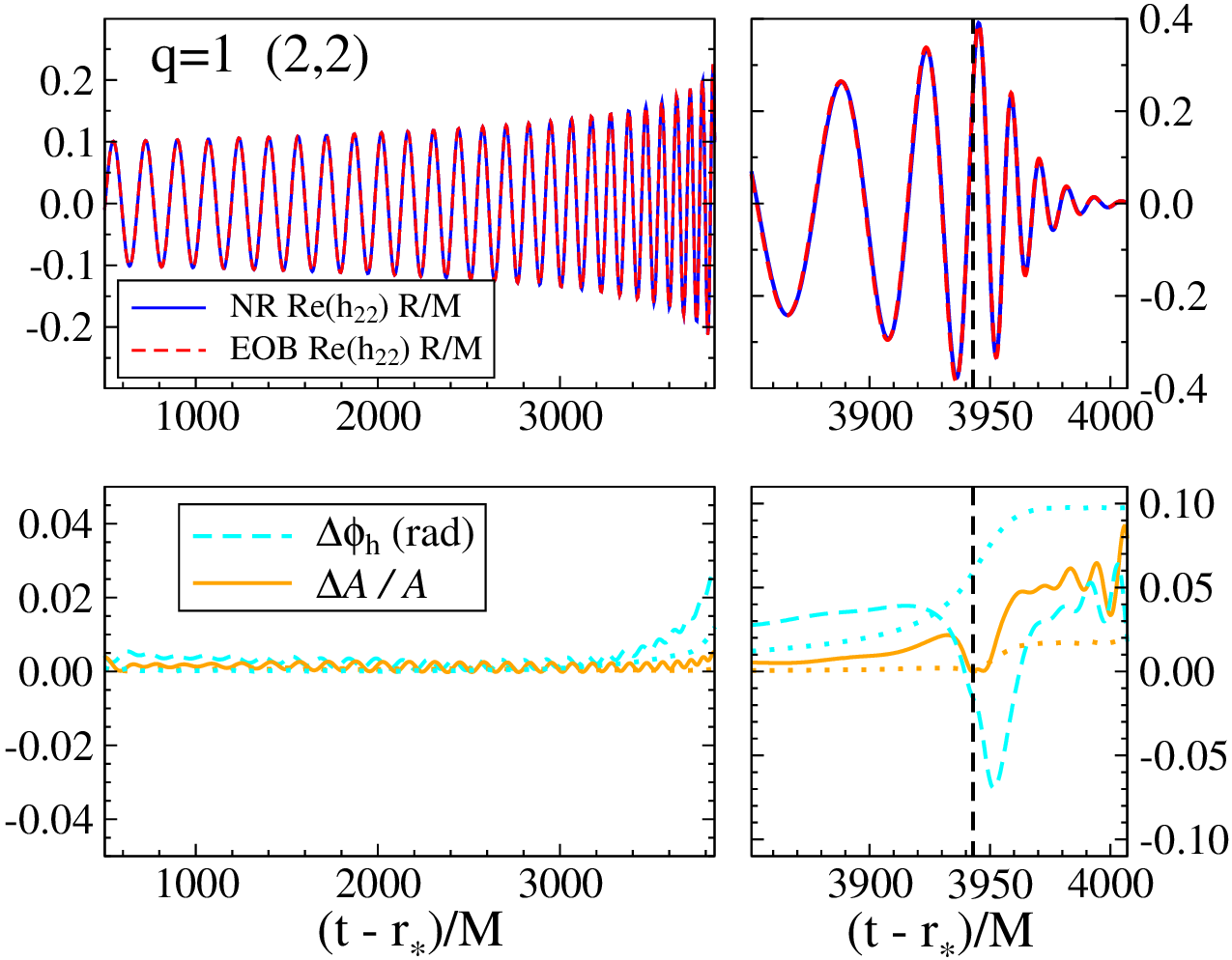}
  \caption{\label{fig:q1} For the equal-mass case, we compare the
    numerical-relativity and calibrated EOB $(2, 2)$ mode. The top
    panels show the real part of numerical and EOB $h_{22}$, the
    bottom panels show amplitude and phase differences between
    them. The horizontal axis is the retarded time in the
    numerical-relativity simulation. The left panels show retarded
    times $t-r_*=0$ to $3850 M$, and the right panels show retarded
    times $t-r_*=3850 M$ to $t-r_*=4070M$ on a different vertical
    scale. The dotted curves are the numerical-relativity errors.}
\end{figure}

\begin{figure*}
  \includegraphics[scale=0.39]{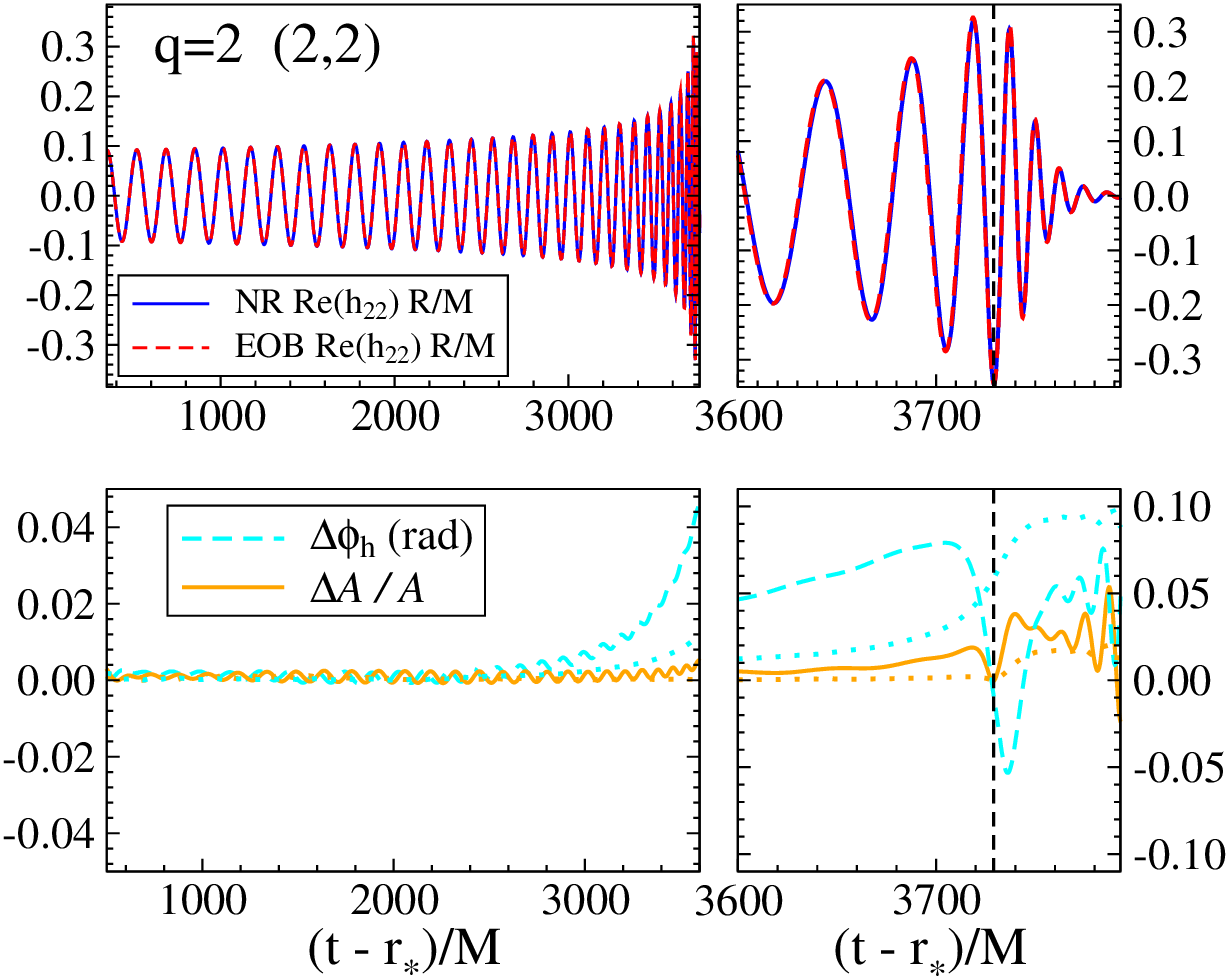} \qquad
  \qquad
  \includegraphics[scale=0.39]{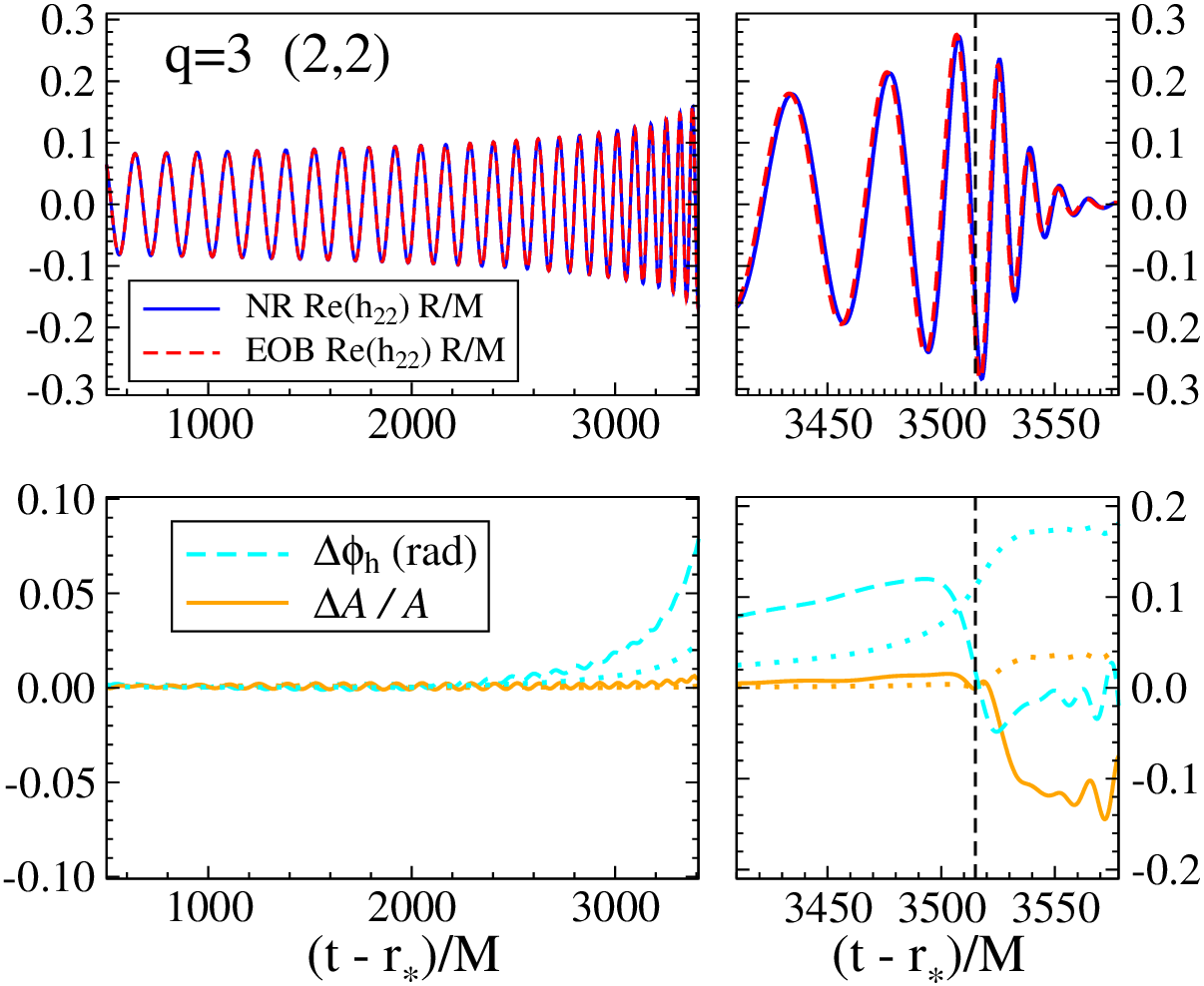}\\[2em]
  \includegraphics[scale=0.39]{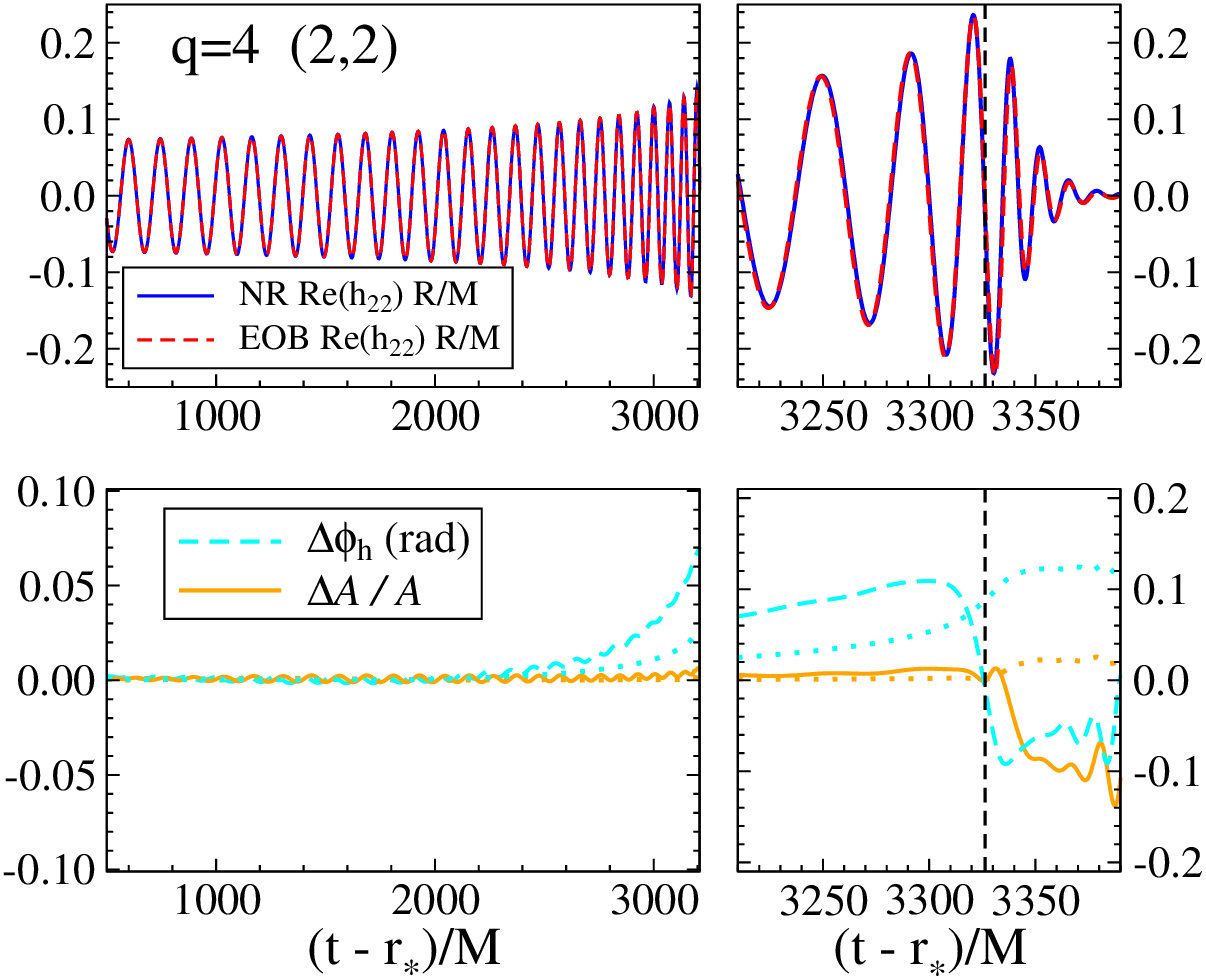} \qquad
  \qquad
  \includegraphics[scale=0.39]{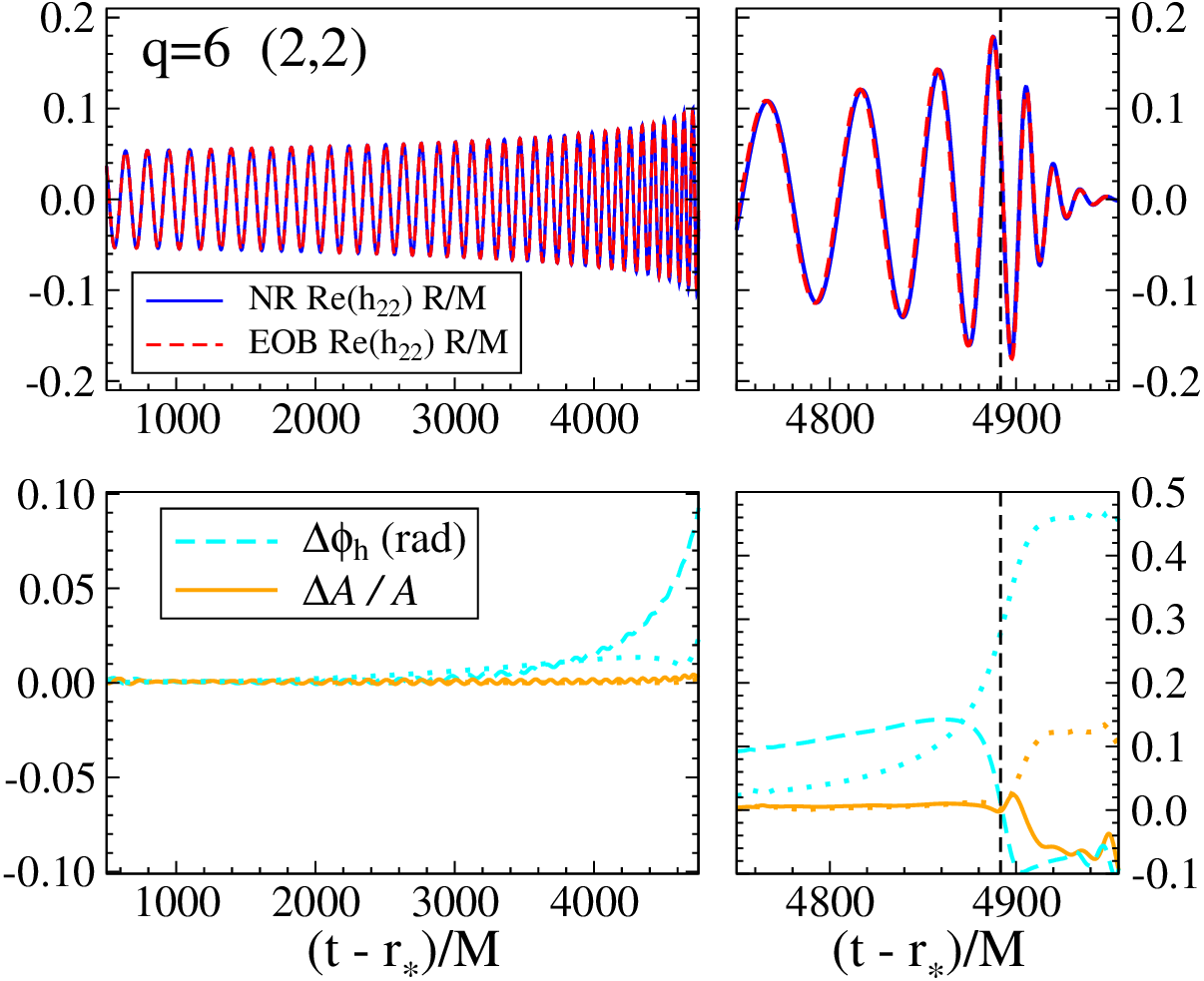}
  \caption{\label{fig:q2346} Comparison between the $(2, 2)$ modes
of the numerical-relativity simulation and the calibrated EOB model
for mass-ratios $q=2, 3, 4, 6$.  In each sub-plot, the top panels show
the real part of the $h_{22}$ mode and the bottom panels show the
phase and amplitude differences between numerical and EOB waveform. 
The dotted curves are the numerical-relativity errors. }
\end{figure*}

We determine the width of the comb $\Delta t_\mathrm{match}^{\ell m}$ by
requiring the best agreement between EOB and numerical $h_{\ell m}$
around merger and ringdown. We find that $\Delta t_\mathrm{match}^{\ell
  m}$ depends moderately on the mass ratio, and we can assume $\Delta
t_\mathrm{match}^{\ell m}$ being a constant. Specifically, we obtain
\begin{subequations}
  \label{combcal}
  \begin{eqnarray}
    &&\Delta t_\mathrm{match}^{22}= 5\, M \,, \\
    &&\Delta t_\mathrm{match}^{33}= 12\, M \,,\qquad\!\! \Delta t_\mathrm{match}^{44}= 9\, M \,, \\
    &&\Delta t_\mathrm{match}^{21}= 8\, M \,,\qquad \Delta t_\mathrm{match}^{55}= 8\, M \,.
  \end{eqnarray}
\end{subequations}
Calibrating the amplitude and phase of the EOB waveform for the
higher-order modes we find
\begin{subequations}
  \label{rhocal}
  \begin{eqnarray}
    &&\rho_{21}^{(6)} = -5\,,\qquad \rho_{33}^{(6)} = -20\,,\\
    \label{rho44cal}
    &&\rho_{44}^{(6)} = -15\,,\qquad \rho_{55}^{(6)} = 4\,,
  \end{eqnarray}
\end{subequations}
%
%
%
%
and
\begin{subequations}
  \label{deltacal}
  \begin{eqnarray}
    &&\delta_{21}^{(7)} = 30 \,,\qquad \delta_{33}^{(7)} = -10 \,,\\
    \label{delta44cal}
    &&\delta_{44}^{(5)} = -70 \,,\qquad \delta_{55}^{(5)} = 40 \,.
  \end{eqnarray}
\end{subequations}

As explained in Sec.~\ref{sec:EOBinspiralwaveforms}, since the
iterative procedure that determines the NQC coefficients
$a_i^{h_{22}}$ usually takes 4 to 5 steps to converge, we give fitting
formulas of $a_i^{h_{22}}$ as quadratic functions of the mass ratio to
save computational cost
\begin{subequations}
  \label{acal}
  \begin{eqnarray}
    a_1^{h_{22}}(\nu) &=& -4.559+18.76\,\nu-24.23\,\nu^2 \,,\\
    a_2^{h_{22}}(\nu) &=& 37.68-201.5\,\nu+324.6\,\nu^2 \,,\\
    a_3^{h_{22}}(\nu) &=& -39.6+228.9\,\nu-387.2\,\nu^2 \,.
  \end{eqnarray}
\end{subequations}

Finally, as discussed in Sec.~\ref{sec:EOBmergerRDwaveform}, to
improve the agreement of the $(4, 4)$ and $(5, 5)$ modes around merger,
we introduce pseudo QNMs having
\begin{subequations}
  \label{modecal}
  \begin{eqnarray}
    M\,\omega_{44}^\mathrm{pQNM} = 0.72\,, \quad M/\tau^\mathrm{pQNM}_{44} = 0.28\,, \\
    M\,\omega_{55}^\mathrm{pQNM} = 0.9\,, \quad M/\tau^\mathrm{pQNM}_{55} = 0.28\,, 
  \end{eqnarray}
\end{subequations}
for all mass ratios considered in this paper. The frequency of these
pseudo QNMs are about the frequency of the inspiral-plunge waveforms
at the matching time $t_\mathrm{match}^{\ell m}$ and the decay time of
these modes are about the same as that of the first overtone of the
physical QNMs.

\subsection{Comparing numerical and EOB \boldmath$(2, 2)$ mode}
\label{sec:eob22}

In Figs.~\ref{fig:q1},~\ref{fig:q2346} we compare the
numerical-relativity and EOB $(2, 2)$ modes for mass ratios $q=1, 2, 3, 4$
and $6$. We find that throughout the evolution the phase difference is
below $\sim 0.1$ rad.  During the inspiral, the relative
amplitude difference is within $2\%$, while during merger and ringdown
it increases to within $12\%$. The numerical errors are also showed in
the figures with dotted lines. We observe that during the inspiral the
phase and amplitude differences can be a factor of a few larger than
the numerical-relativity error, but during the merger and ringdown
they can be comparable or even smaller.  As we shall see in
Sec.~\ref{sec:overlaps}, the mismatch between the numerical and EOB
modes are consistently very small for detection purposes for Advanced
LIGO, and the EOB modes are reasonably accurate for
parameter-estimation purposes.

\subsection{Comparing numerical and EOB \boldmath$(l, m) \neq (2, 2)$ modes}
\label{sec:eobneq22}

\begin{figure}
\includegraphics[scale=0.39]{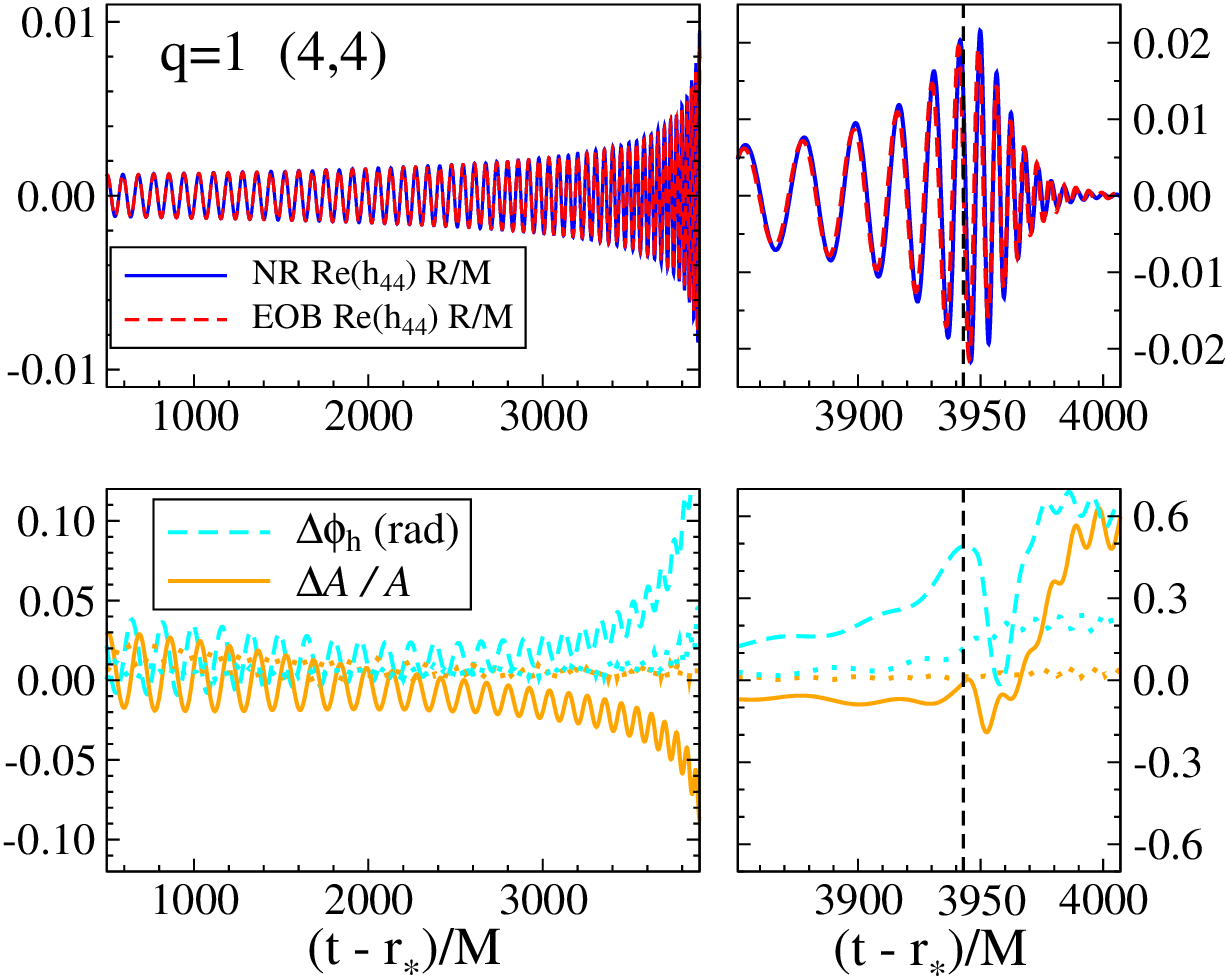} \\[1.5em]
  \includegraphics[scale=0.39]{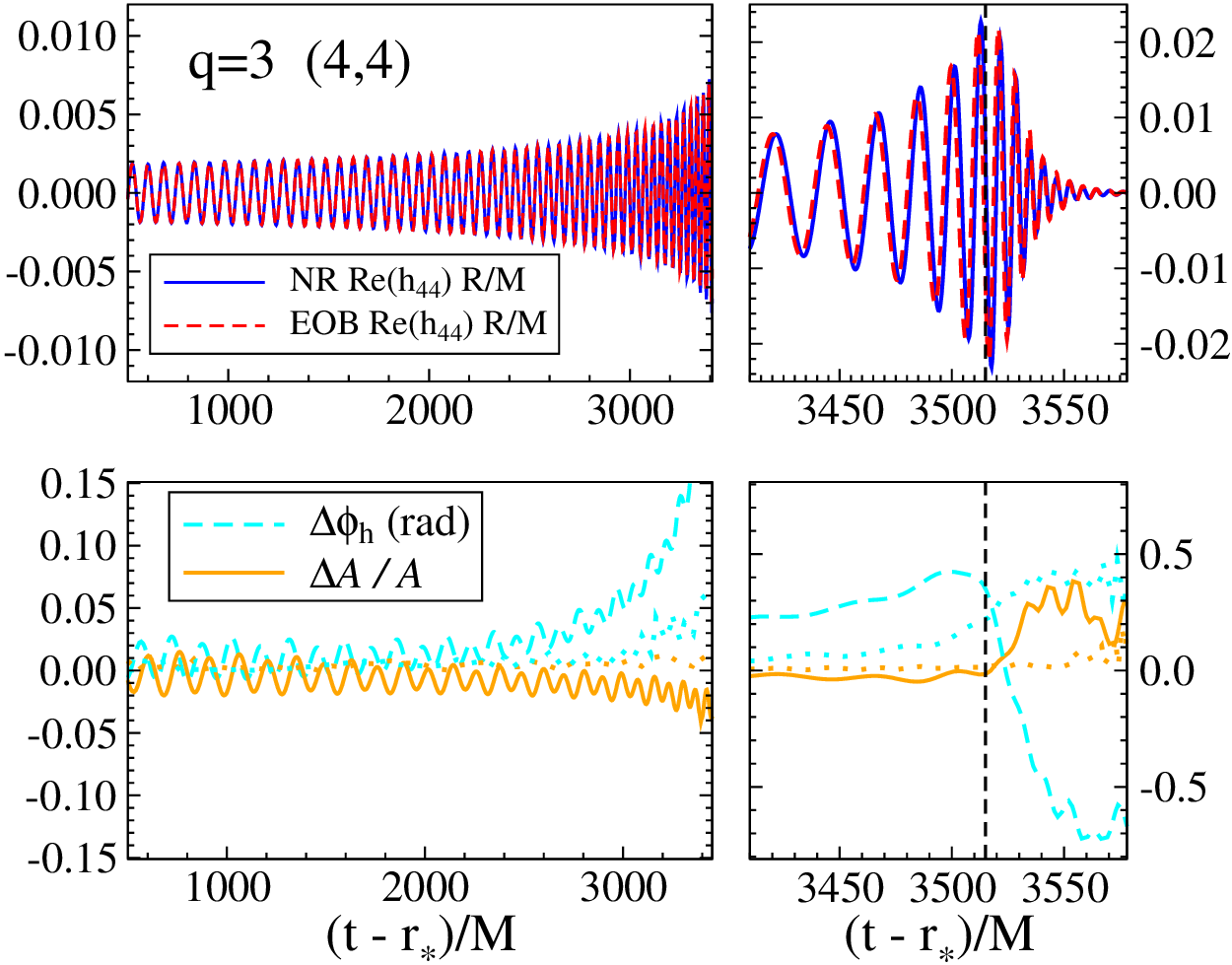} \\[1.5em]
  \includegraphics[scale=0.39]{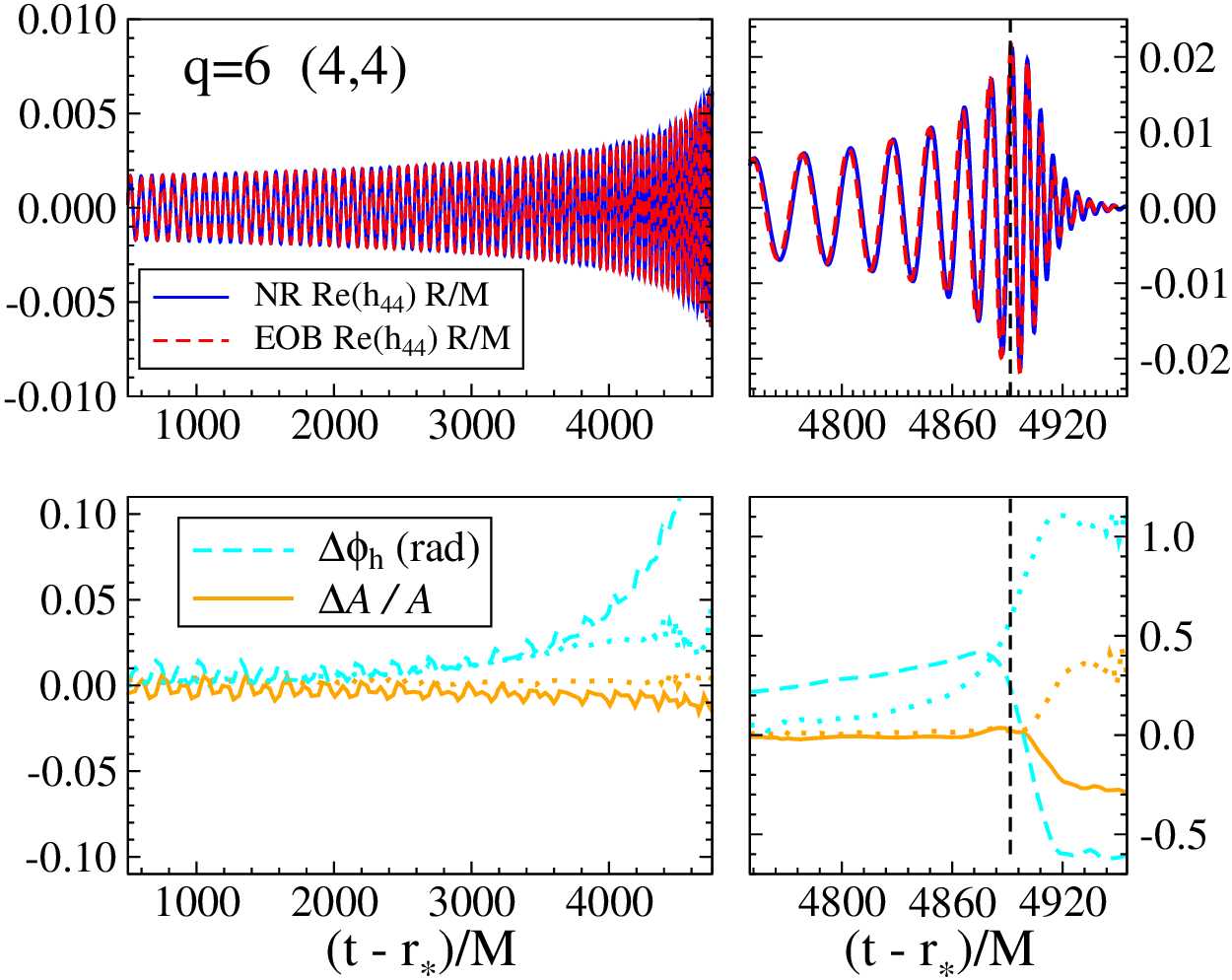}
  \caption{\label{fig:h44q136} Comparison of the $(4, 4)$ mode
      for mass-ratios $q=1, 3, 6$ between numerical and calibrated EOB model.  Plotted data as in Fig.~\ref{fig:q2346}.}
\end{figure}

\begin{figure}
  \includegraphics[scale=0.39]{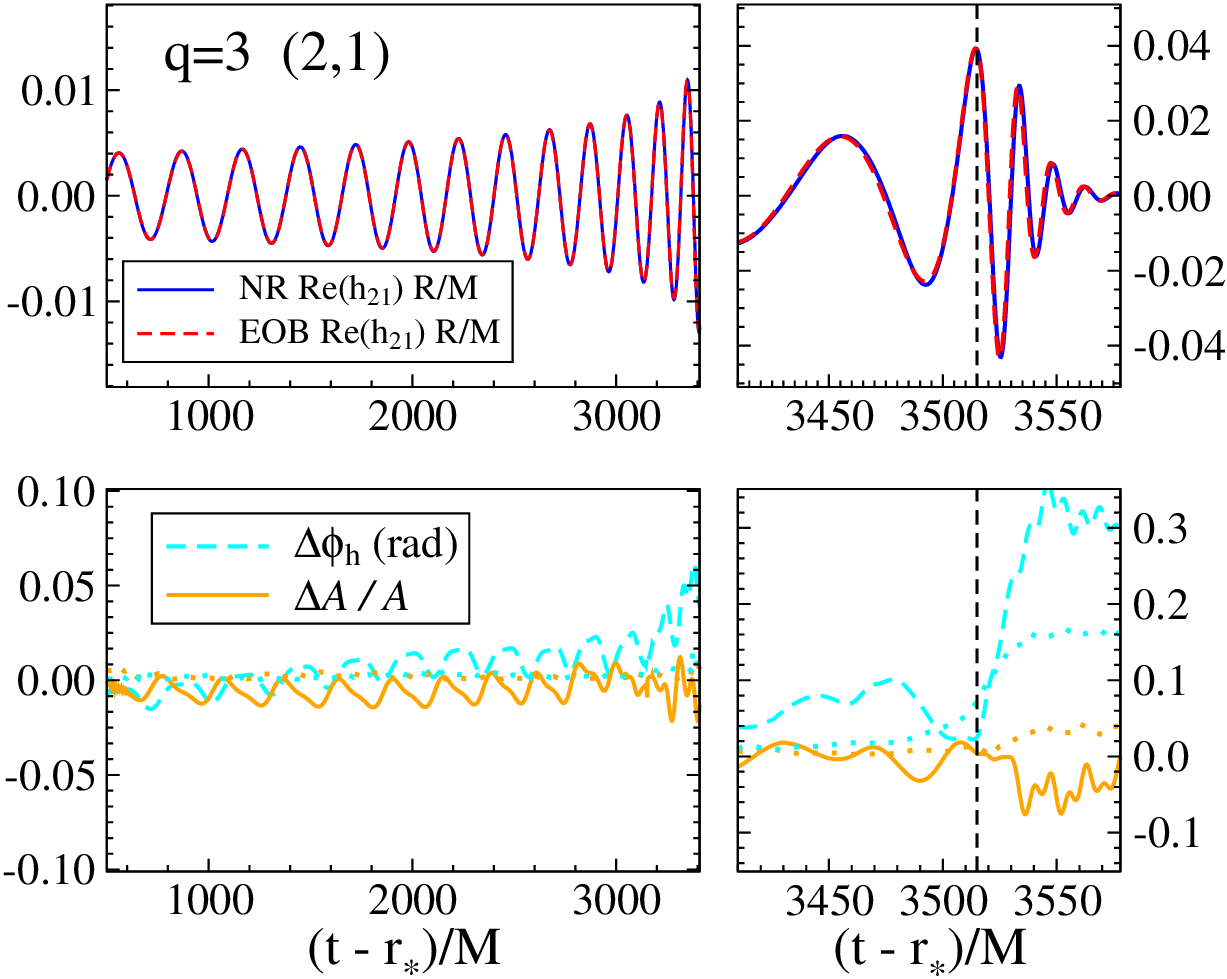} \\[1.5em]
  \includegraphics[scale=0.39]{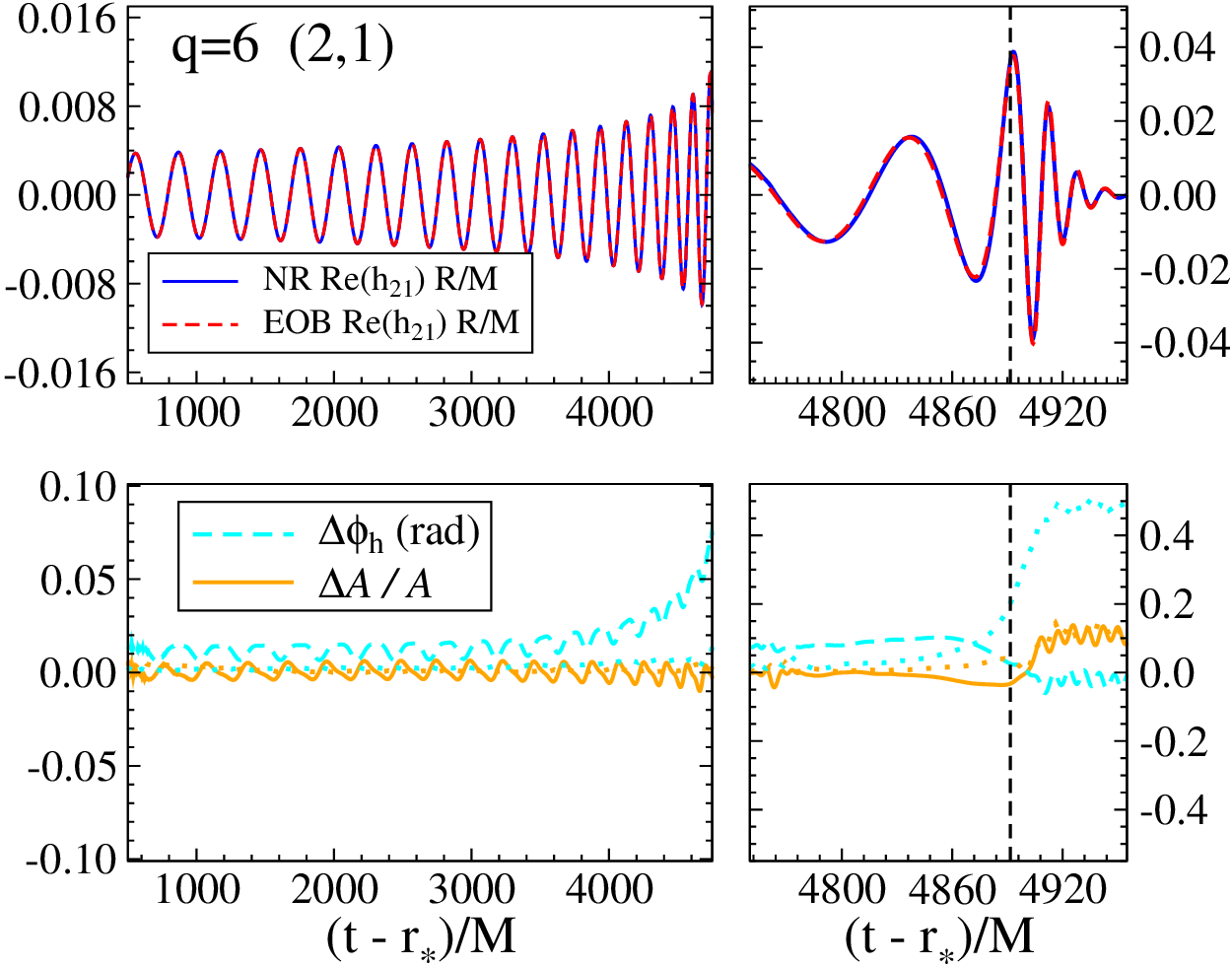}
  \caption{\label{fig:h21q36} Comparison of the $(2, 1)$ mode for mass-ratios $q=3$ (top panel) and
    $q=6$ (bottom panel) between numerical and calibrated
    EOB model. Plotted data as in Fig.~\ref{fig:q2346}.}
\end{figure}

\begin{figure}
  \includegraphics[scale=0.39]{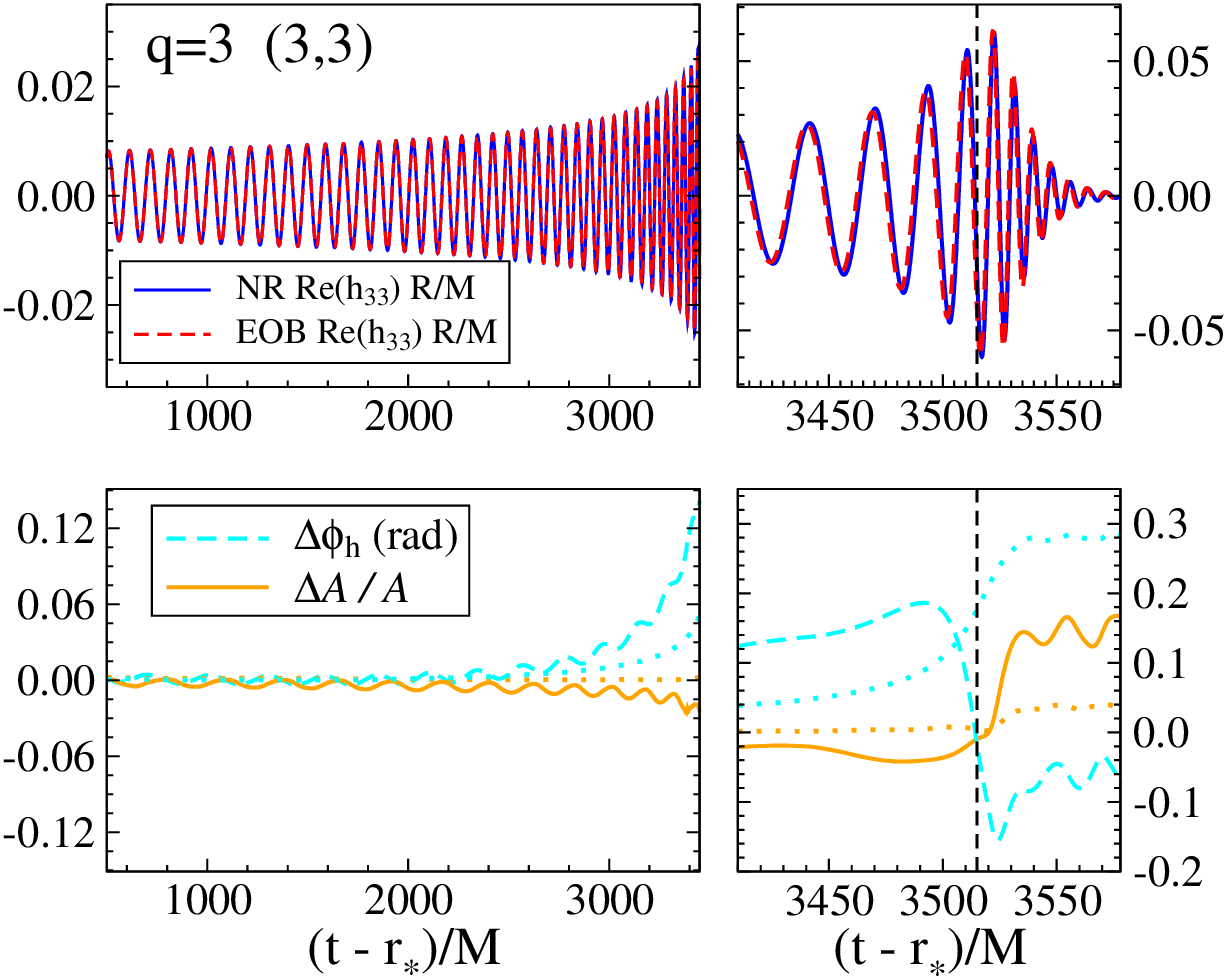} \\[1.5em]
  \includegraphics[scale=0.39]{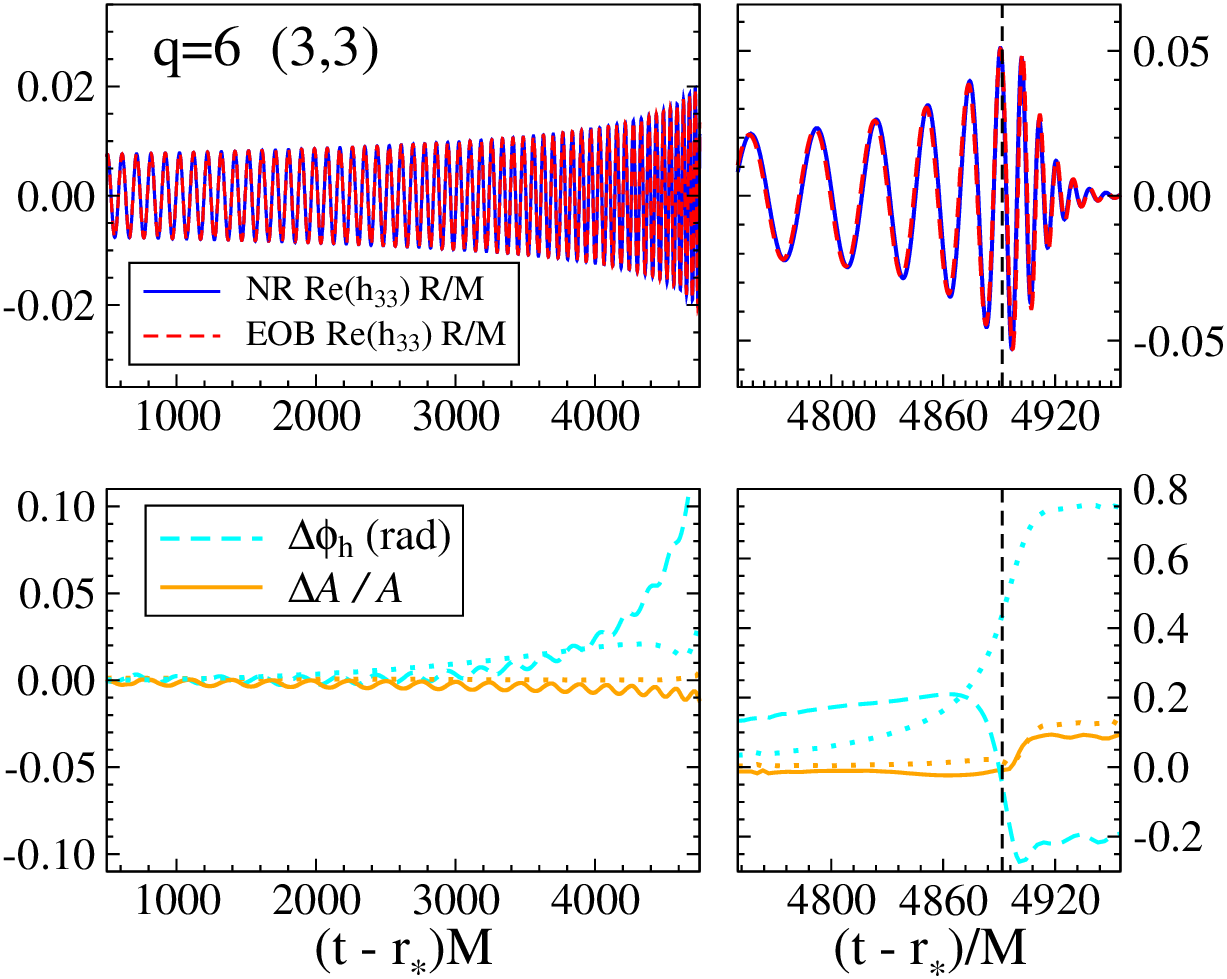}
  \caption{\label{fig:h33q36} Comparison of the $(3, 3)$ mode for mass ratios $q=3$ (top panel) and
    $q=6$ (bottom panel) between numerical and calibrated EOB
    model. Plotted data as in Fig.~\ref{fig:q2346}.}
\end{figure}

In Figs.~\ref{fig:h44q136}, \ref{fig:h21q36} and
\ref{fig:h33q36}, we compare the numerical and EOB subdominant modes
$h_{33}$, $h_{21}$ and $h_{44}$ for the cases $q = 1, 3, 6$. [For mass
ratios $q=2, 4$ the plots look similar, so we do not show them.]
During the inspiral, the numerical and EOB subdominant modes agree
very well, similarly to the agreement we found for the $h_{22}$. This
happens because the numerical frequencies $\omega_{\ell m}$ are well
modeled by a simple multiple of the orbital frequency
$m\,\Omega$. During merger and ringdown, the agreement is very good
for the $h_{33}$ and $h_{21}$ modes, i.e., comparable to the agreement
of the $h_{22}$ mode. Analogous performances hold for the other cases
$q=2$ and $q=4$.  The numerical and EOB $h_{44}$ mode, however, show
larger differences during ringdown. For instance, the phase difference
increases to $\sim 0.6$ rad. There are two reasons for this less
satisfactory result: (i) the larger errors in the numerical mode
$(4, 4)$ and (ii) the EOB QNM matching procedure that generates the
ringdown part. Issue (i) spoils the numerical predictions of the
fitting formulas for the $(4, 4)$ mode (see Table~\ref{tab:NRnufit})
which are essential to model the merger.  Issue (ii) prevents modeling
the ringdown phase of the $h_{44}$ with high accuracy (see
Fig.~\ref{fig:NRattQNM} and discussions therein). Nevertheless, since
as seen in Figs.~\ref{fig:hierarchy}, the $h_{44}$ amplitude is a few
percent of the $h_{22}$ amplitude, the absolute error in $h_{44}$ is generally smaller
than the error with which we currently model the $h_{22}$ EOB
mode. Therefore, the large difference between the numerical and EOB
$h_{44}$ is not the dominant source of systematic error in the
gravitational polarizations.  Since the $h_{55}$ mode comparison is
very similar to that of the $(4, 4)$ mode, except for an even larger
phase difference of $\sim 1$ rad during ringdown, we do not show it
for brevity. Modeling the $h_{55}$ mode is difficult due to the same
two issues discussed above, while on a more severe
level. Nevertheless, we find in Sec.~\ref{sec:overlaps} that there is
substantial benefit in including this mode in the full polarization
waveforms even though its modeling is not fully satisfactory.

We point out that the special treatment of the $(2,1)$ and $(4,4)$
modes in Eq.~\eqref{Veq}, namely the replacement of
$v_\Phi^{(\ell+\epsilon)}$ with $v_\Phi^{(\ell+\epsilon-2)}/r_\Omega$
for these modes, was necessary to improve the agreement of these modes
with the numerical waveforms.  Eq.~\eqref{Veq} was suggested by
similar studies in the test-particle limit~\cite{Barausse:2011a}. The
reason is the following: as shown in Fig.~\ref{fig:hierarchy} the
amplitude of the numerical $(2, 1)$ and $(4, 4)$ modes reaches a peak
quite after the peak of the $(2, 2)$ mode, i.e., they have large,
positive $\Delta t_\mathrm{peak}^{\ell m}$ values. If we want to
impose the condition 1. listed in Sec.~\ref{sec:EOBinspiralwaveforms},
the peak of the EOB mode should be moved to
$t_\mathrm{peak}^{\Omega}+\Delta t_\mathrm{peak}^{\ell m}$. However,
the EOB Newtonian amplitude is proportional to a power of the orbital
frequency and the latter decreases to zero at the EOB horizon, thus
the EOB amplitude drops to an extremely small value at
$t_\mathrm{peak}^{\Omega}+\Delta t_\mathrm{peak}^{\ell m}$.  By
contrast, replacing $v_\Phi^2=(r_\Omega\Omega)^2$ with $1/r_\Omega$,
we slow down the decay of these modes after
$t_\mathrm{peak}^{\Omega}$, and we can successfully move the peak of
the mode to $t_\mathrm{ peak}^{\Omega}+\Delta t_\mathrm{peak}^{\ell
  m}$.  Note that $v_\Phi^2$ and $1/r_\Omega$ are exactly the same
  in the adiabatic Keplerian limit. This replacement introduces higher
  order non-adiabatic non-Keplerian corrections that have negligible
  effects on the inspiral waveform and the energy flux. During plunge
  and merger, perturbative treatments break down and there is no a
  priori justification in describing the mode amplitude with a power
  of either $v_\Phi$ or $1/r_\Omega$. Equation~\eqref{Veq} is adopted since
  combined with the NQC corrections, it is capable of reproducing
  the numerical-relativity waveforms during merger. To make minimal
  adjustments to the Newtonian term of Eq.~\eqref{hlmNewt}, we
  introduce this replacement only when needed, i.e. to the $(2,1)$ and
  $(4,4)$ modes.

\section{Effectualness and measurement accuracy of EOB waveforms}
\label{sec:overlaps}

In this section, we examine the effectualness and measurement accuracy
of the EOB waveforms in matching the numerical-relativity waveforms.

\begin{figure}
  \includegraphics[width=8cm, clip=true]{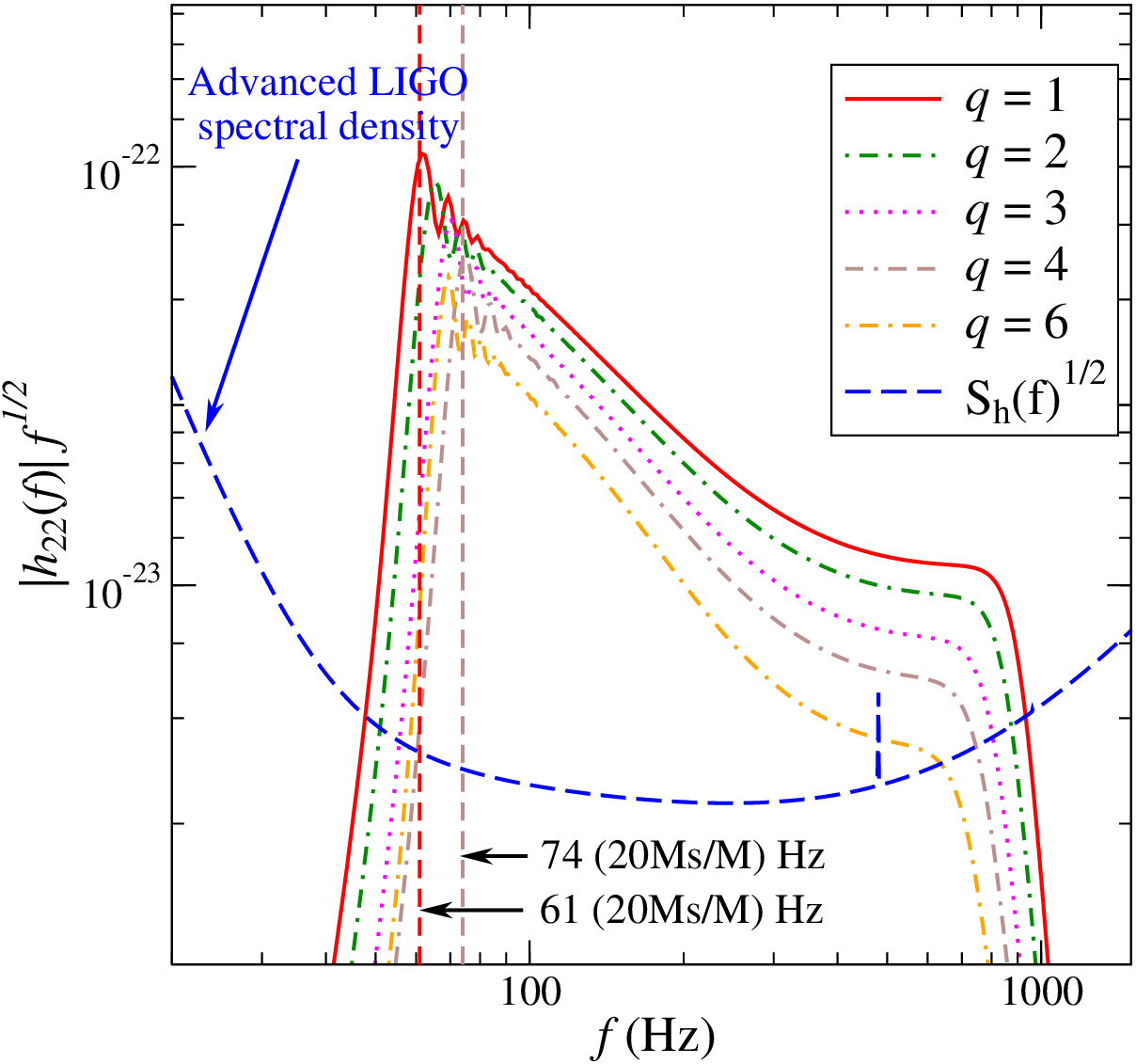}
  \caption{\label{fig:NRspectra} Amplitude of the Fourier transform of
    the (2, 2) mode of the numerical-relativity waveforms, scaled to a
    total mass $M=20M_\odot$.  We also plot the noise spectral density
    of the Advanced LIGO detector. The two vertical lines mark the
    initial gravitational-wave frequency for the numerical waveforms
    with $q=1$ and $q=4$ (lowest and highest start frequency of all
    considered waveforms). }
\end{figure}

These investigations utilize the noise-weighted inner product 
between two waveforms $h_i$, $i=1, 2$:
\begin{equation}
  \langle h_1, h_2\rangle\define4\,\Re
  \int_0^\infty\frac{\tilde{h}_1(f)\tilde{h}^*_2(f)}{S_h(f)}\;df\,,
  \label{InnerProduct}
\end{equation}
where $\tilde h_i(f)$ are the Fourier transforms of $h_i(t)$, and
$S_h(f)$ is the spectral density of noise in the detector.  We choose
one of the Advanced LIGO noise curves, named
\texttt{ZERO\_DET\_HIGH\_P} in Ref.~\cite{Shoemaker2009}.

Figure~\ref{fig:NRspectra} shows the noise curve, and the amplitudes
of the Fourier transforms of the numerical relativity waveforms. For the 
binary's total masses considered, the NR waveforms start in band; to reduce artifacts from this, we taper
the NR waveforms using the Planck-taper window
function~\cite{McKechan:2010kp}. The width of the window function is
set to the length of numerical relativity waveforms, which is about
$0.5(M/20M_\odot)$ seconds. The window function smoothly rises from 0
to 1 in the first 0.0625 seconds and falls from 1 to 0 in the last 0.0125 
seconds. Furthermore,
whenever we compute quantities involving both a NR waveform and an EOB
waveform (e.g., an overlap; see below), we restrict integration of the
EOB waveform to the frequency for which numerical data is available.

Figure~\ref{fig:NRspectra} indicates the initial gravitational-wave frequency of the
$q=1$ and $q=4$ simulations.  The simulation with $q=1$ has the lowest
initial gravitational-wave frequency, while the one with $q=4$ has the
largest initial gravitational-wave frequency. The numerical-relativity
waveforms for mass ratios $q=1, 2, 3, 4, 6$ have $32, 31, 31, 31$, and
$43$ gravitational-wave cycles, respectively, from the initial
frequency to the peak of the $(2, 2)$ mode. Using the EOB model of this
paper we estimate that for a total mass of $20 M_\odot$ and mass
ratios $q=1, 2, 3, 4, 6$, there are $582, 656, 779, 914, 1184$
gravitational-wave cycles between $10$ Hz and the start of the
numerical-relativity simulations.  These missing cycles, which
decrease as the total mass of the binary increases, are not accounted
for when computing mismatches.

\subsection{Subdominant modes}

To investigate the importance of subdominant modes
  $(l, m)$ different from $(2,2)$, we consider the
  gravitational waveform emitted from the binary into a given
  sky direction $(\theta, \phi)$ (as measured relative to the orbital
  plane of the binary 
and note that $\phi$ is degenerate with the initial phase), given as
\begin{equation}\label{eq:Ylm-expansion}
h_+(\theta,\phi;t ) - i h_\times(\theta,\phi) = \sum_{\ell, m} {}_{-\!2}Y_{\ell m}(\theta,\phi)\, h_{\ell m}(t)\,.
\end{equation}
Here ${}_{-\!2}Y_{\ell m}(\theta,\phi)$ is the $-2$ spin-weighted
spherical harmonic, and the summation on $\ell$ and $m$ is over the
available NR or EOB modes.\footnote{In the numerical simulations, 7 modes (14 modes if we count $m<0$) 
  are extracted: $(\ell, m)=(2,2)$, $(2,1)$, $(3,3)$,
  $(3,2)$, $(4,4)$, $(5,5)$, $(6,6)$. In the EOB model, 5 modes (10 modes if we count $m<0$) 
  modes are calibrated in Secs.~\ref{sec:eob22} and
  \ref{sec:eobneq22}: $(\ell, m)=(2,2)$, $(2,1)$,
  $(3,3)$, $(4,4)$, $(5,5)$.}  In the remainder of this
section, we always choose $\theta=\pi/3$ and $\phi=\pi/3$, and assume a
relative binary-detector configuration such that the detector is only
sensitive to $h_+$ (i.e., an antenna pattern $F_+=1, \, F_\times=0$). A
comprehensive study of arbitrary gravitational polarizations for
all sky directions $\theta,\phi$ is left to future work.

\begin{figure}
\includegraphics[width=\linewidth]{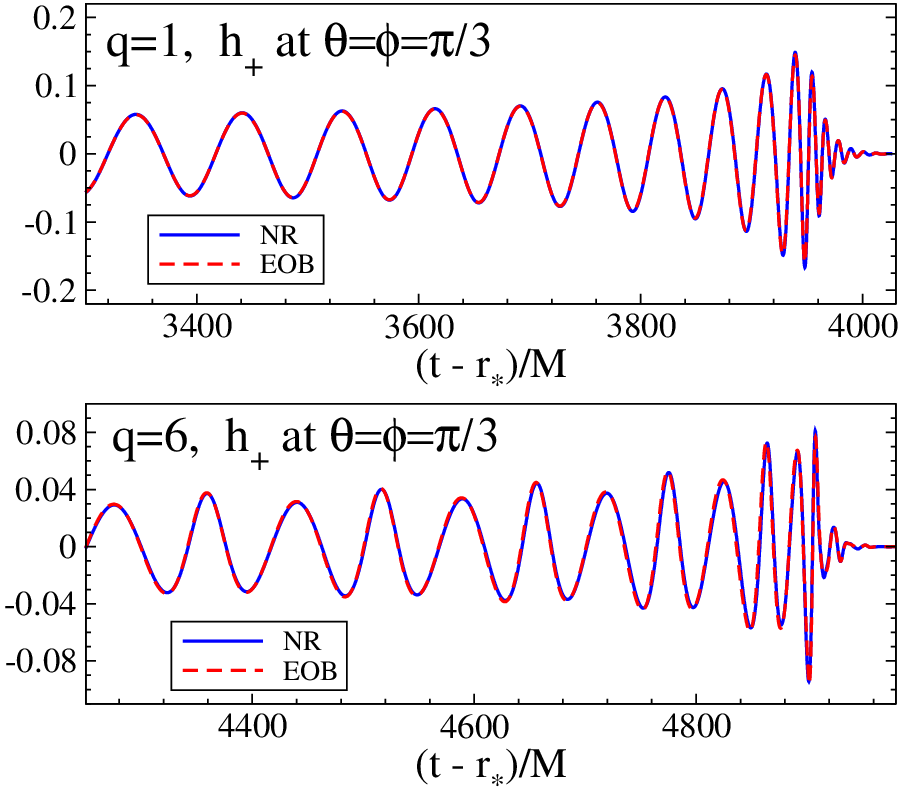}
\caption{
\label{fig:q1_q6_hplus}
The polarization waveform $h_+(\theta,\phi;t)$ as emitted into sky direction
$\theta=\phi=\pi/3$.  \textbf{Top panel:} mass ratio $q=1$; \textbf{bottom
  panel:} mass ratio $q=6$.  The solid blue curve represents the
numerical data, the red dashed curve the EOB model, and only late inspiral, merger and ringdown are shown.}
\end{figure}

Figure~\ref{fig:q1_q6_hplus} shows the resulting waveforms $h_+(\pi/3,
\pi/3; t)$ for the mass-ratios $q=1$ and $q=6$.  This figure clearly
shows that for $q=6$, subdominant modes are more important, and one
immediately expects that disregarding subdominant modes will have a
larger effect for the $q=6$ case. Let us now quantify these
expectations.

\subsection{Effectualness}

The effectualness can be described by the mismatch ($\MM$) between two
time-domain waveforms $h_1$ and
$h_2(t_0,\phi_0,\boldsymbol{\lambda})$.  Here, we consider all
waveforms to be the $+$ polarization evaluated in the direction
$\theta=\phi=\pi/3$ [see Eq.~\eqref{eq:Ylm-expansion}].  We take $h_1$ to
be the numerical relativity waveform at one of the simulated
mass ratios for some given total mass $M$.  The second waveform $h_2$
is taken to be our calibrated EOB model, where we have explicitly
displayed the dependence of this waveform on some reference time $t_0$
and reference phase $\phi_0$, as well as the masses $m_1$ and $m_2$
represented in the vector $\boldsymbol{\lambda}$ of the parameters of
the binary.

The mismatch
is given explicitly by~\cite{Pan2007}
\begin{equation} 
\MM \define 1-\max\;
 \frac{\langle
    h_1, h_2(t_0, \phi_0, \boldsymbol{\lambda})\rangle}
{ \lVert h_1\rVert\;\lVert h_2(t_0,\phi_0,\boldsymbol{\lambda})\rVert }
  \label{FF}
\end{equation}
where 
\begin{equation}\label{eq:norm}
\lVert h_{i} \rVert = \langle h_{i}, h_{i} \rangle^{1/2}
\end{equation}
denotes the norm induced by Eq.~\eqref{InnerProduct}.  When searching for the
signal waveform $h_1$ with the template $h_2(t_0, \phi_0,
\boldsymbol{\lambda})$, the horizon distance is reduced by a factor
$\MM$ relative to searching with the perfect template $h_1$, and the
reduction in event rate is given by $1-(1-\MM)^3\approx
3\MM$.  Ideally, the maximization in Eq.~\eqref{FF} is over
$\{t_0, \phi_0, \boldsymbol{\lambda}\}$; however, sometimes we choose
to neglect maximization over $\boldsymbol{\lambda}$, as detailed below.

\begin{figure}
  \includegraphics[scale=0.38]{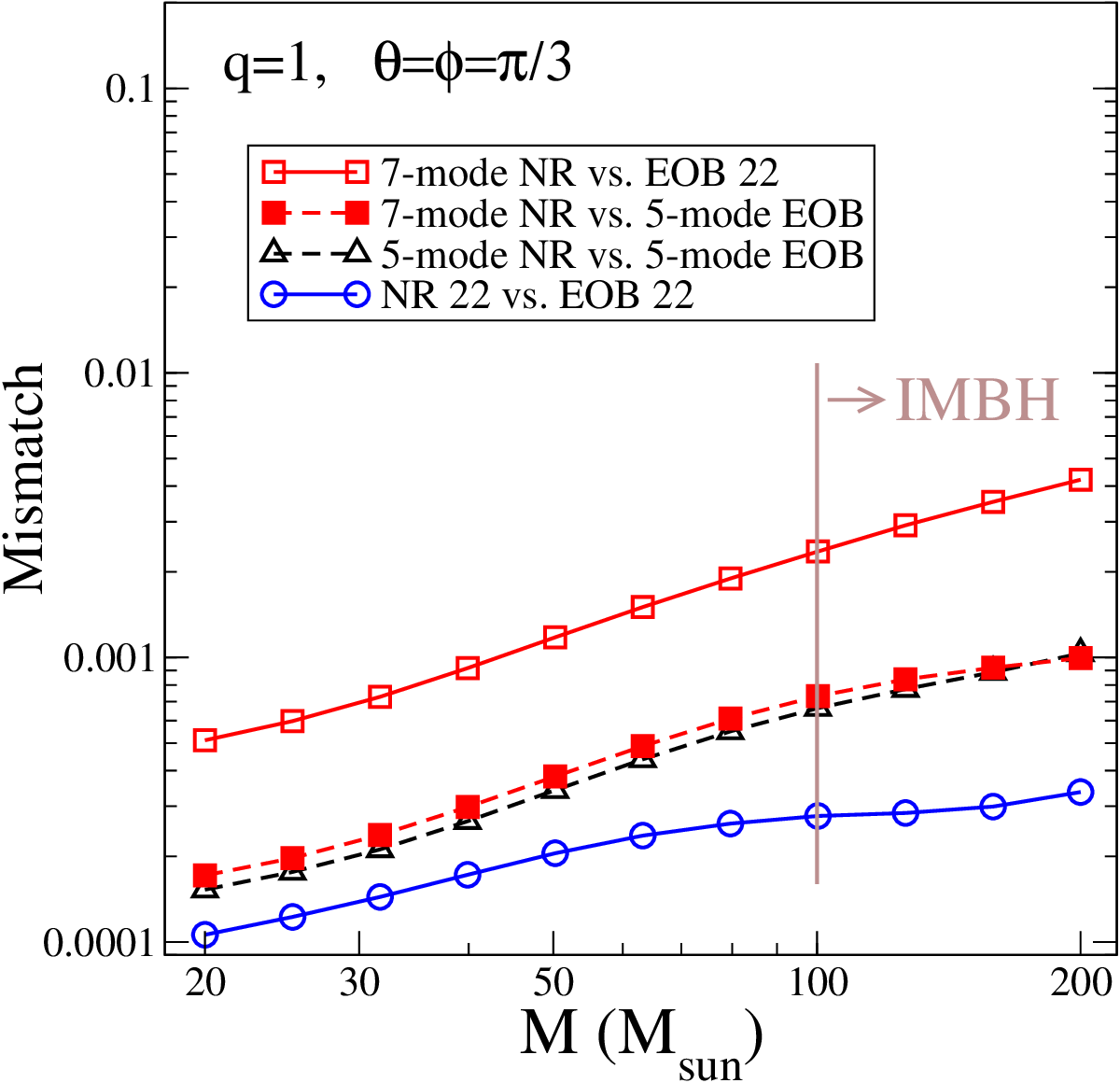}
  \caption{\label{fig:FF1} The mismatch versus binary total mass for
    $q = 1$ using Advanced LIGO noise curve.  Mismatches with the $(2,2)$ mode of the EOB waveform are optimized over $\{t_0,\phi_0,\boldsymbol{\lambda}\}$ whereas mismatches with the 5-mode EOB model are optimized over $\{t_0,\phi_0\}$ only
and represent an upper bound. The vertical line
    represents the maximum total mass for stellar-mass black-hole
    binaries, assuming a maximum black-hole mass of $50 M_\odot$. The
    range of total masses on the right of the vertical line refer to
    intermediate-mass black-hole binaries. }
\end{figure}

Figure~\ref{fig:FF1} presents several mismatch calculations for the
equal-mass case.  The solid lines compare the numerical relativity
data to the leading (2, 2) mode of our EOB model.  For these two
curves, maximization of $\MM$ is performed over $\{t_0,\phi_0,
\boldsymbol{\lambda}\}$.  If the numerical waveform is represented
only by its (2, 2) mode, then the mismatches are very small, reaching
$\sim 10^{-4}$.  However, if the 7-mode numerical waveform is used with
all modes shown in Fig.~\ref{fig:hierarchy}, then the mismatch
increases by about an order of magnitude, showing that subdominant
modes are noticeable even for the $q=1$ case.

The two dashed curves in Fig.~\ref{fig:FF1} use the calibrated EOB model
with all five calibrated modes included. For these two curves, we maximize
the mismatch only over $\{t_0,\phi_0\}$, for technical convenience and
to save computational cost.  Therefore the obtained mismatches are
only upper bounds. We see that the 5-mode EOB model
agrees significantly better with the 7-mode NR waveform than an
EOB model utilizing only the (2, 2) mode.  The line ``5-mode NR
vs. 5-mode EOB'' compares NR with EOB when both waveforms contain only
those five modes for which we calibrate the EOB model.

\begin{figure}
  \includegraphics[scale=0.38]{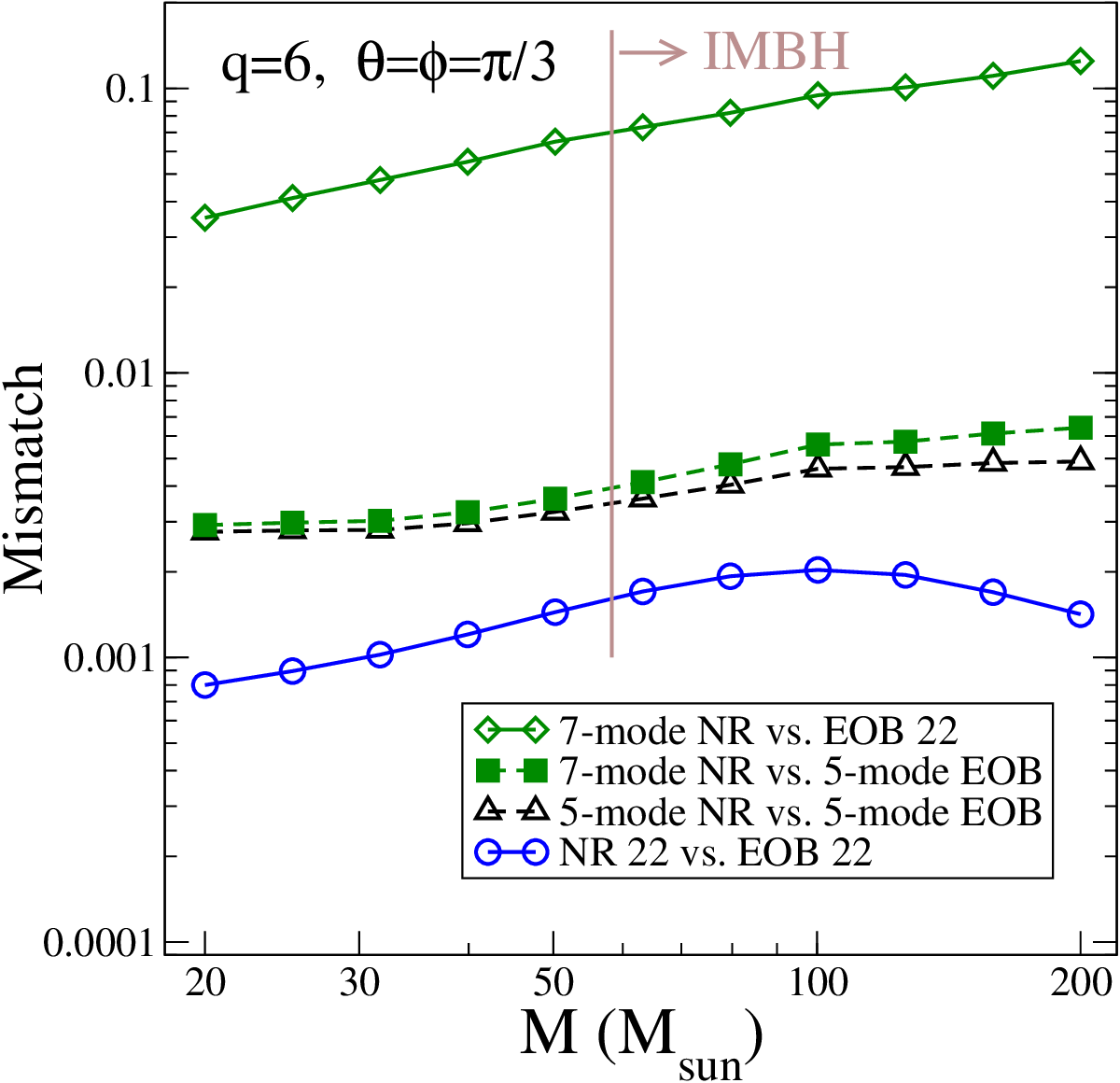}
  \caption{\label{fig:FF6} Mismatch calculation for mass ratio
      $q=6$.  All details as in Fig.~\ref{fig:FF1}.  This figure and
      Fig.~\ref{fig:FF1} use the same vertical axis to ease
      comparisons between them.}
\end{figure}

In Figs.~\ref{fig:FF1},~\ref{fig:FF6} and~\ref{fig:SNRdhdh} we have marked total mass $100 M_\odot$ and $58.3 M_\odot$ for $q=1$ and $q=6$, respectively, the largest masses for \emph{stellar-mass} binary black holes,
assuming that the maximum black-hole mass is $\sim 50
M_\odot$.\footnote{As of today, the heaviest mass of a single black-hole
  source is $23\text{--}34
  M_\odot$~\cite{2008ApJ...678L..17S, 2007ApJ...669L..21P}, but
  considering the possibility of lower metallicity we adopt as maximum
  mass of a black hole $\sim 50 M_\odot$~\cite{2010ApJ...714.1217B}.}
More massive black holes are referred to as \emph{intermediate}-mass
black holes (IMBHs). Their existence and gravitational-wave event rates are
more uncertain~\cite{2004IJMPD..13....1C, 2006ApJ...646L.135F}.

Figure~\ref{fig:FF6} presents the analogous calculations to
Fig.~\ref{fig:FF1} for mass ratio $q=6$.  
The mismatches are generally
larger, owing to the more complex waveform of a $q=6$ binary. 
The numerical $(2,2)$ mode can still be fit by a $(2,2)$ EOB waveform
to $\MM\sim 10^{-3}$. However, trying to
represent the 7-mode NR waveform with only the EOB $(2,2)$ mode results in
mismatches $\sim 7\%$ for total mass $M\sim 58.3 M_\odot$ and above
$10\%$ for total mass $M\sim 200M_\odot$. Thus, it is extremely
important to accurately model higher-order modes when the merger and
ringdown waveforms are in band and binary systems contain intermediate
mass black holes with $\gtrsim 50M_\odot$. So far, higher-order modes
have been largely ignored in the analysis of real detector data (see
for instance Ref.~\cite{Abadie:2011kd}), and by including a few
dominant ones, the event rate or the horizon distance can be
substantially increased, especially for high total masses.

Inclusion of the higher-order modes reduces the mismatches by
  a factor of 10 to $\sim 5\times 10^{-3}$ at low masses and $\sim
  0.01$ at high masses, caused mainly by the error in modeling the
  ringdown waveforms of the higher order modes.  The $(3, 2)$ and
  $(6, 6)$ modes that are not modeled or included in the EOB
  polarizations are not responsible for the comparatively large
  mismatch. To verify this, we show in these figures also the
  mismatches between the 5-mode EOB polarizations and 5-mode numerical
  polarizations that contain the same modes as the EOB
  ones.\footnote{Sometime the mismatches increase by $\sim 0.1\%$
    after we remove the $(3, 2)$ and $(6, 6)$ modes from the
    numerical-relativity polarizations. This happens because the
    numerical precision of our code in optimizing over the initial
    phase is $\sim 0.1\%$}  More diagnostic tests show that the error
  in modeling the $(5, 5)$ mode is responsible for about half the
  increase in $\MM$, because the phase error of the $(5, 5)$ mode
  accumulates faster due to its higher frequency, and because it is
  larger than the phase error of other modes.

As in Fig.~\ref{fig:FF1}, mismatches with the 5-mode EOB model
are not optimized over $\boldsymbol{\lambda}$ and represent therefore
merely upper limits.

We emphasize again that these mismatches measure the difference between
numerical-relativity and EOB waveforms only over the frequency band
where numerical-relativity simulations are available (see Fig.~\ref{fig:NRspectra}).  
For large masses, say $M=100 M_\odot$, the initial frequency of the numerical-relativity $(2, 2)$ mode for
$q=6$ is $15$Hz, and the missing part between $10$Hz and $15$Hz would
likely modify the mismatches only marginally because of the steeply
rising seismic noise wall toward low frequency.  However, for $M=20
M_\odot$ and $q=6$, the initial frequency of the numerical-relativity
$(2, 2)$ mode is $70$Hz which is in the frequency
band of the detector.  In this case we miss the portion of the
numerical-relativity waveform between $10$Hz and $70$Hz. The best
way, though time-consuming, to address this gap problem is to produce
longer numerical-relativity waveforms, or to conduct tests within the
analytical models to assess their reliability below a certain
frequency~\cite{Pan2007, Buonanno:2009, Hannam:2010, Damour:2010,
  MacDonald:2011ne, Boyle:2011dy}.  With the caveat of this frequency gap,
 we conclude that our calibrated EOB model is sufficient for detection purposes.

\subsection{Measurement accuracy}

We are now interested in the question of whether the EOB polarizations
are accurate enough to be used in data analysis for measurement
purpose. We adopt as accuracy requirement for measurement the one
proposed by~\cite{Miller2005, Lindblom2008}
\begin{equation}\label{SNRlimit}
\lVert \delta h \rVert <  \epsilon \,,
\end{equation}
where $\delta h=h^\mathrm{EOB}(t)-h^\mathrm{NR}(t)$ is the error in modeling
the numerical waveforms, and $\epsilon<1$ incorporates a
  safety factor~\cite{Damour:2010} or the effect of a detector
  network~\cite{MacDonald:2011ne}. The left hand side in
Eq.~\eqref{SNRlimit} increases proportionally with the signal-to-noise (SNR) ratio 
$\lVert h^\mathrm{NR}\rVert$ and we calculate the upper bound of the $\mathrm{SNR}^\mathrm{eff}\equiv\mathrm{SNR}/\epsilon$ that satisfies Eq.~\eqref{SNRlimit}. For any SNR below this upper bound, the EOB
waveforms or polarizations are accurate enough for measurement
purposes, i.e., accurate enough not to generate any systematic bias
that is larger than statistical errors in estimating the physical
binary parameters. The upper bound $\mathrm{SNR}^\mathrm{eff}$ is therefore
a very strict accuracy requirement. Unlike the well known fact that
good phase agreement is critical for obtaining good effectualness, to
get high (upper bounds on) $\mathrm{SNR}^\mathrm{eff}$, both the amplitude and
the phase of the templates must agree very well with those of the
exact waveforms.

\begin{figure}
  \includegraphics[width=\linewidth]{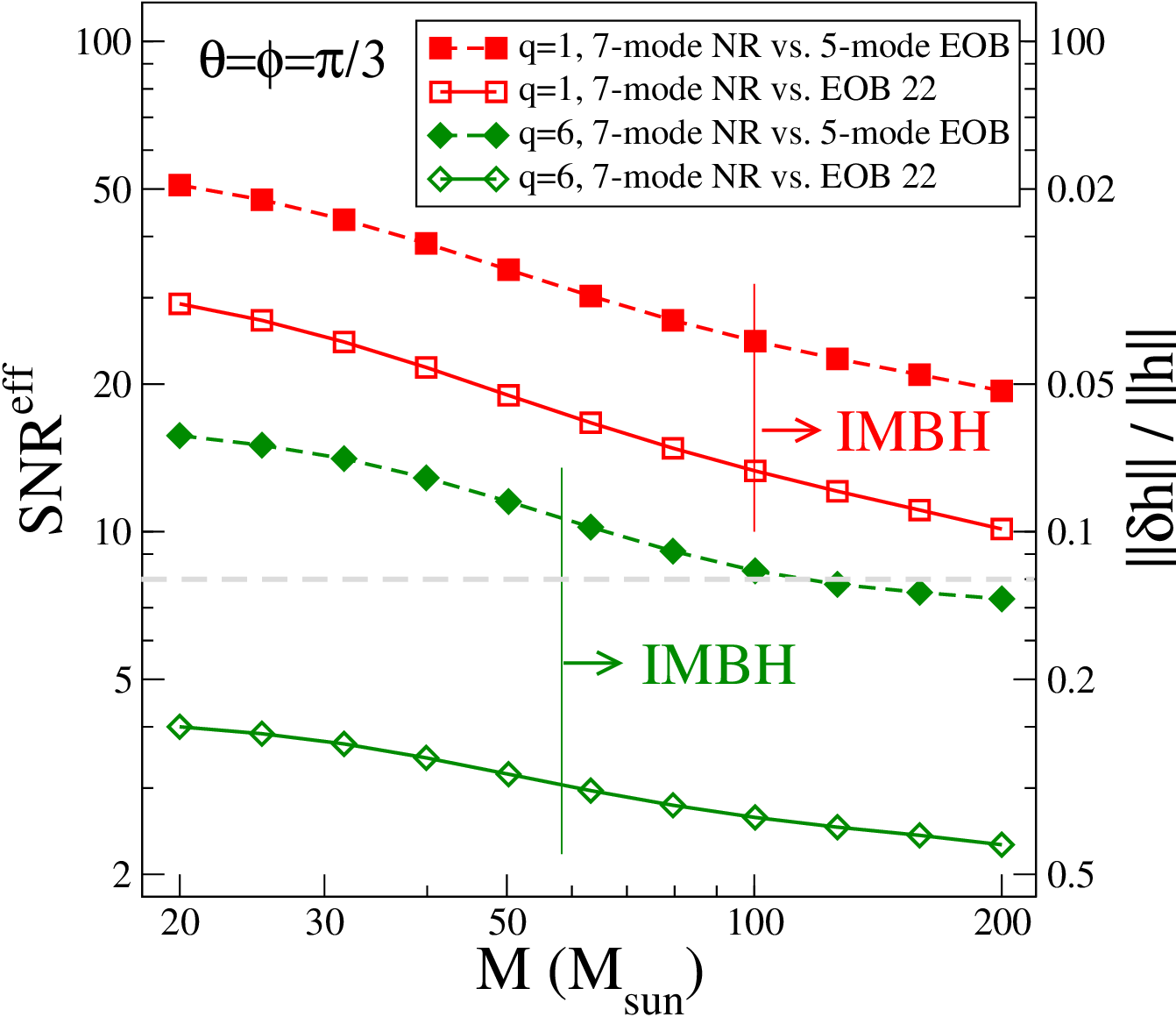}
  \caption{\label{fig:SNRdhdh} 
The upper bound $\mathrm{SNR}^\mathrm{eff}
    \equiv \mathrm{SNR}\,\epsilon$ from the measurement accuracy
    requirement ~\eqref{SNRlimit} versus binary total mass for $q = 1$
    and $6$ using Advanced LIGO noise curve. For any $\mathrm{SNR}^\mathrm{
      eff}$ below the curves, the EOB polarizations are accurate
    enough to avoid a systematic bias that is larger than
    statistical errors when estimating the binary parameters. The
    horizontal line indicates the single detector SNR for Advanced
    LIGO which is 8.  The two vertical lines represent the maximum
    total mass for stellar-mass binary black holes with maximum
    black-hole mass of $50 M_\odot$ and $q=1$ and
    $q=6$.  The range of total masses on the right
    of those vertical lines refer to intermediate-mass black-hole
    binaries. The vertical axis on the right shows $\lVert \delta
  h\rVert / \lVert h\rVert$.}
\end{figure}

In Fig.~\ref{fig:SNRdhdh}, for mass ratios $q=1$ and $q=6$, we show
the upper bound of the $\mathrm{SNR}^\mathrm{eff}$ as a function of the
total mass. These $\mathrm{SNR}^\mathrm{eff}$ are calculated for a single
Advanced LIGO detector. In the $q=1$ case, the 5-mode EOB
polarizations match the 7-mode numerical polarizations 
accurately enough for any $\mathrm{SNR}^\mathrm{eff}<24$ when the total mass is below
$100 M_\odot$, and for any $\mathrm{SNR}^\mathrm{eff}<19$ when the total mass is below
$200M_\odot$.  These upper bounds of the $\mathrm{SNR}^\mathrm{eff}$ may not
seem impressive at the first sight.  However, there is a significant
improvement if we compare them with the upper bounds obtained when
only the EOB $(2, 2)$ mode is used which are also shown in
Fig.~\ref{fig:SNRdhdh}. Note that the right vertical axis in
Fig.~\ref{fig:SNRdhdh} shows 
$\lVert\delta h\rVert / \lVert h\rVert$. 
Once the EOB and NR waveforms are aligned at low frequency, we do not allow 
further time or phase shift in calculating $\delta h$.
This differs from the practice of
  Refs.~\cite{Damour:2010, MacDonald:2011ne}  which plot $\lVert\delta
  h\rVert/\lVert h\rVert$ minimized over time- and phase-shifts.

In the $q=6$ case, the upper-bound SNRs are lower, because of the
larger error in modeling the higher-order modes and the relatively
high contribution to the SNR from higher-order modes for such an
asymmetric binary.  For stellar-mass black holes, $M\lesssim 58.3 M_\odot$, the
5-mode EOB polarizations are accurate for $\mathrm{SNR}^\mathrm{eff}< 10$,
and when the total mass is below $200M_\odot$, the 5-mode EOB
polarizations are accurate for $\mathrm{SNR}^\mathrm{eff} <5$. The higher
order modes in the EOB model, especially their ringdown part, clearly
needs better modeling. Nevertheless, the improvement from using only
the EOB $(2, 2)$ mode as templates is significant.

In closing, we mention a few important caveats of our
  analysis: First, all norms and mismatches computed for
  Figs.~\ref{fig:FF1}--\ref{fig:SNRdhdh} are performed only over
  those frequencies for which numerical data is available.  At lower
  total mass, more gravitational-wave cycles lie in the LIGO frequency band, so this
  restriction becomes more severe.  Therefore, at low masses, our
  analysis might yield an increasingly overoptimistic view.

On the other hand, the requirement \eqref{SNRlimit} on the
accuracy measurement may be unnecessarily stringent. In fact,
although Fig.~\ref{fig:SNRdhdh} would say that for certain mass ratios
and total masses, the EOB waveforms will introduce biases in the
binary parameters, these biases may affect binary parameters that have
little astrophysical relevance---for example the gravitational-wave
phase at coalescence.

Finally, we emphasize again that the results in this section are just
a first step into the problem of including higher-order modes, since
we investigated only one geometrical configuration. A comprehensive
investigation of the gain in effectualness and measurement accuracy,
as well as the impact on estimating the source parameters, when
including these rather accurate higher-order modes will be presented
in a separate paper.

\section{Relation to previous work}
\label{sec:prevwork}

The EOB model we consider here differs from the nonspinning EOB model
employed by Buonanno \etal~\cite{Buonanno:2009qa} in the handling of
the radiation-reaction sector.  Reference~\cite{Buonanno:2009qa}
adopted a \Pade-resummed radiation-reaction force and energy
flux~\cite{Damour98, Boyle2008a}, while here we adopt the
factorized-waveform energy flux of Refs.~\cite{DIN, Fujita:2010xj,
  Pan2010hz}. References~\cite{Pan:2009wj, Pan2010hz} found that when
generalizing the \Pade-resummed flux to the spin case the agreement
with the numerical energy flux is not very satisfactory. Thus, we
concluded that the nonspinning EOB model with \Pade-resummed flux is
not a very good candidate for the generic spin EOB model.

By using the SpEC numerical merger $(2, 2)$ mode with mass ratio $q
=1$, and inspiral $(2, 2)$ mode with $q = 2, 3$, Buonanno
\etal~\cite{Buonanno:2009qa} calibrated the 4PN parameter in the EOB
radial potential $A(r)$ and the parameter $v_\mathrm{pole}$ in the
\Pade-resummed energy flux, such that the EOB model could be also used
outside the range of binary masses employed to calibrate it. We find
that when comparing to the SpEC merger $(2, 2)$ mode of this paper
with mass ratios $q = 1, 2, 3$, the EOB $(2, 2)$ mode of
Ref.~\cite{Buonanno:2009qa} have maximum phase difference until 
merger of $0.12$, $0.22$, and $0.09$ rad, respectively.  In contrast, the model
calibrated here results in smaller or comparable phase differences of
$0.04$, $0.08$ and $0.12$ rad, respectively. Numerical data for
$q=4, 6$ were not available for the calibration in
Ref.~\cite{Buonanno:2009qa}. Comparing the new numerical data
available now with the model from Ref.~\cite{Buonanno:2009qa}, we find
phase differences of $0.43$ and $1.8$ rad, respectively, for $q=4$
and $6$ (in contrast, our new model results in $0.12$ and $0.15$
rad, respectively; see Fig.~\ref{fig:q2346}). Those phase
differences are obtained by using the low-frequency alignment
procedure of Eq.~\eqref{waveshifts}.  If we were adopting the
two-pinching frequency procedure of Ref.~\cite{Damour2009a}, we would
obtain $0.065$ rad with pinching frequencies $M \Omega = 0.052$
and $=0.3$ for $q=4$, and $0.18$ rad with pinching frequencies $M
\Omega = 0.056$ and $=0.15$ for $q=6$.

Before calibration, the radiation-reaction sector of the EOB model used here almost coincides\footnote{Whereas we
  use the factorized waveforms with $\rho_{\ell m}$,
  Ref.~\cite{Damour2009a} employs factorized waveforms where
  $(\rho_{22})^2$ is traded with its Taylor expanded form and then
  \Pade resummed~\cite{DIN}. In the factorized waveforms,
  Ref.~\cite{Damour2009a} also adopts a different value for the
  constant in the tail term $r_0=2$, a different odd-parity source
  term $\hat{S}^{(1)}_\mathrm{
    eff}(r, p_{r_*}, p_{\Phi})=p_{\Phi}\Omega/v_\Omega^2$, and a
  different $N_{\ell m}$ given in Eq. (5) of Ref.~\cite{Damour2009a}.
  Furthermore, while we include radiation reaction force in
  Eq.~\eqref{eq:eobhamthree}, i.e., in the equation of motion of
  $p_{r_*}$, Ref.~\cite{Damour2009a} does not.  Finally,
  Ref.~\cite{Damour2009a} trades the Keplerian velocity with the
  non-Keplerian velocity in $T_{\ell m}$, $\rho_{\ell m}$ but not
  $\delta_{\ell m}$, and does not include the higher-order PN terms
  computed in Ref.~\cite{Pan2010hz}.} with the nonspinning EOB model
used in Damour and Nagar~\cite{Damour2009a} in the radiation-reaction
sector, but only in its uncalibrated version. In fact, the adjustable
parameters used here differ from the ones used in
Ref.~\cite{Damour2009a}. Moreover, in this paper we also introduce
adjustable parameters in the phase of some of the $(\ell, m)$ modes,
and in some cases also in the factorized amplitude. Furthermore, we
also modify the leading Newtonian term for the $(2, 1)$ and $(4, 4)$
modes [see Eq.~\eqref{Veq}].

By using the SpEC numerical-relativity merger $(2, 2)$ mode with mass
ratio $q = 1$, and merger $(2, 2)$ mode with $q = 2, 4$ from the Jena
numerical-relativity group, Damour and Nagar~\cite{Damour2009a}
calibrated the 4PN and 5PN parameters in the radial potential
$A(r)$. We find that when comparing to the SpEC $(2, 2)$ modes of this
paper with mass ratios $q = 1, 2, 4$, the EOB $(2, 2)$ mode of
Ref.~\cite{Damour2009a} has maximum phase difference until 
merger of $0.25$ rad, $0.36$ rad, $1.32$ rad, respectively, when aligning at low frequency
and $0.05$ rad, $0.11$ rad, $0.25$ rad when using the two-frequency
pinching procedure.  In the case of mass ratios $q = 3$ and $6$ where
numerical waveforms were not available for calibration, the phase
differences increase to $0.93$ rad and $2.3$ rad, respectively, when
aligning at low frequency and $0.17$ rad, $0.65$ rad, when using the
two-frequency pinching procedure.

We notice that once we have calibrated the EOB model to a set of
numerical-relativity waveforms using the low-frequency alignment
procedure of Eq.~\eqref{waveshifts}, the phase difference with
numerical-relativity waveforms is not sensitive to the alignment
method---for example we find the same phase differences when using the
two-pinching frequency procedure. By contrast, EOB waveforms
calibrated with the two-frequency pinching procedure do not have the
same phase difference with numerical-relativity waveforms when
aligning them at low frequency.  We deduce then that the low-frequency
alignment procedure is more robust.

Recently, the LIGO/Virgo detectors have completed the first search for
gravitational waves from high-mass black holes using analytical
inspiral, merger and ringdown templates~\cite{Abadie:2011kd}. They
employed the EOB model calibrated to numerical waveforms from
NASA-Goddard of Ref.~\cite{Buonanno2007}. The mismatches between 
the EOB and NASA-Goddard waveforms computed in Ref.~\cite{Buonanno2007} 
are less than $3\%$ when the total mass is $25\text{--}99 M_\odot$, which is the 
mass range used in the LIGO/Virgo search~\cite{Abadie:2011kd}. These mismatches 
were derived maximizing only on the time and phase, but not on the 
binary parameters. We find that when comparing the EOB $(2, 2)$ mode of Ref.~\cite{Buonanno2007}
with the EOB model developed in this paper, the phase and amplitude
differences at merger are at most of $5$ rad and $20\%$, respectively,
when $q = 1, 2, 3, 4$, and $5.8$ rad for $q=6$. 
The EOB higher-order modes $(2, 1)$, $(3, 3)$, and
$(4, 4)$ were also calibrated for the first time in Ref.~\cite{Buonanno2007} 
using the NASA-Goddard numerical waveforms. In this case the matching 
between the inspiral-plunge and merger-ringdown waveforms was chosen 
at the same point in time for all the modes. Those higher-order modes
were not employed in the search of Ref.~\cite{Abadie:2011kd}.

\section{Conclusions}
\label{sec:conclusions}

The first search for gravitational waves from nonspinning high-mass
binary black holes ($M = 25\text{--}99 M_\odot$) with LIGO and Virgo
detectors has been recently completed, setting astrophysically
meaningful upper limits~\cite{Abadie:2011kd}.  The search has used for
the first time templates which include inspiral, merger and
ringdown. Those templates were built by combining numerical-relativity
and analytical-relativity results, either through the EOB waveforms of
Ref.~\cite{Buonanno2007} (see also the most recent improvements in
Refs.~\cite{DN2007b, DN2008, Damour2009a, Buonanno:2009qa, Pan:2009wj, Bernuzzi:2010xj}) or
the phenomenological merger-ringdown waveforms proposed in
Refs.~\cite{Ajith-Babak-Chen-etal:2007b} (see also
Ref.~\cite{Santamaria:2010yb} for an updated version).

In this paper we have built on
Refs.~\cite{Buonanno-Cook-Pretorius:2007, Buonanno2007, Buonanno:2009qa, Pan:2009wj, DN2007b, DN2008, DIN, 
Damour2009a}, and have improved further the EOB model taking advantage of highly
accurate numerical-relativity simulations with mass ratios $q =
1, 2, 3, 4, 6$ from the Caltech-Cornell-CITA
collaboration~\cite{Buchman-etal-in-prep}.

By extracting several numerical-relativity quantities, such as the
mode's amplitude and its second time derivative at the peak, as well
as the frequency and its first time derivative at the peak, we have
improved the agreement of numerical and EOB phase and amplitude very
close to merger, reproducing important features of the numerical
simulations---for example the fact that whereas the $(2, 2)$ mode peaks
at the same position of the EOB light ring, the higher-order modes
peak at late
times~\cite{Buonanno2007, Schnittman2007, Baker2008a, Bernuzzi:2010ty, Bernuzzi:2010xj, Barausse:2011a}.

We have found that the $(2, 2)$ mode can be calibrated with $\MM \leq
0.1\%$ for $M = 20\text{--}200 M_\odot$ and mass ratios $q = 1, 2, 3,
4, 6$, using only three EOB adjustable parameters, notably the 4PN and
5PN terms, $a_5$ and $a_6$, in the radial EOB potential, and the width
of the comb, $\Delta t_\mathrm{match}^{22}$ (see
Table~\ref{tab:adjparams}). We have also found that the strongest
subdominant modes $(2, 1)$, $(3, 3)$, $(4, 4)$ and $(5, 5)$, can be
successfully calibrated by including for each mode four EOB adjustable
parameters, specifically the 3PN terms in $\rho_{21}, \rho_{33}$,
$\rho_{44}$, $\rho_{55}$, the 2.5PN or 3.5PN terms in $\delta_{21},
\delta_{33}$, $\delta_{44}$ and $\delta_{55}$, the width of the comb
$\Delta t_\mathrm{match}^{21}, \Delta t_\mathrm{match}^{33}, \Delta t_\mathrm{
  match}^{44}, \Delta t_\mathrm{match}^{55}$, and, in some cases, a
pseudo QNM (see Table~\ref{tab:adjparams}).  The reason of introducing
more parameters for higher modes rests on the fact that those modes
are known at PN orders lower than the $(2, 2)$ mode.
Furthermore, to achieve this very good agreement of the 
modes' phase and amplitude we have also used the information from numerical-relativity 
about the peak's amplitude and frequency of Table ~\ref{tab:NRnufit}, and the final masses 
and spins of Eq.~(\ref{finalMS}). These data determine the complex 
amplitudes entering the merger-ringdown waveform (\ref{RD}), and the 
NQC coefficients in Eq.~(\ref{Nlm}).

When investigating the effectualness for detection purposes, we have
found that the numerical-relativity polarizations containing the
strongest seven modes have a maximum mismatch of $7\%$ for
stellar-mass binaries, and $10\%$ for intermediate mass binaries, when
only the EOB $(2, 2)$ mode is included for $q = 1, 2, 3, 4, 6$ and
binary total masses $20\text{--}200$ Hz.  However, the mismatches
decrease when all the four subleading EOB modes calibrated in this
paper are taken into account reaching an upper bound of $0.5\%$ for
stellar-mass binaries, and $0.8\%$ for intermediate mass binaries.
Event rates or horizon distances can be substantially increased,
especially for high total masses, if those subleading modes were
included in gravitational-wave searches~\cite{Abadie:2011kd}.

We have also emphasized that when computing the mismatches, we do not
attach any PN waveforms to the numerical-relativity waveforms, because
we do not want to introduce any error due to the procedure of building
hybrid PN--numerical waveforms. Moreover, for binaries with low total mass---say $20\text{--}100 M_\odot$---many more gravitational-wave cycles than
the ones of the numerical simulations used in this paper are in
band. Our mismatches do not take into account these missing
cycles. As a consequence if the EOB model were used to search for
signals of length larger than the one of the numerical waveforms
employed here, the mismatches could become larger.

We have also studied the measurement accuracy of the EOB model using
the accuracy requirement proposed in Ref.~\cite{Lindblom2008}. Using
one single Advanced LIGO detector, we have determined the SNRs below
which the EOB polarizations are accurate enough that systematic biases
are smaller than statistical errors. Unlike the well known fact that
good phase agreement is sufficient for obtaining good effectualness,
to get high upper-bound SNRs, both the amplitude and the phase of the
templates must agree very well with those of the exact
waveforms. Since higher-order modes have non-negligible contribution for large
mass ratios, and those modes have the largest amplitude errors, we have found
that the upper-bound SNRs are lower for the most asymmetric systems.
We stress again the relevance of modeling the higher-order modes,
because using only the $(2, 2)$ mode would decrease significantly the
upper-bound SNRs. However, it is worth to note that the accuracy
requirement that we used may be too stringent since by itself
it does not say which of the binary parameters is going to have
biases~\cite{CutlerV:2007}. It could turn out that the biased parameters have little
relevance in astrophysics or tests of general relativity.

Finally, we have used rather long numerical-relativity waveforms, in
particular the case $q=6$ has forty gravitational-wave cycles before
merger. Confirming previous
studies~\cite{Pan2007, Buonanno:2009, Hannam:2010, Damour:2010, MacDonald:2011ne, Boyle:2011dy}
we have found, that especially for large mass ratios, the addition of
more cycles at low frequency \emph{does} affect the accuracy of the
EOB waveforms (as any other PN waveforms) which were calibrated to a
shorter number of cycles.  So, we had to re-calibrate the EOB
adjustable parameters to achieve very small phase errors around
merger. This re-calibration is crucial for parameter estimation, but
not for detection, and we expect to do it again in the future
when longer or more accurate
numerical-relativity waveforms for asymmetric systems will become
available. Moreover, we plan to improve the matching procedure from
inspiral-plunge to merger-ringdown, since the majority of the phase
and amplitude error is accumulated during this transition, especially
for higher-order modes. Of course, further improvement of the EOB 
higher-order modes also depends on the availability of sufficiently 
accurate numerical-relativity data especially during the last stages 
of inspiral, merger and ringdown.

\begin{acknowledgments} 
  We thank Enrico Barausse and Andrea Taracchini for useful
  discussions, and Ben Lackey and Cole Miller for informative
  interactions.

  A.B. and Y.P. acknowledge support from NSF Grant PHY-0903631.
  A.B. also acknowledges support from NASA grant NNX09AI81G.  M.B.,
  L.B., L.K., H.P., and M.S. are supported in part by grants from the
  Sherman Fairchild Foundation to Caltech and Cornell, and from the
  Brinson Foundation to Caltech; by NSF Grants No. PHY-0601459 and
  No. PHY-0652995 at Caltech; by NASA Grant NNX09AF97G at Caltech; by
  NSF Grants No. PHY-0652952 and No. PHY-0652929 at Cornell; and by
  NASA Grant No. NNX09AF96G at Cornell.  H.P. gratefully acknowledges
  support from the NSERC of Canada, from Canada Research Chairs
  Program, and from the Canadian Institute for Advanced Research.
\end{acknowledgments}

\appendix

\section{The \boldmath$(\ell, m)=(3, 2)$ mode}
\label{app:h32}

Reference~\cite{Buonanno-Cook-Pretorius:2007} found that the
numerical-relativity $(3, 2)$ mode contains QNMs with $(\ell, m)=(3, 2)$
and $(\ell, m)=(2, 2)$. This beating of QNMs with the same $m$ but
different $\ell$ arises in the transformation from the spin-weighted
spheroidal harmonics (which are eigenmodes of the radiation generated
by the perturbed final black hole) to the spin-weighted spherical
harmonics that are used to decompose the multipolar waveform.  This is
a general feature of modes with $\ell > 2$ and $m < \ell$, and since
the $(3, 2)$ mode is the dominant one among such modes, we discuss its
modeling and possible calibration in this section.

\begin{figure}
  \includegraphics[width=8.5cm, clip=true]{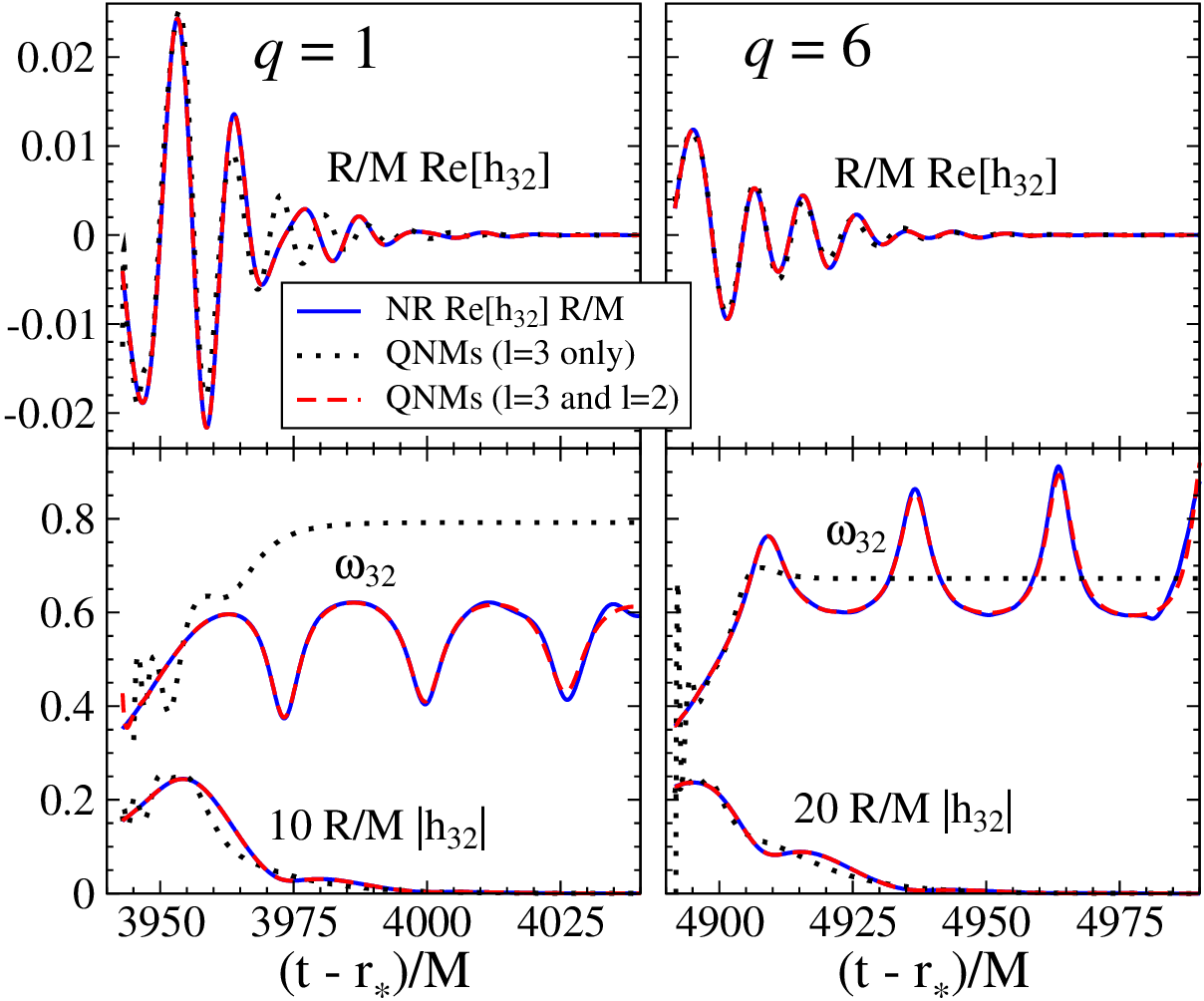}
  \caption{\label{fig:32rdfit} Investigation of the $(3, 2)$ mode
    during ringdown for $q=1$ and 6. The top panels show the mode
    itself, and the lower panels split the mode into amplitude and
    frequency. The continuous lines represent the numerical-relativity
    modes. The dotted lines are a fit to eight QNM modes of ($\ell=3,
    m=2, n=0, 1, \ldots, 7$). The dashed lines are a fit to eight QNMs
    of $(\ell=3, m=2, n=0, 1,\dots, 4)$ and $(\ell=2, m=2, n=0, 1,
    2)$. The horizontal axis is the retarded time in the
    numerical-relativity simulation.}
\end{figure}

First, we confirm the result of
Ref.~\cite{Buonanno-Cook-Pretorius:2007} that the ringdown portion of
the $(3, 2)$ mode can be accurately modeled by a linear superposition
of QNMs with both $(\ell, m)=(3, 2)$ and $(\ell, m)=(2, 2)$ modes.  The
result of Ref.~\cite{Buonanno-Cook-Pretorius:2007} was restricted to
the equal-mass case.  Here we extend this analysis to unequal-masses and 
add more overtones.  We fit the numerical ringdown mode with either
(i) a set of eight QNMs with $(\ell=3, m=2, n=0, 1,\dots, 7)$ or (ii) a
set of eight QNMs with $(\ell=3, m=2, n=0, 1,\dots, 4)$ and
$(\ell=3, m=2, n=0, 1, 2)$. For mass ratios $q=1$ and $6$, we compare the
fitting results with numerical waveforms in
Fig.~\ref{fig:32rdfit}. The results for other mass ratios are
similar. The very different performance of sets (i) and (ii) shows
clearly that the numerical ringdown mode does contain contributions
from $(\ell, m)=(2, 2)$ QNMs.

\begin{figure}
  \includegraphics[width=7cm, clip=true]{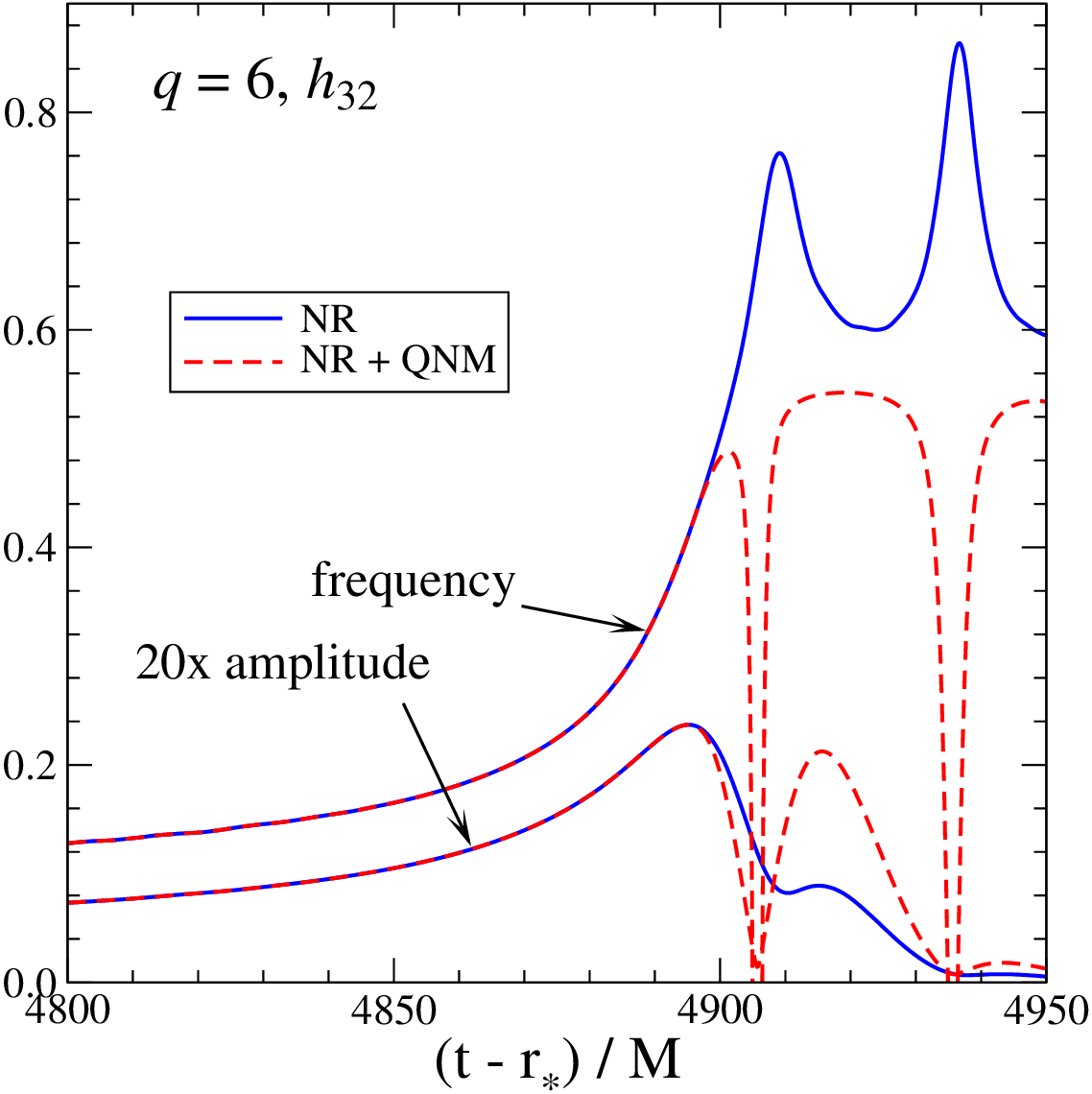}
  \caption{\label{fig:32rdmatch} Amplitude and frequency comparisons
    between full ``NR'' $(3, 2)$ mode and ``NR$+$QNM'' $(3, 2)$ mode
    generated by attaching the QNMs to the inspiral-plunge numerical
    waveform. The ``NR$+$QNM'' waveform is generated by attaching the
    following set of eight QNMs $(\ell=3, m=2, n=0, 1,\dots, 4)$ and
    $(\ell=2, m=2, n=0, 1,2)$. Note that the amplitudes have been
    multiplied by a factor of 20, so that they are more visible. The
    horizontal axis is the retarded time in the numerical-relativity
    simulation.}
\end{figure}

Second, for the case $q=6$, we explore in Fig.~\ref{fig:32rdmatch} the
possibility of modeling the numerical ringdown $(3, 2)$ mode as a
linear superposition of the QNMs of set (ii), using the EOB matching
procedure of Sec.~\ref{sec:EOBmergerRDwaveform}. We observe that
although the beating of the $(\ell, m)=(3, 2)$ and $(\ell, m)=(2, 2)$ QNMs
reproduce qualitatively well the oscillations in the ringdown
amplitude and frequency, the behavior of the two curves ``NR'' and
``NR$+$QNM'' is very different. Therefore, we deduce that the current
matching procedure which use the information of a small segment of
inspiral-plunge waveform before the peak of the amplitude, does not
provide us with the correct QNMs coefficients when different $\ell$
values are present.  Since this matching procedure is applied to build
the calibrated EOB modes, we find that the same problem affects the
calibration of the EOB $(3, 2)$ mode.  We postpone to the future the
solution to this important issue.
  
  \section{Quantities entering the EOB gravitational modes}
  \label{app:eob}

Following Ref.~\cite{DIN}, we introduce the velocity parameter, $\v \equiv (\Omega\, H^\mathrm{real})^{1/3}$. The explicit expressions of the
function $\delta_{\ell m}$ then read~\cite{DIN, Pan2010hz}
\begin{subequations}
  \label{delta2}
  \begin{align}
    \begin{split}
      \label{delta22}
      \delta_{22} &= \dfrac{7}{3}\,\v^{3} + \dfrac{428}{105}\pi\,\v^6
      +
      \left(\frac{1\, 712}{315}\pi^2-\frac{2\, 203}{81}\right)\,\v^9 \\
      &\quad - 24 \,\nu \, v_\Omega^{5}\,,
    \end{split}
    \\
    \begin{split}
      \label{delta21}
      \delta_{21} &= \dfrac{2}{3}\,\v^{3}+ \frac{107}{105}\pi\,\v^6
      +\left(\frac{214}{315}\pi^2-\frac{272}{81}\right)\,\v^9 \\
      &\quad -\dfrac{493}{42}\,\nu\, v_\Omega^{5} +
      \delta_{21}^{(7)}\nu\, v_\Omega^{7}\,,
    \end{split}
  \end{align}
\end{subequations}
\begin{subequations}
  \label{delta3}
  \begin{align}
    \label{delta33}
    \begin{split}
      \delta_{33} &= \dfrac{13}{10}\,\v^{3} + \frac{39}{7}\pi\,\v^6
      +\left(\frac{78}{7}\pi^2-\frac{227\, 827}{3\, 000}\right)\,\v^9 \\
      &\quad -\dfrac{80897}{2430}\,\nu \, v_\Omega^{5}+
      \delta_{33}^{(7)}\nu\, v_\Omega^{7}\,,
    \end{split}
    \\ \label{delta32}
    \delta_{32} &= \dfrac{10+33\nu}{15(1-3\nu)}\,\v^{3} +
    \frac{52}{21}\pi\,\v^6 + \left(\frac{208}{63} \pi^2 -
      \frac{9\, 112} {405} \right)\,\v^9\,, \\
    \begin{split}
      \label{delta31}
      \delta_{31} &= \dfrac{13}{30}\, \v^{3} + \frac{13}{21}\pi\,\v^6
      + \left(\frac{26}{63} \pi^2 -\frac{227\, 827} {81\, 000} \right)\,
      \v^9 \\
      &\quad -\dfrac{17\nu}{10}\, v_\Omega^{5}\,,
    \end{split}
  \end{align}
\end{subequations}

\begin{subequations}
  \label{delta4}
  \begin{align}
    \label{delta44}
    \delta_{44} &=\dfrac{112+219\nu}{120(1-3\nu)}\,\v^{3} +
    \frac{25\, 136}{3\, 465}\pi\,\v^6
    + \delta_{44}^{(5)}\nu\, v_\Omega^{5}\,, \\
    \delta_{43} &= \dfrac{486+4961\nu}{810(1-2\nu)}\,\v^{3} +\frac{1\, 571}{385}\pi\,\v^6,\\
    \delta_{42} &=\dfrac{7(1+6\nu)}{15(1-3\nu)}\,\v^{3} + \frac{6\, 284}{3\, 465}\pi\,\v^6\,,\\
    \delta_{41} &=\dfrac{2+507\nu}{10(1-2\nu)}\,\v^{3} +
    \frac{1\, 571}{3\, 465}\pi\, v^6\,,
  \end{align}
\end{subequations}

\begin{subequations}
  \label{delta5}
  \begin{align}
    \delta_{55} &=\dfrac{96875+857528\nu}{131250(1-2\nu)}\,\v^{3} + \delta_{55}^{(5)}\nu\, v_\Omega^{5}\,, \\
    \delta_{54} &= \frac{8}{15}\,\v^3\,, \\
    \delta_{53} &= \frac{31}{70}\,\v^3\,, \\
    \delta_{52} &= \frac{4}{15}\,\v^3\,, \\
    \delta_{51} &= \frac{31}{210}\,\v^3\,,
  \end{align}
\end{subequations}

\begin{subequations}
  \label{delta6}
  \begin{align}
    \delta_{66} &= \frac{43}{70}\,\v^3\,, \\
    \delta_{65} &= \frac{10}{21}\,\v^3\,, \\
    \delta_{64} &= \frac{43}{105}\,\v^3\,, \\
    \delta_{63} &= \frac{2}{7}\,\v^3\,, \\
    \delta_{62} &= \frac{43}{210}\,\v^3\,, \\
    \delta_{61} &= \frac{2}{21}\,\v^3\,,
  \end{align}
\end{subequations}

\begin{subequations}
  \label{delta7}
  \begin{align}
    \delta_{77} &= \frac{19}{36}\,\v^3\,, \\
    \delta_{75} &= \frac{95}{252}\,\v^3\,, \\
    \delta_{73} &= \frac{19}{84}\,\v^3\,, \\
    \delta_{71} &= \frac{19}{252}\,\v^3\,.
  \end{align}
\end{subequations}
Note that the 2.5PN and 3.5PN coefficients $\delta^{(5)}_{21},
\delta^{(7)}_{33}$ and $\delta^{(7)}_{44}$ in Eqs.~\eqref{delta21},
\eqref{delta33} and \eqref{delta44} are not known in PN theory. They
are determined by calibrating the EOB to numerical-relativity
waveforms. Their explicit expressions for these parameters are given
by Eqs.~\eqref{deltacal}.

The following quantities enter the Newtonian modes in
Eq.~\eqref{hlmNewt}
\begin{subequations}
  \label{n0n1}
  \begin{align}
    n^{(0)}_{\ell m} &= (i\, m)^\ell\frac{8\pi}{(2\ell+1)!!}\sqrt{\frac{(\ell+1)(\ell+2)}{\ell(\ell-1)}}\,, \label{nlmeven} \\
    n^{(1)}_{\ell m} &= -(i\, m)^\ell\frac{16\pi
      i}{(2\ell+1)!!}\sqrt{\frac{(2\ell+1)(\ell+2)(\ell^2-m^2)}{(2\ell-1)(\ell+1)\ell(\ell-1)}}\,,
    \label{nlmodd}
  \end{align}
\end{subequations}
and
\begin{multline}
  \label{cl}
  c_{\ell+\epsilon}(\nu) = \left(\frac{1}{2}-\frac{1}{2}\sqrt{1-4\nu}\right)^{\ell+\epsilon-1} \\
  +(-1)^{\ell+\epsilon}\left(\frac{1}{2}+\frac{1}{2}\sqrt{1-4\nu}\right)^{\ell+\epsilon-1}\,.
\end{multline}

\begin{widetext}

  The odd-parity modes $\rho^L_{\ell m}$ and even-parity modes
  $\rho_{\ell m}$ read~\cite{DIN, Pan2010hz}

  \begin{subequations}
    \label{rho2}
    \begin{align}
      \begin{split}
        \label{rho22}
        \rho_{22}&=1+\left(\frac{55\,\nu}{84}-\frac{43}{42}\right) v_\Omega^2 +\left(\frac{19\, 583\,\nu^2}{42\, 336}-\frac{33\, 025\,\nu}{21\, 168}-\frac{20\, 555}{10\, 584}\right) v_\Omega^4 \\
        &\quad
        + \left(\frac{10\, 620\, 745\,\nu^3}{39\, 118\, 464}-\frac{6\, 292\, 061\,\nu^2}{3\, 259\, 872}+\frac{41\,\pi^2\,\nu}{192}-\frac{48\, 993\, 925\,\nu}{9\, 779\, 616}-\frac{428\,\text{eulerlog}_2(v_\Omega^2)}{105}+\frac{1\, 556\, 919\, 113}{122\, 245\, 200}\right)\, v_\Omega^6 \\
        &\quad +
        \left(\frac{9\, 202\,\text{eulerlog}_2(v_\Omega^2)}{2\, 205}-\frac{387\, 216\, 563\, 023}{160\, 190\, 110\, 080}\right)\, v_\Omega^8
        +
        \left(\frac{439\, 877\,\text{eulerlog}_2(v_\Omega^2)}{55\, 566}-\frac{16\, 094\, 530\, 514\, 677}{533\, 967\, 033\, 600}\right)\, v_\Omega^{10}
        \,,
      \end{split}
      \\
      \begin{split}
        \label{rho21}
        \rho^L_{21}&=1 +\left(\frac{23\,\nu}{84}-\frac{59}{56}\right)v_\Omega^2 +\left(\frac{617\,\nu^2}{4\, 704}-\frac{10\, 993\,\nu}{14\, 112}-\frac{47\, 009}{56\, 448}\right) v_\Omega^4 + \left(\frac{7\, 613\, 184\, 941}{2\, 607\, 897\, 600}-\frac{107\,\text{eulerlog}_1(v_\Omega^2)}{105} \right)\, v_\Omega^6 \\
        &\quad + \rho_{21}^{(6)}\,\nu\, v_\Omega^6 + \left(\frac{6\, 313\,\text{eulerlog}_1(v_\Omega^2)}{5\, 880}-\frac{1\, 168\, 617\, 463\, 883}{911\, 303\, 737\, 344}\right)\, v_\Omega^8 \\
        &\quad +
        \left(\frac{5\, 029\, 963\,\text{eulerlog}_1(v_\Omega^2)}{5\, 927\, 040}-\frac{63\, 735\, 873\, 771\, 463}{16\, 569\, 158\, 860\, 800}\right)\, v_\Omega^{10}
        \,,
      \end{split}
    \end{align}
  \end{subequations}

  \begin{subequations}
    \label{rho3}
    \begin{align}
      \begin{split}
        \label{rho33}
        \rho_{33}&=1+\left(\frac{2\,\nu}{3}-\frac{7}{6}\right) v_\Omega^2 +\left(\frac{149\,\nu^2}{330}-\frac{1\, 861\,\nu}{990}-\frac{6\, 719}{3\, 960}\right) v_\Omega^4 + \left(\frac{3\, 203\, 101\, 567}{227\, 026\, 800}-\frac{26\,\text{eulerlog}_3(v_\Omega^2)}{7}\right)\, v_\Omega^6 + \rho_{33}^{(6)}\,\nu\, v_\Omega^6 \\
        &\quad +
        \left(\frac{13\,\text{eulerlog}_3(v_\Omega^2)}{3}-\frac{57\, 566\, 572\, 157}{8\, 562\, 153\, 600}\right)\, v_\Omega^8
        \,,
      \end{split}
      \\
      \begin{split}
        \label{rho32}
        \rho^L_{32}&=1+\frac{320\,\nu^2-1\, 115\,\nu+328}{270\,(3\,\nu-1)} v_\Omega^2 +\frac{3\, 085\, 640\,\nu^4-20\, 338\, 960\,\nu^3-4\, 725\, 605\,\nu^2+8\, 050\, 045\,\nu-1\, 444\, 528}{1\, 603\, 800\,(1-3\,\nu)^2} v_\Omega^4 \\
        &\quad+
        \left(\frac{5\, 849\, 948\, 554}{940\, 355\, 325}-\frac{104\,\text{eulerlog}_2(v_\Omega^2)}{63}\right)\, v_\Omega^6
        +
        \left(\frac{17\, 056\,\text{eulerlog}_2(v_\Omega^2)}{8\, 505}-\frac{10\, 607\, 269\, 449\, 358}{3\, 072\, 140\, 846\, 775}\right)\, v_\Omega^8
        \,,
      \end{split}
      \\
      \begin{split}
        \label{rho31}
        \rho_{31}&=1-\left(\frac{2\,\nu}{9}+\frac{13}{18}\right) v_\Omega^2+ \left(-\frac{829\,\nu^2}{1\, 782}-\frac{1\, 685\,\nu}{1\, 782}+\frac{101}{7\, 128}\right) v_\Omega^4 + \left(\frac{11\, 706\, 720\, 301}{6\, 129\, 723\, 600}-\frac{26\,\text{eulerlog}_1(v_\Omega^2)}{63}\right)\, v_\Omega^6 \\
        &\quad +
        \left(\frac{169\,\text{eulerlog}_1(v_\Omega^2)}{567}+\frac{2\, 606\, 097\, 992\, 581}{4\, 854\, 741\, 091\, 200}\right)\, v_\Omega^8
        \,.
      \end{split}
    \end{align}
  \end{subequations}

  \begin{subequations}
    \label{rho4}
    \begin{align}
      \begin{split}
        \label{rho44}
        \rho_{44}&=1+\frac{2\, 625\nu^2-5\, 870\,\nu+1\, 614}{1\, 320\,(3\,\nu-1)} v_\Omega^2 \\
        &\quad+\frac{1\, 252\, 563\, 795\,\nu^4-6\, 733\, 146\, 000\,\nu^3-313\, 857\, 376\,\nu^2+2\, 338\, 945\, 704\,\nu-511\, 573\, 572}{317\, 116\, 800\,(1-3\,\nu)^2} v_\Omega^4 \\
        &\quad+
        \left(\frac{16\, 600\, 939\, 332\, 793}{1\, 098\, 809\, 712\, 000}-\frac{12\, 568\,\text{eulerlog}_4(v_\Omega^2)}{3\, 465}\right)\, v_\Omega^6
        + \rho_{44}^{(6)}\,\nu\, v_\Omega^6\,,
      \end{split}
      \\
      \label{rho43}
      \rho^L_{43}&=1+\frac{160\,\nu^2-547\,\nu+222}{176\,(2\,\nu-1)}
      v_\Omega^2-\frac{6\, 894\, 273}{7\, 047\, 040} v_\Omega^4 +
      \left(\frac{1\, 664\, 224\, 207\, 351}{195\, 343\, 948\, 800}-\frac{1\, 571\,\text{eulerlog}_3(v_\Omega^2)}{770}\right)\, v_\Omega^6
      \,,
      \\
      \begin{split}
        \label{rho42}
        \rho_{42}&=1+\frac{285\,\nu^2-3\, 530\,\nu+1\, 146}{1\, 320\,(3\,\nu-1)} v_\Omega^2 \\
        &\quad+\frac{-379\, 526\, 805\,\nu^4-3\, 047\, 981\, 160\,\nu^3+1\, 204\, 388\, 696\,\nu^2+295\, 834\, 536\,\nu-114\, 859\, 044}{317\, 116\, 800\,(1-3\,\nu)^2} v_\Omega^4 \\
        &\quad+
        \left(\frac{848\, 238\, 724\, 511}{219\, 761\, 942\, 400}-\frac{3\, 142\,\text{eulerlog}_2(v_\Omega^2)}{3\, 465}\right)\, v_\Omega^6
        \,,
      \end{split}
      \\
      \begin{split}
        \label{rho41}
        \rho^L_{41}&=1+\frac{288\,\nu^2-1\, 385\,\nu+602}{528\,(2\,\nu-1)}
        v_\Omega^2-\frac{7\, 775\, 491}{21\, 141\, 120} v_\Omega^4 +
        \left(\frac{1\, 227\, 423\, 222\, 031}{1\, 758\, 095\, 539\, 200}-\frac{1571\,\text{eulerlog}_1(v_\Omega^2)}{6\, 930}\right)\, v_\Omega^6
        \,.
      \end{split}
    \end{align}
  \end{subequations}

  \begin{subequations}
    \label{rho5}
    \begin{align}
      \begin{split}
        \label{rho55}
        \rho_{55}&=1+\frac{512\,\nu^2-1\, 298\,\nu+487}{390\,(2\,\nu-1)}
        v_\Omega^2 -\frac{3\, 353\, 747}{2\, 129\, 400} v_\Omega^4 +
        \rho_{55}^{(6)}\,\nu\, v_\Omega^6\,,
      \end{split}
      \\
      \begin{split}
        \label{rho54}
        \rho^L_{54}&=1+\frac{33\, 320\,\nu^3-127\, 610\,\nu^2+96\, 019\,\nu-17\, 448}{13\, 650\,(5\,\nu^2-5\,\nu+1)}
        v_\Omega^2 -\frac{16\, 213\, 384}{15\, 526\, 875} v_\Omega^4 \,,
      \end{split}
      \\
      \begin{split}
        \label{rho53}
        \rho_{53}&=1+\frac{176\,\nu^2-850\,\nu+375}{390\,(2\,\nu-1)}
        v_\Omega^2 -\frac{410\, 833}{709\, 800} v_\Omega^4\,,
      \end{split}
      \\
      \begin{split}
        \label{rho52}
        \rho^L_{52}&=1+\frac{21\, 980\,\nu^3-104\, 930\,\nu^2+84\, 679\,\nu-15\, 828}{13\, 650\,(5\,\nu^2-5\,\nu+1)}
        v_\Omega^2 -\frac{7\, 187\, 914}{15\, 526\, 875} v_\Omega^4 \,,
      \end{split}
      \\
      \begin{split}
        \label{rho51}
        \rho_{51}&=1+\frac{8\,\nu^2-626\,\nu+319}{390\,(2\,\nu-1)}
        v_\Omega^2 -\frac{31\, 877}{304\, 200} v_\Omega^4\,.
      \end{split}
    \end{align}
  \end{subequations}
  
  \begin{subequations}
    \begin{align}
      \rho_{66} &= 1+\frac{273\,\nu^3-861\,\nu^2+602\,\nu-106}
      {84\,(5\,\nu^2-5\,\nu+1)} v_\Omega^2 - \frac{1\, 025\, 435}
      {659\, 736} v_\Omega^4 \,, \label{rho66} \\
      \rho^L_{65}& = 1 + \frac{220\,\nu^3-910\,\nu^2+838\,\nu-185}
      {144\,(3\,\nu^2-4\,\nu+1)} v_\Omega^2\,, \label{rho65} \\
      \rho_{64} &= 1 + \frac{133\,\nu^3-581\,\nu^2+462\,\nu-86}
      {84\,(5\,\nu^2-5\,\nu+1)} v_\Omega^2 - \frac{476\, 887}
      {659\, 736} v_\Omega^4 \,, \label{rho64} \\
      \rho^L_{63} &= 1 + \frac{156\,\nu^3-750\,\nu^2+742\,\nu-169}
      {144\,(3\,\nu^2-4\,\nu+1)} v_\Omega^2\,, \label{rho63} \\
      \rho_{62} &= 1 + \frac{49\,\nu^3-413\,\nu^2+378\,\nu-74}
      {84\,(5\,\nu^2-5\,\nu+1)} v_\Omega^2 - \frac{817\, 991}
      {3\, 298\, 680} v_\Omega^4 \,, \label{rho62} \\
      \rho^L_{61} &= 1 + \frac{124\,\nu^3-670\,\nu^2+694\,\nu-161}
      {144\,(3\,\nu^2-4\,\nu+1)} v_\Omega^2\,, \label{rho61}
    \end{align}
  \end{subequations}

  \begin{subequations}
    \begin{align}
      \rho_{77} &= 1 + \frac{1380 \nu ^3-4963 \nu ^2+4246 \nu -906}
      {714 \left(3 \nu ^2-4 \nu +1\right)} v_{\Omega }^2 \,, \\
      \rho^L_{76} &= 1 + \frac{6104 \nu ^4-29351 \nu ^3+37828 \nu
        ^2-16185 \nu +2144} {1666 \left(7 \nu ^3-14 \nu ^2+7 \nu
          -1\right)} v_{\Omega }^2 \,, \\
      \rho_{75} &= 1 + \frac{804 \nu ^3-3523 \nu ^2+3382 \nu -762}
      {714 \left(3 \nu ^2-4 \nu +1\right)} v_{\Omega }^2 \,, \\
      \rho^L_{74} &= 1 + \frac{41076 \nu ^4-217959 \nu ^3+298872 \nu
        ^2-131805 \nu +17756} {14994 \left(7 \nu ^3-14 \nu ^2+7 \nu
          -1\right)} v_{\Omega }^2 \,, \\
      \rho_{73} &= 1 + \frac{420 \nu ^3-2563 \nu ^2+2806 \nu -666}
      {714 \left(3 \nu ^2-4 \nu +1\right)} v_{\Omega }^2 \,, \\
      \rho^L_{72} &= 1 + \frac{32760 \nu ^4-190239 \nu ^3+273924 \nu
        ^2-123489 \nu +16832} {14994 \left(7 \nu ^3-14 \nu ^2+7 \nu
          -1\right)} v_{\Omega }^2 \,, \\
      \rho_{71} &= 1 + \frac{228 \nu ^3-2083 \nu ^2+2518 \nu -618}
      {714 \left(3 \nu ^2-4 \nu +1\right)} v_{\Omega }^2 \,.
    \end{align}
  \end{subequations}
\end{widetext}
Here we use $\mathrm{eulerlog}_m(v_\Omega^2)=\gamma_E + \log 2 + \log
m + 1/2\log v_\Omega^2$, with $\gamma_E$ being the Euler constant. Note
that the 3PN coefficients $\rho^{(6)}_{21}$, $\rho^{(6)}_{33}$ and
$\rho^{(6)}_{44}$ in Eqs.~\eqref{rho21}, \eqref{rho33} and
\eqref{rho44} are not known in PN theory. They are determined by
calibrating the EOB to numerical-relativity waveforms. The calibrated
expressions are given by Eq.~\eqref{rhocal}.


\begin{thebibliography}{10}%
\makeatletter
\providecommand \@ifxundefined [1]{%
 \ifx #1\undefined \expandafter \@firstoftwo
 \else \expandafter \@secondoftwo
\fi
}%
\providecommand \@ifnum [1]{%
 \ifnum #1\expandafter \@firstoftwo
 \else \expandafter \@secondoftwo
\fi
}%
\providecommand \enquote [1]{``#1''}%
\providecommand \bibnamefont  [1]{#1}%
\providecommand \bibfnamefont [1]{#1}%
\providecommand \citenamefont [1]{#1}%
\providecommand\href[0]{\@sanitize\@href}%
\providecommand\@href[1]{\endgroup\@@startlink{#1}\endgroup\@@href}%
\providecommand\@@href[1]{#1\@@endlink}%
\providecommand \@sanitize [0]{\begingroup\catcode`\&12\catcode`\#12\relax}%
\@ifxundefined \pdfoutput {\@firstoftwo}{%
 \@ifnum{\z@=\pdfoutput}{\@firstoftwo}{\@secondoftwo}%
}{%
 \providecommand\@@startlink[1]{\leavevmode}%
 \providecommand\@@endlink[0]{}%
}{%
 \providecommand\@@startlink[1]{%
  \leavevmode
  \pdfstartlink
   attr{/Border[0 0 1 ]/H/I/C[0 1 1]}%
   user{/Subtype/Link/A<</Type/Action/S/URI/URI(#1)>>}%
  \relax
 }%
 \providecommand\@@endlink[0]{\pdfendlink}%
}%
\providecommand \url  [0]{\begingroup\@sanitize \@url }%
\providecommand \@url [1]{\endgroup\@href {#1}{\urlprefix}}%
\providecommand \urlprefix [0]{URL }%
\providecommand \Eprint[0]{\href }%
\@ifxundefined \urlstyle {%
  \providecommand \doi [1]{doi:\discretionary{}{}{}#1}%
}{%
  \providecommand \doi [0]{doi:\discretionary{}{}{}\begingroup
  \urlstyle{rm}\Url }%
}%
\providecommand \doibase [0]{http://dx.doi.org/}%
\providecommand \Doi[1]{\href{\doibase#1}}%
\providecommand \bibAnnote [3]{%
  \BibitemShut{#1}%
  \begin{quotation}\noindent
    \textsc{Key:}\ #2\\\textsc{Annotation:}\ #3%
  \end{quotation}%
}%
\providecommand \bibAnnoteFile [2]{%
  \IfFileExists{#2}{\bibAnnote {#1} {#2} {\input{#2}}}{}%
}%
\providecommand \typeout [0]{\immediate \write \m@ne }%
\providecommand \selectlanguage [0]{\@gobble}%
\providecommand \bibinfo [0]{\@secondoftwo}%
\providecommand \bibfield [0]{\@secondoftwo}%
\providecommand \translation [1]{[#1]}%
\providecommand \BibitemOpen[0]{}%
\providecommand \bibitemStop [0]{}%
\providecommand \bibitemNoStop [0]{.\EOS\space}%
\providecommand \EOS [0]{\spacefactor3000\relax}%
\providecommand \BibitemShut [1]{\csname bibitem#1\endcsname}%
\bibitem{Abbott:2007}%
  \BibitemOpen
  \bibfield{author}{%
  \bibinfo {author} {\bibfnamefont{B.}~\bibnamefont{Abbott}} \emph{et~al.}
  (\bibinfo {collaboration} {LIGO Scientific Collaboration}),\ }%
  \bibfield{journal}{%
  \bibinfo {journal} {Rep.\ Prog.\ Phys.}\ }%
  \textbf{\bibinfo {volume} {72}},\ \bibinfo {pages} {076901} (\bibinfo {year}
  {2009})%
  \bibAnnoteFile{NoStop}{Abbott:2007}%
\bibitem{Acernese:2008}%
  \BibitemOpen
  \bibfield{author}{%
  \bibinfo {author} {\bibfnamefont{F.}~\bibnamefont{Acernese}} \emph{et~al.}
  (\bibinfo {collaboration} {Virgo Collaboration}),\ }%
  \bibfield{journal}{%
  \bibinfo {journal} {Class.\ Quantum Grav.}\ }%
  \textbf{\bibinfo {volume} {25}},\ \bibinfo {pages} {184001} (\bibinfo {year}
  {2008})%
  \bibAnnoteFile{NoStop}{Acernese:2008}%
\bibitem{Grote:2008zz}%
  \BibitemOpen
  \bibfield{author}{%
  \bibinfo {author} {\bibfnamefont{H.}~\bibnamefont{Grote}} (\bibinfo
  {collaboration} {GEO600 Collaboration}),\ }%
  \bibfield{journal}{%
  \Doi{10.1088/0264-9381/25/11/114043}{\bibinfo {journal} {Class. Quant.
  Grav.}}\ }%
  \textbf{\bibinfo {volume} {25}},\ \bibinfo {pages} {114043} (\bibinfo {year}
  {2008})%
  \bibAnnoteFile{NoStop}{Grote:2008zz}%
\bibitem{Kuroda:2010}%
  \BibitemOpen
  \bibfield{author}{%
  \bibinfo {author} {\bibfnamefont{K.}~\bibnamefont{Kuroda}}\ and\ \bibinfo
  {author} {\bibnamefont{the LCGT~Collaboration}},\ }%
  \bibfield{journal}{%
  \bibinfo {journal} {Class.\ Quantum Grav.}\ }%
  \textbf{\bibinfo {volume} {27}},\ \bibinfo {pages} {084004} (\bibinfo {year}
  {2010})%
  \bibAnnoteFile{NoStop}{Kuroda:2010}%
\bibitem{Blanchet2006}%
  \BibitemOpen
  \bibfield{author}{%
  \bibinfo {author} {\bibfnamefont{L.}~\bibnamefont{Blanchet}},\ }%
  \bibfield{journal}{%
  \bibinfo {journal} {Living Rev.~Rel.}\ }%
  \textbf{\bibinfo {volume} {9}} (\bibinfo {year} {2006})%
  \bibAnnoteFile{NoStop}{Blanchet2006}%
\bibitem{Sasaki:2003xr}%
  \BibitemOpen
  \bibfield{author}{%
  \bibinfo {author} {\bibfnamefont{M.}~\bibnamefont{Sasaki}}\ and\ \bibinfo
  {author} {\bibfnamefont{H.}~\bibnamefont{Tagoshi}},\ }%
  \bibfield{journal}{%
  \bibinfo {journal} {Living Rev.Rel.}\ }%
  \textbf{\bibinfo {volume} {6}},\ \bibinfo {pages} {6} (\bibinfo {year}
  {2003}),\ \Eprint{http://arxiv.org/abs/gr-qc/0306120}{arXiv:gr-qc/0306120
  [gr-qc]}%
  \bibAnnoteFile{NoStop}{Sasaki:2003xr}%
\bibitem{Futamase:2007zz}%
  \BibitemOpen
  \bibfield{author}{%
  \bibinfo {author} {\bibfnamefont{T.}~\bibnamefont{Futamase}}\ and\ \bibinfo
  {author} {\bibfnamefont{Y.}~\bibnamefont{Itoh}},\ }%
  \bibfield{journal}{%
  \bibinfo {journal} {Living Rev. Rel.}\ }%
  \textbf{\bibinfo {volume} {10}},\ \bibinfo {pages} {2} (\bibinfo {year}
  {2007})%
  \bibAnnoteFile{NoStop}{Futamase:2007zz}%
\bibitem{Goldberger:2004jt}%
  \BibitemOpen
  \bibfield{author}{%
  \bibinfo {author} {\bibfnamefont{W.~D.}\ \bibnamefont{Goldberger}}\ and\
  \bibinfo {author} {\bibfnamefont{I.~Z.}\ \bibnamefont{Rothstein}},\ }%
  \bibfield{journal}{%
  \Doi{10.1103/PhysRevD.73.104029}{\bibinfo {journal} {Phys. Rev.}}\ }%
  \textbf{\bibinfo {volume} {D73}},\ \bibinfo {pages} {104029} (\bibinfo {year}
  {2006}),\ \Eprint{http://arxiv.org/abs/hep-th/0409156}{arXiv:hep-th/0409156}%
  \bibAnnoteFile{NoStop}{Goldberger:2004jt}%
\bibitem{Pretorius2005a}%
  \BibitemOpen
  \bibfield{author}{%
  \bibinfo {author} {\bibfnamefont{F.}~\bibnamefont{Pretorius}},\ }%
  \bibfield{journal}{%
  \bibinfo {journal} {Phys.\ Rev.\ Lett.}\ }%
  \textbf{\bibinfo {volume} {95}},\ \bibinfo {eid} {121101} (\bibinfo {year}
  {2005})%
  \bibAnnoteFile{NoStop}{Pretorius2005a}%
\bibitem{Baker2006a}%
  \BibitemOpen
  \bibfield{author}{%
  \bibinfo {author} {\bibfnamefont{J.~G.}\ \bibnamefont{Baker}}, \bibinfo
  {author} {\bibfnamefont{J.}~\bibnamefont{Centrella}}, \bibinfo {author}
  {\bibfnamefont{D.-I.}\ \bibnamefont{Choi}}, \bibinfo {author}
  {\bibfnamefont{M.}~\bibnamefont{Koppitz}},\ and\ \bibinfo {author}
  {\bibfnamefont{J.}~\bibnamefont{van Meter}},\ }%
  \bibfield{journal}{%
  \bibinfo {journal} {Phys.\ Rev.\ Lett.}\ }%
  \textbf{\bibinfo {volume} {96}},\ \bibinfo {eid} {111102} (\bibinfo {year}
  {2006})%
  \bibAnnoteFile{NoStop}{Baker2006a}%
\bibitem{Campanelli2006a}%
  \BibitemOpen
  \bibfield{author}{%
  \bibinfo {author} {\bibfnamefont{M.}~\bibnamefont{Campanelli}}, \bibinfo
  {author} {\bibfnamefont{C.~O.}\ \bibnamefont{Lousto}}, \bibinfo {author}
  {\bibfnamefont{P.}~\bibnamefont{Marronetti}},\ and\ \bibinfo {author}
  {\bibfnamefont{Y.}~\bibnamefont{Zlochower}},\ }%
  \bibfield{journal}{%
  \bibinfo {journal} {Phys.\ Rev.\ Lett.}\ }%
  \textbf{\bibinfo {volume} {96}},\ \bibinfo {eid} {111101} (\bibinfo {year}
  {2006})%
  \bibAnnoteFile{NoStop}{Campanelli2006a}%
\bibitem{Buonanno00}%
  \BibitemOpen
  \bibfield{author}{%
  \bibinfo {author} {\bibfnamefont{A.}~\bibnamefont{Buonanno}}\ and\ \bibinfo
  {author} {\bibfnamefont{T.}~\bibnamefont{Damour}},\ }%
  \bibfield{journal}{%
  \bibinfo {journal} {Phys.\ Rev.\ D}\ }%
  \textbf{\bibinfo {volume} {62}},\ \bibinfo {pages} {064015} (\bibinfo {year}
  {2000})%
  \bibAnnoteFile{NoStop}{Buonanno00}%
\bibitem{Buonanno99}%
  \BibitemOpen
  \bibfield{author}{%
  \bibinfo {author} {\bibfnamefont{A.}~\bibnamefont{Buonanno}}\ and\ \bibinfo
  {author} {\bibfnamefont{T.}~\bibnamefont{Damour}},\ }%
  \bibfield{journal}{%
  \bibinfo {journal} {Phys. Rev. D}\ }%
  \textbf{\bibinfo {volume} {59}},\ \bibinfo {pages} {084006} (\bibinfo {year}
  {1999})%
  \bibAnnoteFile{NoStop}{Buonanno99}%
\bibitem{DJS00}%
  \BibitemOpen
  \bibfield{author}{%
  \bibinfo {author} {\bibfnamefont{T.}~\bibnamefont{Damour}}, \bibinfo {author}
  {\bibfnamefont{P.}~\bibnamefont{Jaranowski}},\ and\ \bibinfo {author}
  {\bibfnamefont{G.}~\bibnamefont{Sch\"afer}},\ }%
  \bibfield{journal}{%
  \Doi{10.1103/PhysRevD.62.084011}{\bibinfo {journal} {Phys.\ Rev.\ D}}\ }%
  \textbf{\bibinfo {volume} {62}},\ \bibinfo {pages} {084011} (\bibinfo {month}
  {Sep}\ \bibinfo {year} {2000})%
  \bibAnnoteFile{NoStop}{DJS00}%
\bibitem{Damour01c}%
  \BibitemOpen
  \bibfield{author}{%
  \bibinfo {author} {\bibfnamefont{T.}~\bibnamefont{Damour}},\ }%
  \bibfield{journal}{%
  \bibinfo {journal} {Phys.\ Rev.\ D}\ }%
  \textbf{\bibinfo {volume} {64}},\ \bibinfo {pages} {124013} (\bibinfo {year}
  {2001})%
  \bibAnnoteFile{NoStop}{Damour01c}%
\bibitem{Buonanno06}%
  \BibitemOpen
  \bibfield{author}{%
  \bibinfo {author} {\bibfnamefont{A.}~\bibnamefont{Buonanno}}, \bibinfo
  {author} {\bibfnamefont{Y.}~\bibnamefont{Chen}},\ and\ \bibinfo {author}
  {\bibfnamefont{T.}~\bibnamefont{Damour}},\ }%
  \bibfield{journal}{%
  \Doi{10.1103/PhysRevD.74.104005}{\bibinfo {journal} {Phys.\ Rev.\ D}}\ }%
  \textbf{\bibinfo {volume} {74}},\ \bibinfo {eid} {104005} (\bibinfo {year}
  {2006})%
  \bibAnnoteFile{NoStop}{Buonanno06}%
\bibitem{Damour:024009}%
  \BibitemOpen
  \bibfield{author}{%
  \bibinfo {author} {\bibfnamefont{T.}~\bibnamefont{Damour}}, \bibinfo {author}
  {\bibfnamefont{P.}~\bibnamefont{Jaranowski}},\ and\ \bibinfo {author}
  {\bibfnamefont{G.}~\bibnamefont{Sch\"{a}fer}},\ }%
  \bibfield{journal}{%
  \bibinfo {journal} {Phys. Rev. D}\ }%
  \textbf{\bibinfo {volume} {78}},\ \bibinfo {pages} {024009} (\bibinfo {year}
  {2008})%
  \bibAnnoteFile{NoStop}{Damour:024009}%
\bibitem{Barausse:2009xi}%
  \BibitemOpen
  \bibfield{author}{%
  \bibinfo {author} {\bibfnamefont{E.}~\bibnamefont{Barausse}}\ and\ \bibinfo
  {author} {\bibfnamefont{A.}~\bibnamefont{Buonanno}},\ }%
  \bibfield{journal}{%
  \bibinfo {journal} {Phys. Rev.}\ }%
  \textbf{\bibinfo {volume} {D81}},\ \bibinfo {pages} {084024} (\bibinfo {year}
  {2010})%
  \bibAnnoteFile{NoStop}{Barausse:2009xi}%
\bibitem{Buonanno-Cook-Pretorius:2007}%
  \BibitemOpen
  \bibfield{author}{%
  \bibinfo {author} {\bibfnamefont{A.}~\bibnamefont{Buonanno}}, \bibinfo
  {author} {\bibfnamefont{G.~B.}\ \bibnamefont{Cook}},\ and\ \bibinfo {author}
  {\bibfnamefont{F.}~\bibnamefont{Pretorius}},\ }%
  \bibfield{journal}{%
  \Doi{10.1103/PhysRevD.75.124018}{\bibinfo {journal} {Phys.\ Rev.\ D}}\ }%
  \textbf{\bibinfo {volume} {75}},\ \bibinfo {eid} {124018} (\bibinfo {year}
  {2007}),\ \Eprint{http://arxiv.org/abs/gr-qc/0610122}{gr-qc/0610122},\
  \url{http://link.aps.org/abstract/PRD/v75/e124018}%
  \bibAnnoteFile{NoStop}{Buonanno-Cook-Pretorius:2007}%
\bibitem{Buonanno2007}%
  \BibitemOpen
  \bibfield{author}{%
  \bibinfo {author} {\bibfnamefont{A.}~\bibnamefont{Buonanno}}, \bibinfo
  {author} {\bibfnamefont{Y.}~\bibnamefont{Pan}}, \bibinfo {author}
  {\bibfnamefont{J.~G.}\ \bibnamefont{Baker}}, \bibinfo {author}
  {\bibfnamefont{J.}~\bibnamefont{Centrella}}, \bibinfo {author}
  {\bibfnamefont{B.~J.}\ \bibnamefont{Kelly}}, \bibinfo {author}
  {\bibfnamefont{S.~T.}\ \bibnamefont{McWilliams}},\ and\ \bibinfo {author}
  {\bibfnamefont{J.~R.}\ \bibnamefont{van Meter}},\ }%
  \bibfield{journal}{%
  \bibinfo {journal} {Phys.\ Rev.\ D}\ }%
  \textbf{\bibinfo {volume} {76}},\ \bibinfo {pages} {104049} (\bibinfo {year}
  {2007})%
  \bibAnnoteFile{NoStop}{Buonanno2007}%
\bibitem{Pan2007}%
  \BibitemOpen
  \bibfield{author}{%
  \bibinfo {author} {\bibfnamefont{Y.}~\bibnamefont{Pan}}, \bibinfo {author}
  {\bibfnamefont{A.}~\bibnamefont{Buonanno}}, \bibinfo {author}
  {\bibfnamefont{J.~G.}\ \bibnamefont{Baker}}, \bibinfo {author}
  {\bibfnamefont{J.}~\bibnamefont{Centrella}}, \bibinfo {author}
  {\bibfnamefont{B.~J.}\ \bibnamefont{Kelly}}, \bibinfo {author}
  {\bibfnamefont{S.~T.}\ \bibnamefont{McWilliams}}, \bibinfo {author}
  {\bibfnamefont{F.}~\bibnamefont{Pretorius}},\ and\ \bibinfo {author}
  {\bibfnamefont{J.~R.}\ \bibnamefont{van Meter}},\ }%
  \bibfield{journal}{%
  \Doi{10.1103/PhysRevD.77.024014}{\bibinfo {journal} {Phys.\ Rev.\ D}}\ }%
  \textbf{\bibinfo {volume} {77}},\ \bibinfo {eid} {024014} (\bibinfo {year}
  {2008})%
  \bibAnnoteFile{NoStop}{Pan2007}%
\bibitem{Boyle2008a}%
  \BibitemOpen
  \bibfield{author}{%
  \bibinfo {author} {\bibfnamefont{M.}~\bibnamefont{Boyle}}, \bibinfo {author}
  {\bibfnamefont{A.}~\bibnamefont{Buonanno}}, \bibinfo {author}
  {\bibfnamefont{L.~E.}\ \bibnamefont{Kidder}}, \bibinfo {author}
  {\bibfnamefont{A.~H.}\ \bibnamefont{Mrou\'e}}, \bibinfo {author}
  {\bibfnamefont{Y.}~\bibnamefont{Pan}}, \bibinfo {author}
  {\bibfnamefont{H.~P.}\ \bibnamefont{Pfeiffer}},\ and\ \bibinfo {author}
  {\bibfnamefont{M.~A.}\ \bibnamefont{Scheel}},\ }%
  \bibfield{journal}{%
  \Doi{10.1103/PhysRevD.78.104020}{\bibinfo {journal} {Phys.\ Rev.\ D}}\ }%
  \textbf{\bibinfo {volume} {78}},\ \bibinfo {eid} {104020} (\bibinfo {year}
  {2008}),\ \Eprint{http://arxiv.org/abs/0804.4184}{arXiv:0804.4184 [gr-qc]}%
  \bibAnnoteFile{NoStop}{Boyle2008a}%
\bibitem{Buonanno:2009qa}%
  \BibitemOpen
  \bibfield{author}{%
  \bibinfo {author} {\bibfnamefont{A.}~\bibnamefont{Buonanno}}, \bibinfo
  {author} {\bibfnamefont{Y.}~\bibnamefont{Pan}}, \bibinfo {author}
  {\bibfnamefont{H.~P.}\ \bibnamefont{Pfeiffer}}, \bibinfo {author}
  {\bibfnamefont{M.~A.}\ \bibnamefont{Scheel}}, \bibinfo {author}
  {\bibfnamefont{L.~T.}\ \bibnamefont{Buchman}},\ and\ \bibinfo {author}
  {\bibfnamefont{L.~E.}\ \bibnamefont{Kidder}},\ }%
  \bibfield{journal}{%
  \bibinfo {journal} {Phys.\ Rev.\ D}\ }%
  \textbf{\bibinfo {volume} {79}},\ \bibinfo {pages} {124028} (\bibinfo {year}
  {2009})%
  \bibAnnoteFile{NoStop}{Buonanno:2009qa}%
\bibitem{Racine2008}%
  \BibitemOpen
  \bibfield{author}{%
  \bibinfo {author} {\bibfnamefont{E.}~\bibnamefont{Racine}}, \bibinfo {author}
  {\bibfnamefont{A.}~\bibnamefont{Buonanno}},\ and\ \bibinfo {author}
  {\bibfnamefont{L.~E.}\ \bibnamefont{Kidder}},\ }%
  \bibfield{journal}{%
  \bibinfo {journal} {Phys. Rev. D}\ }%
  \textbf{\bibinfo {volume} {80}},\ \bibinfo {pages} {044010} (\bibinfo {year}
  {2009})%
  \bibAnnoteFile{NoStop}{Racine2008}%
\bibitem{Pan:2009wj}%
  \BibitemOpen
  \bibfield{author}{%
  \bibinfo {author} {\bibfnamefont{Y.}~\bibnamefont{{Pan}}}, \bibinfo {author}
  {\bibfnamefont{A.}~\bibnamefont{{Buonanno}}}, \bibinfo {author}
  {\bibfnamefont{L.}~\bibnamefont{Buchman}}, \bibinfo {author}
  {\bibfnamefont{T.}~\bibnamefont{Chu}}, \bibinfo {author}
  {\bibfnamefont{L.}~\bibnamefont{Kidder}}, \bibinfo {author}
  {\bibfnamefont{H.}~\bibnamefont{Pfeiffer}},\ and\ \bibinfo {author}
  {\bibfnamefont{M.}~\bibnamefont{Scheel}},\ }%
  \bibfield{journal}{%
  \bibinfo {journal} {Phys. Rev.}\ }%
  \textbf{\bibinfo {volume} {D81}},\ \bibinfo {pages} {084041} (\bibinfo {year}
  {2010})%
  \bibAnnoteFile{NoStop}{Pan:2009wj}%
\bibitem{Pan2010hz}%
  \BibitemOpen
  \bibfield{author}{%
  \bibinfo {author} {\bibfnamefont{Y.}~\bibnamefont{Pan}}, \bibinfo {author}
  {\bibfnamefont{A.}~\bibnamefont{Buonanno}}, \bibinfo {author}
  {\bibfnamefont{R.}~\bibnamefont{Fujita}}, \bibinfo {author}
  {\bibfnamefont{E.}~\bibnamefont{Racine}},\ and\ \bibinfo {author}
  {\bibfnamefont{H.}~\bibnamefont{Tagoshi}}}%
   (\bibinfo {year} {2010}),\
  \Eprint{http://arxiv.org/abs/1006.0431}{arXiv:1006.0431 [gr-qc]}%
  \bibAnnoteFile{NoStop}{Pan2010hz}%
\bibitem{Damour2007a}%
  \BibitemOpen
  \bibfield{author}{%
  \bibinfo {author} {\bibfnamefont{T.}~\bibnamefont{Damour}}\ and\ \bibinfo
  {author} {\bibfnamefont{A.}~\bibnamefont{Nagar}},\ }%
  \bibfield{journal}{%
  \Doi{10.1103/PhysRevD.77.024043}{\bibinfo {journal} {Phys.\ Rev.\ D}}\ }%
  \textbf{\bibinfo {volume} {77}},\ \bibinfo {eid} {024043} (\bibinfo {year}
  {2008})%
  \bibAnnoteFile{NoStop}{Damour2007a}%
\bibitem{DN2007b}%
  \BibitemOpen
  \bibfield{author}{%
  \bibinfo {author} {\bibfnamefont{T.}~\bibnamefont{Damour}}, \bibinfo {author}
  {\bibfnamefont{A.}~\bibnamefont{Nagar}}, \bibinfo {author}
  {\bibfnamefont{E.~N.}\ \bibnamefont{Dorband}}, \bibinfo {author}
  {\bibfnamefont{D.}~\bibnamefont{Pollney}},\ and\ \bibinfo {author}
  {\bibfnamefont{L.}~\bibnamefont{Rezzolla}},\ }%
  \bibfield{journal}{%
  \Doi{10.1103/PhysRevD.77.084017}{\bibinfo {journal} {Phys.\ Rev.\ D}}\ }%
  \textbf{\bibinfo {volume} {77}},\ \bibinfo {eid} {084017} (\bibinfo {year}
  {2008})%
  \bibAnnoteFile{NoStop}{DN2007b}%
\bibitem{DN2008}%
  \BibitemOpen
  \bibfield{author}{%
  \bibinfo {author} {\bibfnamefont{T.}~\bibnamefont{{Damour}}}, \bibinfo
  {author} {\bibfnamefont{A.}~\bibnamefont{{Nagar}}}, \bibinfo {author}
  {\bibfnamefont{M.}~\bibnamefont{{Hannam}}}, \bibinfo {author}
  {\bibfnamefont{S.}~\bibnamefont{{Husa}}},\ and\ \bibinfo {author}
  {\bibfnamefont{B.}~\bibnamefont{{Br{\"u}gmann}}},\ }%
  \bibfield{journal}{%
  \bibinfo {journal} {Phys.\ Rev.\ D}\ }%
  \textbf{\bibinfo {volume} {78}},\ \bibinfo {pages} {044039} (\bibinfo {year}
  {2008})%
  \bibAnnoteFile{NoStop}{DN2008}%
\bibitem{DIN}%
  \BibitemOpen
  \bibfield{author}{%
  \bibinfo {author} {\bibfnamefont{T.}~\bibnamefont{{Damour}}}, \bibinfo
  {author} {\bibfnamefont{B.~R.}\ \bibnamefont{{Iyer}}},\ and\ \bibinfo
  {author} {\bibfnamefont{A.}~\bibnamefont{{Nagar}}},\ }%
  \bibfield{journal}{%
  \bibinfo {journal} {Phys.\ Rev.\ D}\ }%
  \textbf{\bibinfo {volume} {79}},\ \bibinfo {pages} {064004} (\bibinfo {year}
  {2009})%
  \bibAnnoteFile{NoStop}{DIN}%
\bibitem{Damour2009a}%
  \BibitemOpen
  \bibfield{author}{%
  \bibinfo {author} {\bibfnamefont{T.}~\bibnamefont{Damour}}\ and\ \bibinfo
  {author} {\bibfnamefont{A.}~\bibnamefont{Nagar}},\ }%
  \bibfield{journal}{%
  \bibinfo {journal} {Phys.\ Rev.\ D}\ }%
  \textbf{\bibinfo {volume} {79}},\ \bibinfo {pages} {081503} (\bibinfo {year}
  {2009})%
  \bibAnnoteFile{NoStop}{Damour2009a}%
\bibitem{Bernuzzi:2010xj}%
  \BibitemOpen
  \bibfield{author}{%
  \bibinfo {author} {\bibfnamefont{S.}~\bibnamefont{Bernuzzi}}, \bibinfo
  {author} {\bibfnamefont{A.}~\bibnamefont{Nagar}},\ and\ \bibinfo {author}
  {\bibfnamefont{A.}~\bibnamefont{Zenginoglu}},\ }%
  \bibfield{journal}{%
  \bibinfo {journal} {Phys.Rev.}\ }%
  \textbf{\bibinfo {volume} {D83}},\ \bibinfo {pages} {064010} (\bibinfo {year}
  {2011}),\ \Eprint{http://arxiv.org/abs/1012.2456}{arXiv:1012.2456 [gr-qc]}%
  \bibAnnoteFile{NoStop}{Bernuzzi:2010xj}%
\bibitem{Abadie:2011kd}%
  \BibitemOpen
  \bibfield{author}{%
  \bibinfo {author} {\bibfnamefont{J.}~\bibnamefont{Abadie}} \emph{et~al.}
  (\bibinfo {collaboration} {The LIGO Scientific and the Virgo})}%
   (\bibinfo {year} {2011}),\
  \Eprint{http://arxiv.org/abs/1102.3781}{arXiv:1102.3781 [gr-qc]}%
  \bibAnnoteFile{NoStop}{Abadie:2011kd}%
\bibitem{Scheel2009}%
  \BibitemOpen
  \bibfield{author}{%
  \bibinfo {author} {\bibnamefont{{M. Scheel, M. Boyle, T. Chu, L. Kidder, K.
  Matthews and H. Pfeiffer}}},\ }%
  \bibfield{journal}{%
  \bibinfo {journal} {Phys.\ Rev.\ D}\ }%
  \textbf{\bibinfo {volume} {79}},\ \bibinfo {pages} {024003} (\bibinfo {year}
  {2009}),\
  \Eprint{http://arxiv.org/abs/arXiv:gr-qc/0810.1767}{arXiv:gr-qc/0810.1767}%
  \bibAnnoteFile{NoStop}{Scheel2009}%
\bibitem{Szilagyi:2009qz}%
  \BibitemOpen
  \bibfield{author}{%
  \bibinfo {author} {\bibfnamefont{B.}~\bibnamefont{Szilagyi}}, \bibinfo
  {author} {\bibfnamefont{L.}~\bibnamefont{Lindblom}},\ and\ \bibinfo {author}
  {\bibfnamefont{M.~A.}\ \bibnamefont{Scheel}},\ }%
  \bibfield{journal}{%
  \bibinfo {journal} {Phys.\ Rev.\ D}\ }%
  \textbf{\bibinfo {volume} {80}},\ \bibinfo {pages} {124010} (\bibinfo {year}
  {2009}),\ \Eprint{http://arxiv.org/abs/0909.3557}{arXiv:0909.3557 [gr-qc]}%
  \bibAnnoteFile{NoStop}{Szilagyi:2009qz}%
\bibitem{Boyle2007}%
  \BibitemOpen
  \bibfield{author}{%
  \bibinfo {author} {\bibfnamefont{M.}~\bibnamefont{Boyle}}, \bibinfo {author}
  {\bibfnamefont{D.~A.}\ \bibnamefont{Brown}}, \bibinfo {author}
  {\bibfnamefont{L.~E.}\ \bibnamefont{Kidder}}, \bibinfo {author}
  {\bibfnamefont{A.~H.}\ \bibnamefont{Mrou{\'e}}}, \bibinfo {author}
  {\bibfnamefont{H.~P.}\ \bibnamefont{Pfeiffer}}, \bibinfo {author}
  {\bibfnamefont{M.~A.}\ \bibnamefont{Scheel}}, \bibinfo {author}
  {\bibfnamefont{G.~B.}\ \bibnamefont{Cook}},\ and\ \bibinfo {author}
  {\bibfnamefont{S.~A.}\ \bibnamefont{Teukolsky}},\ }%
  \bibfield{journal}{%
  \Doi{10.1103/PhysRevD.76.124038}{\bibinfo {journal} {Phys.\ Rev.\ D}}\ }%
  \textbf{\bibinfo {volume} {76}},\ \bibinfo {eid} {124038} (\bibinfo {year}
  {2007})%
  \bibAnnoteFile{NoStop}{Boyle2007}%
\bibitem{Pfeiffer-Brown-etal:2007}%
  \BibitemOpen
  \bibfield{author}{%
  \bibinfo {author} {\bibfnamefont{H.~P.}\ \bibnamefont{Pfeiffer}}, \bibinfo
  {author} {\bibfnamefont{D.~A.}\ \bibnamefont{Brown}}, \bibinfo {author}
  {\bibfnamefont{L.~E.}\ \bibnamefont{Kidder}}, \bibinfo {author}
  {\bibfnamefont{L.}~\bibnamefont{Lindblom}}, \bibinfo {author}
  {\bibfnamefont{G.}~\bibnamefont{Lovelace}},\ and\ \bibinfo {author}
  {\bibfnamefont{M.~A.}\ \bibnamefont{Scheel}},\ }%
  \bibfield{journal}{%
  \bibinfo {journal} {Class.\ Quantum Grav.}\ }%
  \textbf{\bibinfo {volume} {24}},\ \bibinfo {pages} {S59} (\bibinfo {year}
  {2007})%
  \bibAnnoteFile{NoStop}{Pfeiffer-Brown-etal:2007}%
\bibitem{Lindblom2006}%
  \BibitemOpen
  \bibfield{author}{%
  \bibinfo {author} {\bibfnamefont{L.}~\bibnamefont{Lindblom}}, \bibinfo
  {author} {\bibfnamefont{M.~A.}\ \bibnamefont{Scheel}}, \bibinfo {author}
  {\bibfnamefont{L.~E.}\ \bibnamefont{Kidder}}, \bibinfo {author}
  {\bibfnamefont{R.}~\bibnamefont{Owen}},\ and\ \bibinfo {author}
  {\bibfnamefont{O.}~\bibnamefont{Rinne}},\ }%
  \bibfield{journal}{%
  \bibinfo {journal} {Class.\ Quantum Grav.}\ }%
  \textbf{\bibinfo {volume} {23}},\ \bibinfo {pages} {S447} (\bibinfo {year}
  {2006})%
  \bibAnnoteFile{NoStop}{Lindblom2006}%
\bibitem{Scheel2006}%
  \BibitemOpen
  \bibfield{author}{%
  \bibinfo {author} {\bibfnamefont{M.~A.}\ \bibnamefont{Scheel}}, \bibinfo
  {author} {\bibfnamefont{H.~P.}\ \bibnamefont{Pfeiffer}}, \bibinfo {author}
  {\bibfnamefont{L.}~\bibnamefont{Lindblom}}, \bibinfo {author}
  {\bibfnamefont{L.~E.}\ \bibnamefont{Kidder}}, \bibinfo {author}
  {\bibfnamefont{O.}~\bibnamefont{Rinne}},\ and\ \bibinfo {author}
  {\bibfnamefont{S.~A.}\ \bibnamefont{Teukolsky}},\ }%
  \bibfield{journal}{%
  \bibinfo {journal} {Phys.\ Rev.\ D}\ }%
  \textbf{\bibinfo {volume} {74}},\ \bibinfo {pages} {104006} (\bibinfo {year}
  {2006})%
  \bibAnnoteFile{NoStop}{Scheel2006}%
\bibitem{SpECwebsite}%
  \BibitemOpen
  \bibinfo {howpublished} {\url{http://www.black-holes.org/SpEC.html}}%
  \bibAnnoteFile{NoStop}{SpECwebsite}%
\bibitem{Buchman-etal-in-prep}%
  \BibitemOpen
  \bibfield{author}{%
  \bibinfo {author} {\bibfnamefont{L.~T.}\ \bibnamefont{Buchman}}, \bibinfo
  {author} {\bibfnamefont{H.~P.}\ \bibnamefont{Pfeiffer}},\ and\ \bibinfo
  {author} {\bibfnamefont{M.~A.}\ \bibnamefont{Scheel}},\ }%
  \enquote{\bibinfo {title} {Simulations of non-equal mass black hole
  binaries},}\ \bibinfo {note} {In preparation}%
  \bibAnnoteFile{NoStop}{Buchman-etal-in-prep}%
\bibitem{Rinne2008b}%
  \BibitemOpen
  \bibfield{author}{%
  \bibinfo {author} {\bibfnamefont{O.}~\bibnamefont{Rinne}}, \bibinfo {author}
  {\bibfnamefont{L.~T.}\ \bibnamefont{Buchman}}, \bibinfo {author}
  {\bibfnamefont{M.~A.}\ \bibnamefont{Scheel}},\ and\ \bibinfo {author}
  {\bibfnamefont{H.~P.}\ \bibnamefont{Pfeiffer}},\ }%
  \bibfield{journal}{%
  \bibinfo {journal} {Class.\ Quantum Grav.}\ }%
  \textbf{\bibinfo {volume} {26}},\ \bibinfo {pages} {075009} (\bibinfo {year}
  {2009})%
  \bibAnnoteFile{NoStop}{Rinne2008b}%
\bibitem{Boyle-Mroue:2008}%
  \BibitemOpen
  \bibfield{author}{%
  \bibinfo {author} {\bibfnamefont{M.}~\bibnamefont{Boyle}}\ and\ \bibinfo
  {author} {\bibfnamefont{A.~H.}\ \bibnamefont{Mrou{\'{e}}}},\ }%
  \bibfield{journal}{%
  \Doi{10.1103/PhysRevD.80.124045}{\bibinfo {journal} {Phys.\ Rev.\ D}}\ }%
  \textbf{\bibinfo {volume} {80}},\ \bibinfo {pages} {124045} (\bibinfo {month}
  {Dec.}\ \bibinfo {year} {2009}),\
  \Eprint{http://arxiv.org/abs/0905.3177}{arXiv:0905.3177 [gr-qc]},\
  \url{http://link.aps.org/abstract/PRD/v80/e124045}%
  \bibAnnoteFile{NoStop}{Boyle-Mroue:2008}%
\bibitem{Damour03}%
  \BibitemOpen
  \bibfield{author}{%
  \bibinfo {author} {\bibfnamefont{T.}~\bibnamefont{Damour}}, \bibinfo {author}
  {\bibfnamefont{B.~R.}\ \bibnamefont{Iyer}}, \bibinfo {author}
  {\bibfnamefont{P.}~\bibnamefont{Jaranowski}},\ and\ \bibinfo {author}
  {\bibfnamefont{B.~S.}\ \bibnamefont{Sathyaprakash}},\ }%
  \bibfield{journal}{%
  \Doi{10.1103/PhysRevD.67.064028}{\bibinfo {journal} {Phys.\ Rev.\ D}}\ }%
  \textbf{\bibinfo {volume} {67}},\ \bibinfo {pages} {064028} (\bibinfo {year}
  {2003})%
  \bibAnnoteFile{NoStop}{Damour03}%
\bibitem{Damour2007}%
  \BibitemOpen
  \bibfield{author}{%
  \bibinfo {author} {\bibfnamefont{T.}~\bibnamefont{Damour}}\ and\ \bibinfo
  {author} {\bibfnamefont{A.}~\bibnamefont{Nagar}},\ }%
  \bibfield{journal}{%
  \Doi{10.1103/PhysRevD.76.064028}{\bibinfo {journal} {Phys.\ Rev.\ D}}\ }%
  \textbf{\bibinfo {volume} {76}},\ \bibinfo {eid} {064028} (\bibinfo {year}
  {2007})%
  \bibAnnoteFile{NoStop}{Damour2007}%
\bibitem{Damour:2007cb}%
  \BibitemOpen
  \bibfield{author}{%
  \bibinfo {author} {\bibfnamefont{T.}~\bibnamefont{Damour}}\ and\ \bibinfo
  {author} {\bibfnamefont{A.}~\bibnamefont{Nagar}},\ }%
  \bibfield{journal}{%
  \bibinfo {journal} {Phys.\ Rev.\ D}\ }%
  \textbf{\bibinfo {volume} {76}},\ \bibinfo {pages} {044003} (\bibinfo {year}
  {2007})%
  \bibAnnoteFile{NoStop}{Damour:2007cb}%
\bibitem{Damour:2009sm}%
  \BibitemOpen
  \bibfield{author}{%
  \bibinfo {author} {\bibfnamefont{T.}~\bibnamefont{Damour}},\ }%
  \bibfield{journal}{%
  \Doi{10.1103/PhysRevD.81.024017}{\bibinfo {journal} {Phys. Rev.}}\ }%
  \textbf{\bibinfo {volume} {D81}},\ \bibinfo {pages} {024017} (\bibinfo {year}
  {2010}),\ \Eprint{http://arxiv.org/abs/0910.5533}{arXiv:0910.5533 [gr-qc]}%
  \bibAnnoteFile{NoStop}{Damour:2009sm}%
\bibitem{Blanchet:2010zd}%
  \BibitemOpen
  \bibfield{author}{%
  \bibinfo {author} {\bibfnamefont{L.}~\bibnamefont{Blanchet}}, \bibinfo
  {author} {\bibfnamefont{S.~L.}\ \bibnamefont{Detweiler}}, \bibinfo {author}
  {\bibfnamefont{A.}~\bibnamefont{Le~Tiec}},\ and\ \bibinfo {author}
  {\bibfnamefont{B.~F.}\ \bibnamefont{Whiting}},\ }%
  \bibfield{journal}{%
  \Doi{10.1103/PhysRevD.81.084033}{\bibinfo {journal} {Phys. Rev.}}\ }%
  \textbf{\bibinfo {volume} {D81}},\ \bibinfo {pages} {084033} (\bibinfo {year}
  {2010}),\ \Eprint{http://arxiv.org/abs/1002.0726}{arXiv:1002.0726 [gr-qc]}%
  \bibAnnoteFile{NoStop}{Blanchet:2010zd}%
\bibitem{Yunes:2009ef}%
  \BibitemOpen
  \bibfield{author}{%
  \bibinfo {author} {\bibfnamefont{N.}~\bibnamefont{Yunes}}, \bibinfo {author}
  {\bibfnamefont{A.}~\bibnamefont{Buonanno}}, \bibinfo {author}
  {\bibfnamefont{S.~A.}\ \bibnamefont{Hughes}}, \bibinfo {author}
  {\bibfnamefont{M.}~\bibnamefont{Coleman~Miller}},\ and\ \bibinfo {author}
  {\bibfnamefont{Y.}~\bibnamefont{Pan}},\ }%
  \bibfield{journal}{%
  \bibinfo {journal} {Phys. Rev. Lett.}\ }%
  \textbf{\bibinfo {volume} {104}},\ \bibinfo {pages} {091102} (\bibinfo {year}
  {2010}),\ \Eprint{http://arxiv.org/abs/0909.4263}{arXiv:0909.4263 [gr-qc]}%
  \bibAnnoteFile{NoStop}{Yunes:2009ef}%
\bibitem{Yunes:2010zj}%
  \BibitemOpen
  \bibfield{author}{%
  \bibinfo {author} {\bibfnamefont{N.}~\bibnamefont{Yunes}} \emph{et~al.},\ }%
  \bibfield{journal}{%
  \bibinfo {journal} {Phys. Rev.}\ }%
  \textbf{\bibinfo {volume} {D83}},\ \bibinfo {pages} {044044} (\bibinfo {year}
  {2011}),\ \Eprint{http://arxiv.org/abs/1009.6013}{arXiv:1009.6013 [gr-qc]}%
  \bibAnnoteFile{NoStop}{Yunes:2010zj}%
\bibitem{Schnittman2007}%
  \BibitemOpen
  \bibfield{author}{%
  \bibinfo {author} {\bibfnamefont{J.~D.}\ \bibnamefont{{Schnittman}}},
  \bibinfo {author} {\bibfnamefont{A.}~\bibnamefont{{Buonanno}}}, \bibinfo
  {author} {\bibfnamefont{J.~R.}\ \bibnamefont{{van Meter}}}, \bibinfo {author}
  {\bibfnamefont{J.~G.}\ \bibnamefont{{Baker}}}, \bibinfo {author}
  {\bibfnamefont{W.~D.}\ \bibnamefont{{Boggs}}}, \bibinfo {author}
  {\bibfnamefont{J.}~\bibnamefont{{Centrella}}}, \bibinfo {author}
  {\bibfnamefont{B.~J.}\ \bibnamefont{{Kelly}}},\ and\ \bibinfo {author}
  {\bibfnamefont{S.~T.}\ \bibnamefont{{McWilliams}}},\ }%
  \bibfield{journal}{%
  \bibinfo {journal} {Phys.\ Rev.\ D}\ }%
  \textbf{\bibinfo {volume} {77}},\ \bibinfo {pages} {044031} (\bibinfo {year}
  {2008})%
  \bibAnnoteFile{NoStop}{Schnittman2007}%
\bibitem{Baker2008a}%
  \BibitemOpen
  \bibfield{author}{%
  \bibinfo {author} {\bibfnamefont{J.~G.}\ \bibnamefont{{Baker}}}, \bibinfo
  {author} {\bibfnamefont{W.~D.}\ \bibnamefont{{Boggs}}}, \bibinfo {author}
  {\bibfnamefont{J.}~\bibnamefont{{Centrella}}}, \bibinfo {author}
  {\bibfnamefont{B.~J.}\ \bibnamefont{{Kelly}}}, \bibinfo {author}
  {\bibfnamefont{S.~T.}\ \bibnamefont{{McWilliams}}},\ and\ \bibinfo {author}
  {\bibfnamefont{J.~R.}\ \bibnamefont{{van Meter}}},\ }%
  \bibfield{journal}{%
  \bibinfo {journal} {Phys.\ Rev.\ D}\ }%
  \textbf{\bibinfo {volume} {78}},\ \bibinfo {pages} {044046} (\bibinfo {month}
  {Aug.}\ \bibinfo {year} {2008})%
  \bibAnnoteFile{NoStop}{Baker2008a}%
\bibitem{Bernuzzi:2010ty}%
  \BibitemOpen
  \bibfield{author}{%
  \bibinfo {author} {\bibfnamefont{S.}~\bibnamefont{Bernuzzi}}\ and\ \bibinfo
  {author} {\bibfnamefont{A.}~\bibnamefont{Nagar}},\ }%
  \bibfield{journal}{%
  \bibinfo {journal} {Phys.Rev.}\ }%
  \textbf{\bibinfo {volume} {D81}},\ \bibinfo {pages} {084056} (\bibinfo {year}
  {2010}),\ \Eprint{http://arxiv.org/abs/1003.0597}{arXiv:1003.0597 [gr-qc]}%
  \bibAnnoteFile{NoStop}{Bernuzzi:2010ty}%
\bibitem{Damour06}%
  \BibitemOpen
  \bibfield{author}{%
  \bibinfo {author} {\bibfnamefont{T.}~\bibnamefont{Damour}}\ and\ \bibinfo
  {author} {\bibfnamefont{A.}~\bibnamefont{Gopakumar}},\ }%
  \bibfield{journal}{%
  \Doi{10.1103/PhysRevD.73.124006}{\bibinfo {journal} {Phys.\ Rev.\ D}}\ }%
  \textbf{\bibinfo {volume} {73}},\ \bibinfo {eid} {124006} (\bibinfo {year}
  {2006})%
  \bibAnnoteFile{NoStop}{Damour06}%
\bibitem{Berti2006a}%
  \BibitemOpen
  \bibfield{author}{%
  \bibinfo {author} {\bibfnamefont{E.}~\bibnamefont{Berti}}, \bibinfo {author}
  {\bibfnamefont{V.}~\bibnamefont{Cardoso}},\ and\ \bibinfo {author}
  {\bibfnamefont{C.~M.}\ \bibnamefont{Will}},\ }%
  \bibfield{journal}{%
  \bibinfo {journal} {Phys.\ Rev.\ D}\ }%
  \textbf{\bibinfo {volume} {73}},\ \bibinfo {pages} {064030} (\bibinfo {year}
  {2006})%
  \bibAnnoteFile{NoStop}{Berti2006a}%
\bibitem{ReggeWheeler1957}%
  \BibitemOpen
  \bibfield{author}{%
  \bibinfo {author} {\bibfnamefont{T.}~\bibnamefont{Regge}}\ and\ \bibinfo
  {author} {\bibfnamefont{J.~A.}\ \bibnamefont{Wheeler}},\ }%
  \bibfield{journal}{%
  \bibinfo {journal} {Phys.\ Rev.}\ }%
  \textbf{\bibinfo {volume} {108}},\ \bibinfo {pages} {1063} (\bibinfo {year}
  {1957})%
  \bibAnnoteFile{NoStop}{ReggeWheeler1957}%
\bibitem{Zerilli1970b}%
  \BibitemOpen
  \bibfield{author}{%
  \bibinfo {author} {\bibfnamefont{F.~J.}\ \bibnamefont{Zerilli}},\ }%
  \bibfield{journal}{%
  \bibinfo {journal} {Phys.\ Rev.\ Lett.}\ }%
  \textbf{\bibinfo {volume} {24}},\ \bibinfo {pages} {737} (\bibinfo {year}
  {1970})%
  \bibAnnoteFile{NoStop}{Zerilli1970b}%
\bibitem{Sarbach2001}%
  \BibitemOpen
  \bibfield{author}{%
  \bibinfo {author} {\bibfnamefont{O.}~\bibnamefont{Sarbach}}\ and\ \bibinfo
  {author} {\bibfnamefont{M.}~\bibnamefont{Tiglio}},\ }%
  \bibfield{journal}{%
  \Doi{10.1103/PhysRevD.64.084016}{\bibinfo {journal} {Phys. Rev. D}}\ }%
  \textbf{\bibinfo {volume} {64}},\ \bibinfo {pages} {084016} (\bibinfo {month}
  {Sep}\ \bibinfo {year} {2001})%
  \bibAnnoteFile{NoStop}{Sarbach2001}%
\bibitem{MacDonald:2011ne}%
  \BibitemOpen
  \bibfield{author}{%
  \bibinfo {author} {\bibfnamefont{I.}~\bibnamefont{{MacDonald}}}, \bibinfo
  {author} {\bibfnamefont{S.}~\bibnamefont{Nissanke}},\ and\ \bibinfo {author}
  {\bibfnamefont{H.~P.}\ \bibnamefont{Pfeiffer}},\ }%
  \bibfield{journal}{%
  \Doi{10.1088/0264-9381/28/13/134002}{\bibinfo {journal} {Class.\ Quantum
  Grav.}}\ }%
  \textbf{\bibinfo {volume} {28}},\ \bibinfo {pages} {134002} (\bibinfo {month}
  {Jul.}\ \bibinfo {year} {2011}),\ ISSN \bibinfo {issn} {0264-9381},\
  \Eprint{http://arxiv.org/abs/1102.5128}{arXiv:1102.5128 [gr-qc]},\
  \url{http://iopscience.iop.org/0264-9381/28/13/134002/}%
  \bibAnnoteFile{NoStop}{MacDonald:2011ne}%
\bibitem{Barausse:2011a}%
  \BibitemOpen
  \bibfield{author}{%
  \bibinfo {author} {\bibfnamefont{E.}~\bibnamefont{Barausse}} \emph{et~al.}}%
   (\bibinfo {year} {2011}),\ \bibinfo {note} {in preparation}%
  \bibAnnoteFile{NoStop}{Barausse:2011a}%
\bibitem{BarackSago09}%
  \BibitemOpen
  \bibfield{author}{%
  \bibinfo {author} {\bibfnamefont{L.}~\bibnamefont{Barack}}\ and\ \bibinfo
  {author} {\bibfnamefont{N.}~\bibnamefont{Sago}},\ }%
  \bibfield{journal}{%
  \bibinfo {journal} {Phys. Rev. Lett.}\ }%
  \textbf{\bibinfo {volume} {102}},\ \bibinfo {pages} {191101} (\bibinfo {year}
  {2009})%
  \bibAnnoteFile{NoStop}{BarackSago09}%
\bibitem{Shoemaker2009}%
  \BibitemOpen
  \bibfield{author}{%
  \bibinfo {author} {\bibfnamefont{D.}~\bibnamefont{Shoemaker}} (\bibinfo
  {collaboration} {{LIGO} Collaboration}),\ }%
  \enquote{\bibinfo {title} {Advanced {LIGO} anticipated sensitivity curves},}\
   (\bibinfo {year} {2010}),\ \bibinfo {note} {{LIGO} Document T0900288-v3},\
  \url{https://dcc.ligo.org/cgi-bin/DocDB/ShowDocument?docid=2974}%
  \bibAnnoteFile{NoStop}{Shoemaker2009}%
\bibitem{McKechan:2010kp}%
  \BibitemOpen
  \bibfield{author}{%
  \bibinfo {author} {\bibfnamefont{D.}~\bibnamefont{McKechan}}, \bibinfo
  {author} {\bibfnamefont{C.}~\bibnamefont{Robinson}},\ and\ \bibinfo {author}
  {\bibfnamefont{B.}~\bibnamefont{Sathyaprakash}},\ }%
  \bibfield{journal}{%
  \Doi{10.1088/0264-9381/27/8/084020}{\bibinfo {journal} {Class.Quant.Grav.}}\
  }%
  \textbf{\bibinfo {volume} {27}},\ \bibinfo {pages} {084020} (\bibinfo {year}
  {2010}),\ \Eprint{http://arxiv.org/abs/1003.2939}{arXiv:1003.2939 [gr-qc]}%
  \bibAnnoteFile{NoStop}{McKechan:2010kp}%
\bibitem{2008ApJ...678L..17S}%
  \BibitemOpen
  \bibfield{author}{%
  \bibinfo {author} {\bibfnamefont{J.~M.}\ \bibnamefont{{Silverman}}}\ and\
  \bibinfo {author} {\bibfnamefont{A.~V.}\ \bibnamefont{{Filippenko}}},\ }%
  \bibfield{journal}{%
  \bibinfo {journal} {Astrophys.\ J.\ Lett.}\ }%
  \textbf{\bibinfo {volume} {678}},\ \bibinfo {pages} {L17} (\bibinfo {year}
  {2008}),\ \Eprint{http://arxiv.org/abs/0802.2716}{arXiv:0802.2716}%
  \bibAnnoteFile{NoStop}{2008ApJ...678L..17S}%
\bibitem{2007ApJ...669L..21P}%
  \BibitemOpen
  \bibfield{author}{%
  \bibinfo {author} {\bibfnamefont{A.~H.}\ \bibnamefont{{Prestwich}}}, \bibinfo
  {author} {\bibfnamefont{R.}~\bibnamefont{{Kilgard}}}, \bibinfo {author}
  {\bibfnamefont{P.~A.}\ \bibnamefont{{Crowther}}}, \bibinfo {author}
  {\bibfnamefont{S.}~\bibnamefont{{Carpano}}}, \bibinfo {author}
  {\bibfnamefont{A.~M.~T.}\ \bibnamefont{{Pollock}}}, \bibinfo {author}
  {\bibfnamefont{A.}~\bibnamefont{{Zezas}}}, \bibinfo {author}
  {\bibfnamefont{S.~H.}\ \bibnamefont{{Saar}}}, \bibinfo {author}
  {\bibfnamefont{T.~P.}\ \bibnamefont{{Roberts}}},\ and\ \bibinfo {author}
  {\bibfnamefont{M.~J.}\ \bibnamefont{{Ward}}},\ }%
  \bibfield{journal}{%
  \Doi{10.1086/523755}{\bibinfo {journal} {Astrophys.\ J.\ Lett.}}\ }%
  \textbf{\bibinfo {volume} {669}},\ \bibinfo {pages} {L21} (\bibinfo {month}
  {Nov.}\ \bibinfo {year} {2007}),\
  \Eprint{http://arxiv.org/abs/0709.2892}{arXiv:0709.2892}%
  \bibAnnoteFile{NoStop}{2007ApJ...669L..21P}%
\bibitem{2010ApJ...714.1217B}%
  \BibitemOpen
  \bibfield{author}{%
  \bibinfo {author} {\bibfnamefont{K.}~\bibnamefont{{Belczynski}}}, \bibinfo
  {author} {\bibfnamefont{T.}~\bibnamefont{{Bulik}}}, \bibinfo {author}
  {\bibfnamefont{C.~L.}\ \bibnamefont{{Fryer}}}, \bibinfo {author}
  {\bibfnamefont{A.}~\bibnamefont{{Ruiter}}}, \bibinfo {author}
  {\bibfnamefont{F.}~\bibnamefont{{Valsecchi}}}, \bibinfo {author}
  {\bibfnamefont{J.~S.}\ \bibnamefont{{Vink}}},\ and\ \bibinfo {author}
  {\bibfnamefont{J.~R.}\ \bibnamefont{{Hurley}}},\ }%
  \bibfield{journal}{%
  \bibinfo {journal} {apj}\ }%
  \textbf{\bibinfo {volume} {714}},\ \bibinfo {pages} {1217} (\bibinfo {year}
  {2010}),\ \Eprint{http://arxiv.org/abs/0904.2784}{arXiv:0904.2784
  [astro-ph.SR]}%
  \bibAnnoteFile{NoStop}{2010ApJ...714.1217B}%
\bibitem{2004IJMPD..13....1C}%
  \BibitemOpen
  \bibfield{author}{%
  \bibinfo {author} {\bibfnamefont{M.}~\bibnamefont{{Coleman Miller}}}\ and\
  \bibinfo {author} {\bibfnamefont{E.~J.~M.}\ \bibnamefont{{Colbert}}},\ }%
  \bibfield{journal}{%
  \bibinfo {journal} {Int. J. Mod. Phys. D}\ }%
  \textbf{\bibinfo {volume} {13}},\ \bibinfo {pages} {1} (\bibinfo {year}
  {2004}),\
  \Eprint{http://arxiv.org/abs/arXiv:astro-ph/0308402}{arXiv:astro-ph/0308402}%
  \bibAnnoteFile{NoStop}{2004IJMPD..13....1C}%
\bibitem{2006ApJ...646L.135F}%
  \BibitemOpen
  \bibfield{author}{%
  \bibinfo {author} {\bibfnamefont{J.~M.}\ \bibnamefont{{Fregeau}}}, \bibinfo
  {author} {\bibfnamefont{S.~L.}\ \bibnamefont{{Larson}}}, \bibinfo {author}
  {\bibfnamefont{M.~C.}\ \bibnamefont{{Miller}}}, \bibinfo {author}
  {\bibfnamefont{R.}~\bibnamefont{{O'Shaughnessy}}},\ and\ \bibinfo {author}
  {\bibfnamefont{F.~A.}\ \bibnamefont{{Rasio}}},\ }%
  \bibfield{journal}{%
  \bibinfo {journal} {Astrophys.\ J.\ Lett.}\ }%
  \textbf{\bibinfo {volume} {646}},\ \bibinfo {pages} {L135} (\bibinfo {year}
  {2006}),\
  \Eprint{http://arxiv.org/abs/arXiv:astro-ph/0605732}{arXiv:astro-ph/0605732}%
  \bibAnnoteFile{NoStop}{2006ApJ...646L.135F}%
\bibitem{Buonanno:2009}%
  \BibitemOpen
  \bibfield{author}{%
  \bibinfo {author} {\bibfnamefont{A.}~\bibnamefont{Buonanno}}, \bibinfo
  {author} {\bibfnamefont{B.~R.}\ \bibnamefont{Iyer}}, \bibinfo {author}
  {\bibfnamefont{E.}~\bibnamefont{Ochsner}}, \bibinfo {author}
  {\bibfnamefont{Y.}~\bibnamefont{Pan}},\ and\ \bibinfo {author}
  {\bibfnamefont{B.~S.}\ \bibnamefont{Sathyaprakash}},\ }%
  \bibfield{journal}{%
  \Doi{10.1103/PhysRevD.80.084043}{\bibinfo {journal} {Phys. Rev. D}}\ }%
  \textbf{\bibinfo {volume} {80}},\ \bibinfo {pages} {084043} (\bibinfo {month}
  {Oct}\ \bibinfo {year} {2009})%
  \bibAnnoteFile{NoStop}{Buonanno:2009}%
\bibitem{Hannam:2010}%
  \BibitemOpen
  \bibfield{author}{%
  \bibinfo {author} {\bibfnamefont{M.}~\bibnamefont{{Hannam}}}, \bibinfo
  {author} {\bibfnamefont{S.}~\bibnamefont{{Husa}}}, \bibinfo {author}
  {\bibfnamefont{F.}~\bibnamefont{{Ohme}}},\ and\ \bibinfo {author}
  {\bibfnamefont{P.}~\bibnamefont{{Ajith}}},\ }%
  \bibfield{journal}{%
  \Doi{10.1103/PhysRevD.82.124052}{\bibinfo {journal} {Phys. Rev. D}}\ }%
  \textbf{\bibinfo {volume} {82}},\ \bibinfo {pages} {124052} (\bibinfo {year}
  {2010})%
  \bibAnnoteFile{NoStop}{Hannam:2010}%
\bibitem{Damour:2010}%
  \BibitemOpen
  \bibfield{author}{%
  \bibinfo {author} {\bibfnamefont{T.}~\bibnamefont{{Damour}}}, \bibinfo
  {author} {\bibfnamefont{A.}~\bibnamefont{{Nagar}}},\ and\ \bibinfo {author}
  {\bibfnamefont{M.}~\bibnamefont{{Trias}}},\ }%
  \bibfield{journal}{%
  \Doi{10.1103/PhysRevD.83.024006}{\bibinfo {journal} {Phys. Rev. D}}\ }%
  \textbf{\bibinfo {volume} {83}},\ \bibinfo {pages} {024006} (\bibinfo {year}
  {2011})%
  \bibAnnoteFile{NoStop}{Damour:2010}%
\bibitem{Boyle:2011dy}%
  \BibitemOpen
  \bibfield{author}{%
  \bibinfo {author} {\bibfnamefont{M.}~\bibnamefont{Boyle}},\ }%
  \bibfield{journal}{%
  \bibinfo {journal} {Phys.\ Rev.\ D}\ }%
  \textbf{\bibinfo {volume} {84}} (\bibinfo {month} {Sep.}\ \bibinfo {year}
  {2011}),\ ISSN \bibinfo {issn} {1550-7998, 1550-2368},\ \doi{\bibinfo {doi}
  {10.1103/PhysRevD.84.064013}},\
  \url{http://link.aps.org/doi/10.1103/PhysRevD.84.064013}%
  \bibAnnoteFile{NoStop}{Boyle:2011dy}%
\bibitem{Miller2005}%
  \BibitemOpen
  \bibfield{author}{%
  \bibinfo {author} {\bibfnamefont{M.~A.}\ \bibnamefont{Miller}},\ }%
  \bibfield{journal}{%
  \bibinfo {journal} {Phys.\ Rev.\ D}\ }%
  \textbf{\bibinfo {volume} {71}},\ \bibinfo {pages} {104016} (\bibinfo {year}
  {2005})%
  \bibAnnoteFile{NoStop}{Miller2005}%
\bibitem{Lindblom2008}%
  \BibitemOpen
  \bibfield{author}{%
  \bibinfo {author} {\bibfnamefont{L.}~\bibnamefont{Lindblom}}, \bibinfo
  {author} {\bibfnamefont{B.~J.}\ \bibnamefont{Owen}},\ and\ \bibinfo {author}
  {\bibfnamefont{D.~A.}\ \bibnamefont{Brown}},\ }%
  \bibfield{journal}{%
  \bibinfo {journal} {Phys. Rev. D}\ }%
  \textbf{\bibinfo {volume} {78}},\ \bibinfo {pages} {124020} (\bibinfo {year}
  {2008})%
  \bibAnnoteFile{NoStop}{Lindblom2008}%
\bibitem{Damour98}%
  \BibitemOpen
  \bibfield{author}{%
  \bibinfo {author} {\bibfnamefont{T.}~\bibnamefont{Damour}}, \bibinfo {author}
  {\bibfnamefont{B.~R.}\ \bibnamefont{Iyer}},\ and\ \bibinfo {author}
  {\bibfnamefont{B.~S.}\ \bibnamefont{Sathyaprakash}},\ }%
  \bibfield{journal}{%
  \Doi{10.1103/PhysRevD.57.885}{\bibinfo {journal} {Phys. Rev. D}}\ }%
  \textbf{\bibinfo {volume} {57}},\ \bibinfo {pages} {885} (\bibinfo {month}
  {Jan}\ \bibinfo {year} {1998})%
  \bibAnnoteFile{NoStop}{Damour98}%
\bibitem{Fujita:2010xj}%
  \BibitemOpen
  \bibfield{author}{%
  \bibinfo {author} {\bibfnamefont{R.}~\bibnamefont{Fujita}}\ and\ \bibinfo
  {author} {\bibfnamefont{B.~R.}\ \bibnamefont{Iyer}},\ }%
  \bibfield{journal}{%
  \bibinfo {journal} {Phys.Rev.}\ }%
  \textbf{\bibinfo {volume} {D82}},\ \bibinfo {pages} {044051} (\bibinfo {year}
  {2010}),\ \Eprint{http://arxiv.org/abs/1005.2266}{arXiv:1005.2266 [gr-qc]}%
  \bibAnnoteFile{NoStop}{Fujita:2010xj}%
\bibitem{Ajith-Babak-Chen-etal:2007b}%
  \BibitemOpen
  \bibfield{author}{%
  \bibinfo {author} {\bibfnamefont{P.}~\bibnamefont{Ajith}}, \bibinfo {author}
  {\bibfnamefont{S.}~\bibnamefont{Babak}}, \bibinfo {author}
  {\bibfnamefont{Y.}~\bibnamefont{Chen}}, \bibinfo {author}
  {\bibfnamefont{M.}~\bibnamefont{Hewitson}}, \bibinfo {author}
  {\bibfnamefont{B.}~\bibnamefont{Krishnan}}, \bibinfo {author}
  {\bibfnamefont{A.~M.}\ \bibnamefont{Sintes}}, \bibinfo {author}
  {\bibfnamefont{J.~T.}\ \bibnamefont{Whelan}}, \bibinfo {author}
  {\bibfnamefont{B.}~\bibnamefont{Br\"{u}gmann}}, \bibinfo {author}
  {\bibfnamefont{P.}~\bibnamefont{Diener}}, \bibinfo {author}
  {\bibfnamefont{N.}~\bibnamefont{Dorband}}, \bibinfo {author}
  {\bibfnamefont{J.}~\bibnamefont{Gonzalez}}, \bibinfo {author}
  {\bibfnamefont{M.}~\bibnamefont{Hannam}}, \bibinfo {author}
  {\bibfnamefont{S.}~\bibnamefont{Husa}}, \bibinfo {author}
  {\bibfnamefont{D.}~\bibnamefont{Pollney}}, \bibinfo {author}
  {\bibfnamefont{L.}~\bibnamefont{Rezzolla}}, \bibinfo {author}
  {\bibfnamefont{L.}~\bibnamefont{Santamar\'{\i}a}}, \bibinfo {author}
  {\bibfnamefont{U.}~\bibnamefont{Sperhake}},\ and\ \bibinfo {author}
  {\bibfnamefont{J.}~\bibnamefont{Thornburg}},\ }%
  \bibfield{journal}{%
  \bibinfo {journal} {Phys.\ Rev.\ D}\ }%
  \textbf{\bibinfo {volume} {77}},\ \bibinfo {eid} {104017} (\bibinfo {year}
  {2008})%
  \bibAnnoteFile{NoStop}{Ajith-Babak-Chen-etal:2007b}%
\bibitem{Santamaria:2010yb}%
  \BibitemOpen
  \bibfield{author}{%
  \bibinfo {author} {\bibfnamefont{L.}~\bibnamefont{Santamar\'{i}a}}, \bibinfo
  {author} {\bibfnamefont{F.}~\bibnamefont{Ohme}}, \bibinfo {author}
  {\bibfnamefont{P.}~\bibnamefont{Ajith}}, \bibinfo {author}
  {\bibfnamefont{B.}~\bibnamefont{Br{\"u}gmann}}, \bibinfo {author}
  {\bibfnamefont{N.}~\bibnamefont{Dorband}}, \bibinfo {author}
  {\bibfnamefont{M.}~\bibnamefont{Hannam}}, \bibinfo {author}
  {\bibfnamefont{S.}~\bibnamefont{Husa}}, \bibinfo {author}
  {\bibfnamefont{P.}~\bibnamefont{M{\"o}sta}}, \bibinfo {author}
  {\bibfnamefont{D.}~\bibnamefont{Pollney}}, \bibinfo {author}
  {\bibfnamefont{C.}~\bibnamefont{Reisswig}}, \bibinfo {author}
  {\bibfnamefont{E.~L.}\ \bibnamefont{Robinson}}, \bibinfo {author}
  {\bibfnamefont{J.}~\bibnamefont{Seiler}},\ and\ \bibinfo {author}
  {\bibfnamefont{B.}~\bibnamefont{Krishnan}},\ }%
  \bibfield{journal}{%
  \Doi{10.1103/PhysRevD.82.064016}{\bibinfo {journal} {Phys.\ Rev.\ D}}\ }%
  \textbf{\bibinfo {volume} {82}},\ \bibinfo {pages} {064016} (\bibinfo {year}
  {2010})%
  \bibAnnoteFile{NoStop}{Santamaria:2010yb}%
\bibitem{CutlerV:2007}%
  \BibitemOpen
  \bibfield{author}{%
  \bibinfo {author} {\bibfnamefont{C.}~\bibnamefont{{Cutler}}}\ and\ \bibinfo
  {author} {\bibfnamefont{M.}~\bibnamefont{{Vallisneri}}},\ }%
  \bibfield{journal}{%
  \Doi{10.1103/PhysRevD.76.104018}{\bibinfo {journal} {Phys. Rev. D}}\ }%
  \textbf{\bibinfo {volume} {76}},\ \bibinfo {pages} {104018} (\bibinfo {month}
  {Nov.}\ \bibinfo {year} {2007}),\
  \Eprint{http://arxiv.org/abs/0707.2982}{arXiv:0707.2982 [gr-qc]}%
  \bibAnnoteFile{NoStop}{CutlerV:2007}%
\end{thebibliography}
%
\end{document}